\documentclass{aa}

\usepackage{amsmath}
\usepackage{txfonts}
\usepackage{float}
\usepackage{graphicx}
\usepackage{subcaption}
\usepackage{longtable}
\usepackage{natbib}
\usepackage{psfig}
\usepackage{multirow}

\bibpunct{(}{)}{;}{a}{}{,}

\usepackage[colorinlistoftodos]{todonotes}
\usepackage[colorlinks=true, allcolors=blue]{hyperref}
\usepackage{siunitx}

\newcommand{\rms}{r{.}m{.}s{.}~}
\newcommand{\resp}{resp{.}}
\newcommand{\ie}{i{.}e{.}~}

\newcommand{\eq}{Eq{.}~}                 
\newcommand{\eqs}{Eqs{.}~}
  
\newcommand{\fg}{Fig{.}~}
\newcommand{\fgs}{Figs{.}~}
\newcommand{\sct}{Sect{.}~}
\newcommand{\scts}{Sects{.}~}

\definecolor{darkgreen}{rgb}{0.0,0.5,0.0}
\definecolor{darkred}{rgb}{0.7,0.0,0.0}
\definecolor{brown}{rgb}{0.65,.35,0.}
\definecolor{grey}{rgb}{0.4,0.5,0.55}
\definecolor{royalblue}{rgb}{0,0.5,1}
\definecolor{violet}{rgb}{0.5,0,0.7}
\definecolor{lightgrey}{rgb}{0.85,0.9,0.95}

\usepackage{soul}
\begin{document}

\title{
  Tailoring galaxies: Size-luminosity-surface brightness relations of bulges and disks along the morphological sequence}
   \author{Quilley, L.\thanks{email: louis.quilley@iap.fr} de Lapparent, V.\thanks{email: valerie.de\_lapparent@iap.fr}}

\titlerunning{Revisiting the scaling relations of galaxies}
\authorrunning{Quilley, de Lapparent}

   \institute{Institut d'Astrophysique de Paris, CNRS,
    Sorbonne Universit\'e, 98 bis boulevard Arago, 75014 Paris, France
}
  
   \date{Received 28 April 2023 / Accepted 22 August 2023}
  \abstract
% context
{}
% aims 
{We revisit the scaling relations between size, luminosity, and surface brightness as a function of morphology, for the bulge and disk components of the 3106 weakly inclined galaxies of the ``Extraction de Formes Id\'ealis\'ees de Galaxies en Imagerie'' (EFIGI) sample, in the nearby Universe.}
% methods
{The luminosity profiles from the Sloan Digital Sky Survey (SDSS) $gri$ images were modeled as the sum of a S\'ersic (bulge) and an exponential (disk) component for cD, elliptical (E), lenticular, and spiral galaxies, or as a single S\'ersic profile for cD, E, dE, and irregular (Im) galaxies, by controlled profile fitting with the SourceXtractor++ software.}
% results
{ For the EFIGI sample, we remeasured the Kormendy (1977) relation between effective surface brightness $\langle\mu\rangle_e$ and effective radius $R_e$ of elliptical galaxies, and show that it is also valid for the bulges (or S\'ersic components) of galaxy types Sb and earlier. In contrast, there is a progressive departure toward fainter and smaller bulges for later Hubble types, as well as with decreasing bulge-to-total ratios ($B/T$) and S\'ersic indices. This depicts  a continuous transition from pseudo-bulges to classical ones, which we suggest to occur for absolute $g$ magnitudes $M_g$ between $-17.8$ and $-19.1$. We also obtain partial agreement with the Binggeli \textit{et al{.}} (1984) relations between effective radius and $M_g$ (known as ``size-luminosity'' relations, in log-log scale) for E and dE galaxies. There is a convex size-luminosity relation for the bulges of all EFIGI types. Both $\langle\mu\rangle_e-R_e$ and $R_e-M_g$ scaling relations are projections of a plane in which bulges are located according to their value of $B/T$, which partly determines the morphological type. Analogous scaling relations were derived for the disks of lenticular and spiral types, and the irregulars. The curvature of the size-luminosity relation for disks is such that while they grow, they first brighten and then stabilize in surface brightness. Moreover, we obtain the unprecedented result that the effective radii of both the bulges and disks of lenticular and spiral galaxies increase as power laws of $B/T$, with a steeper increase for the bulges. Both bulges and disks of lenticular galaxies have a similar and largely steeper increase with $B/T$ than those for spirals. These relations propagate into a single scaling relation for the disk-to-bulge ratio of effective radii across $\sim 2$ orders of magnitude in $B/T$, and for all types. We provide the parameters of all of these relations that can be used to build realistic mock images of nearby galaxies. The new convex size-luminosity relations are more reliable estimates of bulge, disk, and galaxy sizes at all magnitudes in the nearby Universe.}
% conclusion
{This analysis describes the joint size and luminosity variations of bulges and disks along the Hubble sequence. The characteristics of the successive phases of disk and bulge size growth strengthen a picture of morphological evolution in which irregulars and late spirals merge to form earlier spirals, lenticulars, and eventually ellipticals.}


\maketitle

\section{Introduction
\label{sct-intro}}

In the Hubble sequence \citep{Hubble-1926-extragalactic-nebulae}, spiral and lenticular galaxies are made of two components: a bulge and a disk, with the former being in the center of the latter. In order to better understand the formation and evolution of these galaxies, astronomers have naturally looked at the properties of each component, as well as the relations between the two. In order to do so, the bulge and disk decomposition have become a standard approach to model galaxy images \citep{Simien-1986-BT-ratios, De-Jong-1996, Allen-2006-MGC-BD-decomp, Simard-2011-BD-decomp-SDSS}. Bulges are usually composed of old stellar populations, hence their redder colors, while disks are bluer because they are the major loci of star formation. Spiral and lenticular disks differ due to the presence or absence of spiral arms, respectively, as well as their significant or weak star formation, respectively. Another difference that partly characterizes galaxy types are the fractions of total light of the galaxy that bulges and disks enclose, with earlier Hubble types (from late spirals to lenticular) having more prominent bulges on average \citep{Quilley-2022-bimodality}.

Kinematics analyses have shown that early-type galaxies can be separated into two classes of objects between slow and fast rotators \citep{Emsellem-2007-kinematics-classification-ETG, Emsellem-2011-fast-slow-rotators}, depending on the overall level of rotation of their stellar orbits. Similarly, the bulges nested at the centers of lenticular and spiral galaxies can also be differentiated based on their kinematics. Classical bulges are nearly spherical in shape, dominantly pressure-supported, and built by violent relaxation events such as mergers, while pseudo-bulges are flatter in shape and mainly rotation-supported \citep{Kormendy-Kennicutt-2004-bulges-disk-galaxies-review}. Among the pseudo-bulges, one can further differentiate disky bulges that are within the disk and built through secular evolution, from boxy/peanut bulges that also show circular orbits but are vertically more extended than the disk, and thought to be built by the buckling of the bar \citep{Athanassoula-2005-nature-of-bulges-n-body-simul, Athanassoula-2008-pseudo-bulges, Athanassoula-2013-bars-secular-evolution, Debattista-2006-secular-evolution-disk}. 

Classical bulges show similar properties to elliptical galaxies, which are also pressure-supported systems \citep{Davies-1983-dynamics-ellipticals-bulges}, and tend to be oblate \citep{Costantin-2018-shape-bulges-CALIFA}. Elliptical galaxies can be characterized by a scaling relation between their effective surface brightness and effective radius across a large range in luminosities, as shown by the so-called Kormendy relation \citep{Kormendy-1977-II-kormendy-relation}: if the $G$ absolute magnitude interval is limited to $\sim[-22.1;-19.75]$ (see \sct\ref{sct-results-kormendy} for details), and the interval of effective radius only spans $\sim1.2$ dex, the surface brightness describes a very large interval of nearly 4 magnitudes per arcsec$^2$ ($\sim[19.5;23.5]$). Across a much wider range in absolute $B_T$ magnitude ($\sim[-23.5;-12.0]$\footnote{If one excludes dwarf spheroidal galaxies.}), and a similar interval of $\sim1.5$ dex in effective radius, \citet{Binggeli-1984} derived another relation between the absolute magnitude and radius of elliptical galaxies, hence its name, the size-luminosity relation\footnote{Hereafter, although not mentioned, the size-luminosity relation is always considered in log-log scale, as it may be linear in this coordinate plane.}. The characteristic surface brightnesses of the disks of lenticular and spiral galaxies span a similar $\sim3-4$ magnitude range, with indications of a scaling with effective radius \citep{De-Jong-1996, Simard-2011-BD-decomp-SDSS}.

In this article, we revisit these relations with larger statistics using the sample of nearby, well-resolved galaxies extracted from the Sloan Digital Sky Survey (SDSS) images to create the ``Extraction de Formes Id\'ealis\'ees de Galaxies en Imagerie'' (EFIGI) catalog with visual morphological classification \citep{Baillard-2011-EFIGI}. Our goal is to examine whether the scaling relations of \citet{Kormendy-1977-II-kormendy-relation} and \citet{Binggeli-1984} also apply for bulges of lenticulars and spirals, and whether there are similar scaling relations for their disks. In \citet{Quilley-2022-bimodality}, we performed bulge and disk decomposition of all EFIGI galaxies in order to obtain a reliable description of both galaxy components and to study the role of morphology in galaxy evolution. We suggest that the Hubble sequence is an inverse sequence of galaxy physical evolution driven by mergers and dominated by bulge growth and disk quenching, which can be characterized quantitatively using the bulge-to-total ratio and the disk color. We also show that as galaxies evolve along the Hubble sequence, they cross the Green Valley (the transition region between the Blue Cloud and the Red Sequence) which spreads between Sab up to S0$^+$ morphological types, in which the star formation of galaxies fades the fastest with respect to the change in morphological type, and the bulge profiles become steeper \citep{Quilley-2022-bimodality}.

The new analysis allows us to further examine how the bulges and disks of the various Hubble types change in their light profile as they grow in mass and progressively halt their star formation (commonly referred to as ``quenching''). \citet{Drory-Fisher-2007-bulge-bimodality} and \citet{Gadotti-2009-bulge-structure-SDSS} have proposed  using the bulge light profiles to differentiate among the classical and pseudo-bulges (steeper for the former than for the latter) in lenticular and spiral galaxies. As these changes are symptomatic of the different processes driving the evolution of bulges and disks, they may shed further light on the evolution of the Hubble morphological sequence. Examining the evolution of the size-luminosity relation with cosmic time can provide insight into these issues by confronting the mass growth of galaxies with their size growth \citep{Trujillo-2004-size-luminosity-mass-z-3, Brooks-2011-size-luminosity-z-0-1, Grazian-2012-size-luminosity-z-7-CANDELS-reionization, Kawamata-2018-UVLF-size-lum-relation-HubbleFF, Yang-2022-size-lum-rel-lensed-galaxies, Yang-2022-size-lum-rel-JWST}. 
Time evolution of the size-luminosity relations are directly accessible from numerical simulations, and several studies have highlighted discrepancies between the sizes of observed and simulated galaxies \citep{Joung-2009-simulated-galaxies-too-small-z-3, Bottrell-2017-Illustris-size-luminosity}.

To fully describe the light profile of a galaxy, or its bulge or disk, one needs at least parameters characterizing the size of the profile, its level of flux, its steepness, and its ellipticity. We here focus on relations between parameters describing the light profiles of galaxies; however, for the sake of interpretation, it is important to note that this still carries information about the stellar mass distribution due to their strong correlation \citep{Quilley-2022-bimodality}. Replacing the absolute magnitude by the stellar mass or the surface brightness by the stellar mass density does indeed lead to similar scaling laws for elliptical galaxies and bulges \citep{Gao-2022-bulges-CIGS-mass-scaling-relations}.

The present article is structured as follows. In \sct \ref{sct-data}, we present the data used in this study. In \sct \ref{sct-methodo}, we detail the methodology used to perform disk and bulge decomposition (\sct\ref{sct-methodo-srx}) with the SourceXtractor++ software \citep{Bertin-2020-SourceXtractor-plus-plus}, as well as of the spectral energy distribution (SED) model-fitting (\sct\ref{sct-methodo-zpeg}) using the ZPEG software \citep{Le-Borgne-2002-ZPEG}. We then present the analytical expressions for deriving the surface brightnesses and physical effective radii of the S\'ersic and exponential profiles used to model the galaxy components (\sct\ref{sct-methodo-math}), and the technical approaches used to perform the fits (\scts\ref{sct-methodo-uncertainties} and \ref{sct-methodo-odr}). In \sct\ref{sct-results}, we then analyze our results for the EFIGI scaling relations of surface brightness versus effective radii (\sct\ref{sct-results-kormendy}), and effective radii versus magnitude (\sct\ref{sct-results-binggeli}), and compare them with the original relations of \citet{Kormendy-1977-II-kormendy-relation} and \citet{Binggeli-1984}, respectively. We also show how these three quantities lie on a plane in the 3D parameter space (\sct\ref{sct-results-3D-space}). We bring to light a size-luminosity relation for bulges of lenticular and spiral galaxies (\sct\ref{sct-results-binggeli-bulge}), as well as for their disks (\sct\ref{sct-results-disk-scaling-relations}), and for the latter, we compare in \sct\ref{sct-results-disk-bivar} their bi-variate luminosity-radius distribution to the modeled function derived by \cite{de-Jong-Lacey-2000-spiral-galaxies-functions}. We show in \sct\ref{sct-results-size-evol} how bulge and disk radii vary along the Hubble sequence, and explain how we derived unprecedented power-law dependencies with the bulge-to-total ratio. Based on this analysis, we make the connection between the phases of mass and morphological evolution of galaxies along the Hubble sequence and the size variation of their bulges and disks (\sct\ref{sct-discussion-B-D-sizes}). Moreover, we discuss the distinction between pseudo- and classical bulges (\sct\ref{sct-discussion-bulge-types}), as well as the variations in the volume density of spheroids and in the surface density of disks (\sct\ref{sct-discussion-diffuse}). Lastly, we provide the parameters to all fits in \sct\ref{sct-discussion-mock}, so that they can be used to generate realistic galaxy mock catalogs of the full diversity of morphological types in the observed Universe at $z\sim0$. In this article, we use the standard $\Lambda$CDM cosmology with parameters $H_0=\SI{70}{km.s^{-1}.Mpc^{-1}}$ \citep{Freedman-2001-Hubble-constant}, $\Omega_m = 0.258 \pm 0.030$, $\Omega_\Lambda = 0.742 \pm 0.030$ \citep{Dunkley-2009-omega-cosmo-wmap}.

\section{Data         \label{sct-data}}

We use the EFIGI morphological catalog \citep{Baillard-2011-EFIGI} of 4458 galaxies which were visually classified based on $g$, $r$, $i$ Sloan Digital Sky Survey (SDSS) images, by their Hubble type, as well as 16 morphological attributes \citep{Baillard-2011-EFIGI,de-Lapparent-2011-EFIGI-stats} , taking integer values between 0 and 4. Here, we only use the {\tt Incl-Elong} attribute, measuring the apparent elongation of objects, and the {\tt VisibleDust} attribute, measuring the strength of the diverse features indicating the presence of dust in galaxies. Our profile fits are based on images extracted from the SDSS in the five optical bands $u$, $g$, $r$, $i$ and $z$.

EFIGI is a subsample of the Morphological Catalog (MorCat ; see Quilley \& de Lapparent, \textit{in prep.}), which is complete in apparent magnitude to $g\leq 15.5$. EFIGI is not magnitude-limited because it was designed  with the goal of having, when possible, several hundreds of galaxies of each Hubble type. Therefore, it is not a representative sample of the Universe. Because it mostly includes galaxies with apparent diameter $\ge 1$ arcmin, it is well suited for profile-fitting, and allows an in-depth study of the role of morphology on other galaxy properties.

In the current analysis, we limit the sample to the 3106 EFIGI galaxies with the EFIGI attribute {\tt Incl-Elong} $\leq 2 $: this corresponds to face-on or moderate inclination of galaxies when they have a disk, that is $\leq 70^\circ$, and elongation $\leq0.7$ for disk-less galaxies; this removes highly inclined disks, but keeps all E galaxies as their values of {\tt Incl-Elong} are between 0 and 2 \citep{Baillard-2011-EFIGI}.

\section{Methodology    \label{sct-methodo}}

\subsection{Luminosity profile fitting using SourceXtractor++   \label{sct-methodo-srx}}

\subsubsection{Generalities}

We use the new SourceXtractor++ software \citep{Bertin-2020-SourceXtractor-plus-plus} to decompose the 2D projected galaxy light profiles with the sum of two components, aimed at modeling the bulge and disk in lenticular and spiral galaxies, using a S\'ersic \citep{Sersic-1963-sersic-model} and exponential profile, respectively. This model-fitting is performed simultaneously in the $g$, $r$, and $i$ bands (further details on the SourceXtractor++ fits of the EFIGI galaxies can be found in \citealt{Quilley-2022-bimodality}). This model-fitting is preceded by multiple steps to measure bulge properties and use them as priors, thus leading to more reliable bulge and disk decompositions (suffering less degeneracies, see Quilley \& de Lapparent, \textit{in prep.}). We also model some Hubble types (E, cE, cD, dE, Im) with a single S\'ersic profile, for reasons described in \sct\ref{sct-methodo-BD-vs-1p}.

Although the SourceXtractor++ profiles have elliptical symmetry (as galaxies are frequently seen as elongated), in the following, we provide for simplicity the functional forms in the case of circular symmetry\footnote{See \cite{Graham-2005-sersic-considerations} for an exhaustive description of the circular S\'ersic profile}. The S\'ersic profile fitted to the galaxy bulges (and full galaxies for some types) is:
\begin{equation}
    I(r) = I_{\rm e} \exp\left\{ -b_n\left[\left( \frac{r}{r_{\rm e}}\right) ^{1/n} -1\right]\right\}
    \label{eq-sersic}
\end{equation}
where $r$ is the angular radius to the profile center, and $r_e$ the effective radius that encloses half of the total light of the profile, that is
\begin{equation}
    \displaystyle \int_{0}^{r_e} I(r) \, 2 \pi r \mathrm{d}r = \frac{1}{2} \int_{0}^{+\infty} I(r) \, 2 \pi r \mathrm{d}r
\end{equation}
for a profile with circular symmetry (in the case of an elliptical profile, it is the ellipse of semi-major axis $r_e$ and semi-minor axis $\frac{b}{a} r_e$ that encloses half of the total light).
In \eq\ref{eq-sersic}, $I_e = I(r_e)$ is the intensity at $r_e$, $n$ is the S\'ersic index that defines the steepness of the profile, with higher $n$ corresponding to steeper profiles, and $b_n$ is a normalization parameter depending on $n$ only.

The exponential profile used for galaxy disks can be written, in the case of circular symmetry, as
\begin{equation}
    I(r) = I_0 \exp{\frac{-r}{\mathfrak{h}}}
    \label{eq-exp-h}
\end{equation}
where $\mathfrak{h}$ is the angular scale-length. The exponential profile is a S\'ersic profile with $n=1$, which can be written, using \eq\ref{eq-sersic}, as
\begin{equation}
    I(r) = I_{\rm e} \exp\left[ -b_1\left( \frac{r}{\mathfrak{h}_{\rm e}} -1\right)\right]
    \label{eq-exp-sersic}
\end{equation}
where $\mathfrak{h}_e$ is the angular effective radius of the profile used for modeling the disks, which allows one to make the correspondence and perform comparisons with $r_e$, used in the bulge (S\'ersic) profile (see \eq\ref{eq-sersic}). 
From \eqs \ref{eq-exp-h} and \ref{eq-exp-sersic}, we infer that the effective radius $\mathfrak{h}_e$ and scale-length $\mathfrak{h}$ only differ by a multiplicative factor:
\begin{equation}
    \mathfrak{h}_e = b_1 \mathfrak{h} = 1.678 \mathfrak{h}
\end{equation}
which remains unchanged when converting to physical distances $h$ and $h_e$ (see \sct\ref{sct-methodo-math}). 

The model-fitting with SourceXtractor++ provides us with a set of parameters for the bulge (S\'ersic) and disk (exponential) components fitted to each galaxy which are: the total integrated apparent magnitude $m$, the corresponding bulge and disk integrated apparent magnitudes $m_{bulge}$ and $m_{disk}$ \resp, the $n$ index of the S\'ersic profile, the bulge and disk semi-major effective radii $r_e$ and $\mathfrak{h}_e$ \resp, the position angle of the major axis, and the elongation of the profile $b/a$ (where a is $r_e$ or $\mathfrak{h}_e$, and b is the semi-minor axis of the physical or angular profile, respectively).

\subsubsection{Bulge and disk decomposition versus single-profile modeling {\label{sct-methodo-BD-vs-1p}}}

In the current analysis, the bulge and disk decomposition is applied to all types including E, cE, cD, and dE. Indeed, even though E galaxies do not show evidence for a disk in optical images, kinematic studies have shown that stellar disk components are present in many of them \citep{Krajnovic-2008-disks-in-fast-rotators-SAURON, Krajnovic-2011-ATLAS-3D-morpho-kinemetric-features-ETG, Emsellem-2011-fast-slow-rotators}, leading to improved profile fits with a 2-component model \citep{Krajnovic-2013-ATLAS-3D-photo-kine-stellar-disks-in-ETG}. Fitting a bulge and disk profile to elliptical galaxies also allows us to compare their parameters to those of lenticular and spiral types via a common modeling method. Nevertheless, \cite{Bernardi-2014-model-fitting-effects-size-luminosity-relation} showed that the choice between a single S\'ersic profile and a bulge and disk decomposition conditions the resulting ranges and properties of the derived parameters. A single S\'ersic profile is therefore fitted to E, cE, cD and dE galaxies in order to compare our results with the historical relations that were derived using a single S\'ersic or de Vaucouleurs profile (see \scts\ref{sct-results-kormendy} and \ref{sct-results-binggeli}).
In the following, we mention the ``S\'ersic component'' and the ``exponential component'' when referring to corresponding components of the bulge and disk decomposition applied to E and cD types, to be differentiated from the ``single S\'ersic'' profile.

In the bulge and disk decomposition, if the S\'ersic component aims at adjusting a central concentration within the disk, it  sometimes fails, and such objects need to be identified in order to minimize biases in the measured bulge radii. Indeed, because of the very faint bulges of the latest Sd and Sm spiral types (and even more so in the bulgeless Im types), the S\'ersic bulge component might be inappropriately used to model either the whole galaxy in addition to the exponential profile intended for the disk, or any other kind of excess light, such as an HII region. To identify these erroneous fits, we compare the flux of their bulge component to the one from the zoom-in process mentioned in \sct\ref{sct-methodo-srx}, estimated as the excess flux in the center of the galaxy isophotal print above an approximate 2D background calculated from the inner disk, and providing an estimated $B/T_{zoom}$ when compared to the galaxy isophotal flux. We discard the fits for Sc and later morphological types for which the $B/T$ value from the bulge and disk decomposition in any of the $g$, $r$ or $i$ bands verifies $B/T>f(B/T_{zoom})$, with the $f$ threshold function empirically defined as a second degree polynomial: 32\% Sc, 41\% Scd, 59\% Sd, 80\% Sdm, and 83\% Sm fits are removed.

%f est un polynome de 2nd degré de coeffs 60, 0.5, 0.001

We tested this bulge validating procedure on the extreme case of Im galaxies, that do not host a bulge. Among the 179 Im with {\tt Inclination}$\leq 2$ in EFIGI, only 31 have a bulge modeling that verifies the previously described criterion, and among them, only 6 have a $B/T > 1\%$ (caused by an HII region or a contaminating star), confirming that these types do not host a bulge. We therefore only use the results of the single-S\'ersic profile fits for Im galaxies. As their S\'ersic index distribution peaks near 1, in the following, we examine the scaling relations for Im galaxies with those for the disks of the lenticulars and spirals.

\subsection{Correcting extinction effects \label{sct-methodo-extinction}}

We correct all magnitudes obtained from the luminosity profile-fitting for both atmospheric and galactic extinction. For the atmospheric correction, we use the $kk$ coefficients multiplied by the air masses provided for the SDSS\footnote{\url{https://classic.sdss.org/dr5/algorithms/fluxcal.html}}.
We base the galactic correction on the dust reddening galactic maps from \cite{SFD_1998_gal_extinction} from which we obtain $E(B-V)$ values for each galaxy with its sky coordinates, as well as on the conversions to extinction in the $g$, $r$, $i$ bands listed in Table~6 of \cite{S_and_F_2011_gal_extinction} using a Milky Way reddening law with an extinction to reddening ratio $A_V/E(B-V)=3.1$.

\subsection{SED fitting with ZPEG         \label{sct-methodo-zpeg}}

We use the ZPEG software \citep{Le-Borgne-2002-ZPEG} to fit Spectral Energy Distributions (SED) to the apparent magnitudes of the EFIGI galaxies, in order to derive their absolute (rest-frame) magnitudes and colors. This software receives as inputs the apparent magnitudes in the $g$, $r$, $i$ for EFIGI galaxies measured by SourceXtractor++ (\sct\ref{sct-methodo-srx}) and corrected for extinction (\sct\ref{sct-methodo-extinction}), as well as the HyperLeda redshifts corrected for Virgocentric infall (see \sct 2.2 of \citealt{de-Lapparent-2011-EFIGI-stats}). ZPEG adjusts to these apparent magnitudes the SEDs of families of templates from the PEGASE.2 library, that were determined from the major galaxy types (E, S0, Sa, Sab, Sb, Sbc, Sc, Sd, Im, starburt), and are characterized by specific functions for the evolution of the star formation rate with time \citep{Fioc-1999-PEGASE2}. We also obtain from this SED model-fitting the age of the scenario corresponding to the best-fit template, as well as several galaxy parameters including most notably the stellar mass $M_\ast$ and the star formation rate (SFR). 
This analysis is then repeated separately for the magnitudes of both bulges and disks, in order to also obtain the absolute magnitudes (hence colors), as well as the stellar masses of these components. Further details, robustness checks and results regarding these fits are given in \cite{Quilley-2022-bimodality}.

\subsection{Computing the mean effective surface brightness       \label{sct-methodo-math}}

The (major axis) effective radii $r_e$ or $\mathfrak{h}_e$ provided by the model-fitting with SourceXtractor++ for the bulge and disk components \resp, are angular distances. To deduct from them the physical effective radii $R_e$ and $h_e$ \resp, we multiply by the angular diameter distance $D_{ang}$ derived from the HyperLeda redshifts (see \sct\ref{sct-methodo-zpeg}). This yields
\begin{equation}
    R_e = \frac{\pi}{180} \ r_e \ D_{ang}
    \label{eq-radius}
\end{equation}
where the distances $R_e$ and $D_{ang}$ are in parsec, and $r_e$ in degree. The equation relating $h_e$ and $\mathfrak{h}_e$ is identical to \eq\ref{eq-radius}. In the rest of this section, we only refer to the bulge parameters $r_e$ and $R_e$ for simplicity (but equations also apply for the disk component with parameters $\mathfrak{h}_e$ and $h_e$).

The mean effective surface brightness $\langle\mu\rangle_e$ is defined as the mean surface brightness in the central region of an object above the isophote containing half of the object total flux. For a circular object, it is the mean surface brigthness in a disk of radius $r_e$: $\langle\mu\rangle_e = \langle\mu(r\leq r_e)\rangle$.
The mean surface brightness of a galaxy in a photometric band and within a projected area $\mathcal{A}$ on the sky measured in square arcseconds, is derived from its apparent magnitude $m(r\in \mathcal{A})$ in this area using
\begin{equation}
    \langle\mu(r\in \mathcal{A})\rangle = m(r\in \mathcal{A}) + 2.5\log(\mathcal{A})
    \label{eq-mu}
.\end{equation}
In the case of the bulge and disk elliptical profiles defined here, the mean effective surface brightness $\langle\mu\rangle_e$ is computed over the area of the ellipse with major axis the angular effective radius $r_e$ and of elongation $b/a$. It corresponds to half the total flux, hence to a magnitude $m+2.5\log(2)$ if $m$ is an estimate of the total apparent magnitude, so for $\mathcal{A}$ the area of the effective ellipse, one can write $\langle\mu\rangle_e$ as 
\begin{equation}
    \langle\mu\rangle_e = m + 2.5\log\left(2 \pi r_e^2 \frac{b}{a}\right)
    \label{eq-mu-def}
,\end{equation}
which is the equation we use to compute the mean effective surface brightness from the SourceXtractor++ parameters $m$, $r_e$ and $b/a$.

As far as disks are concerned, one often considers the central surface brightness $\mu_0=-2.5\log I_0$, with $I_0$ the central intensity of the exponential disk profile defined in \eq\ref{eq-exp-h}. \cite{Graham-2005-sersic-considerations} indicate in their \eqs 7 and 9 that the difference between $\mu_0$ and $\mu_e$ and between $\mu_e$ and $\langle \mu \rangle_e$ respectively, are only dependent on $n$. In particular, for the $n=1$  S\'ersic index of an exponential profile, one can write
\begin{equation}
    \mu_0 = \langle \mu \rangle_e -1.123
    \label{eq-mu-n-1}
.\end{equation}
These considerations show that the choices between $h_e$ or $h$ (see \eqs\ref{eq-exp-h} and \ref{eq-exp-sersic}) and between $\langle \mu \rangle_e$ and $\mu_0$ (see \eq\ref{eq-mu-n-1}) do not affect the measured correlations between disk radii in logarithmic scale and surface brigthness as it merely shifts all data points by a constant value. Here, we then use \eq\ref{eq-mu-n-1} to derive the central surface brightness $\mu_0$ of the disks of EFIGI galaxies from the mean effective surface brightness $\langle \mu \rangle_e$, in turn computed using \eq\ref{eq-mu-def}, and which takes into account the moderate inclinations of disks in the considered EFIGI subsample (see \sct\ref{sct-data}).

To rewrite the mean effective surface brightness in terms of the physical effective radius $R_e$ and the absolute magnitude $M$, hence to estimate how these 3 quantities characterizing galaxy fluxes and profiles scale with each other, we use
\begin{equation}
    M = m - 5\log D_{\mathrm{lum}}(z) + 5 - K_{\mathrm{cor}}(z)
    \label{eq-mag-def}
,\end{equation}
where $m$ and $M$ are in the same band, $D_{\mathrm{lum}}(z)$ is the luminosity distance at redshift $z$, and $K_{\mathrm{cor}}$ the k-correction in the considered band at that redshift. Using $D_{\mathrm{lum}}(z)=(1+z)^2 D_A(z)$ and \eqs\ref{eq-radius} and \ref{eq-mag-def}, we then rewrite the mean effective surface brightness in \eq\ref{eq-mu-def} as
\begin{align}
\begin{split}
    \langle \mu \rangle_e &= M + 5\log R_e + 2.5\log(1+z)^4 + K_{\mathrm{cor}}(z) \\
    & \quad - 5 + 5\log\frac{648000}{\pi} + 2.5\log\left(2\pi \frac{b}{a}\right)
    \label{eq-full}
\end{split}
\end{align}
with $\langle \mu \rangle_e$ in mag arcsec$^{-2}$, and $R_e$ in parsec. The term $2.5\log(1+z)^4$ corresponds to the surface brightness dimming with redshift.

Because most galaxies in the EFIGI sample have $z\lesssim0.05$, the surface brightness dimming term is $\lesssim 0.2$, and the k-correction term is $\lesssim 0.2$ in the $g$ band and 0.1 in $r$ and $i$ bands. As we have limited our sample to low values of the {\tt Incl-Elong} attribute (see \sct\ref{sct-data}), the elongation term $2.5\log\frac{b}{a}$ is $\lesssim0.4$. Added in quadrature, all three terms would add a total dispersion of $\lesssim0.5$, which is negligible compared to the $\sim9$-magnitude interval for bulges and disks magnitudes, and the $\sim15$ and $\sim7$ dex intervals for $5\log R_e$ and $5\log h_e$ respectively (see \fgs\ref{binggeli-cmap-BT} and \ref{binggeli-disk}). These considerations therefore justify the following approximate scaling relation between absolute magnitude $M$, $R_e$ and $\langle \mu \rangle_e$ for nearby galaxies:
\begin{equation}
    \langle \mu \rangle_e \simeq M + 5\log R_e + \kappa,
    \label{eq-mu-scaling}
\end{equation}
where $\kappa$ is a constant only for galaxies of equal redshift, k-correction and elongation. Otherwise, when considering the whole EFIGI sample, and more generally a sample of galaxies of different Hubble types, distances and elongations, $\kappa$ undergoes a limited dispersion of $0.5$ at maximum (as indicated above).

\subsection{Uncertainties in effective radii\label{sct-methodo-uncertainties}}

\begin{figure*}
\includegraphics[width=\columnwidth]{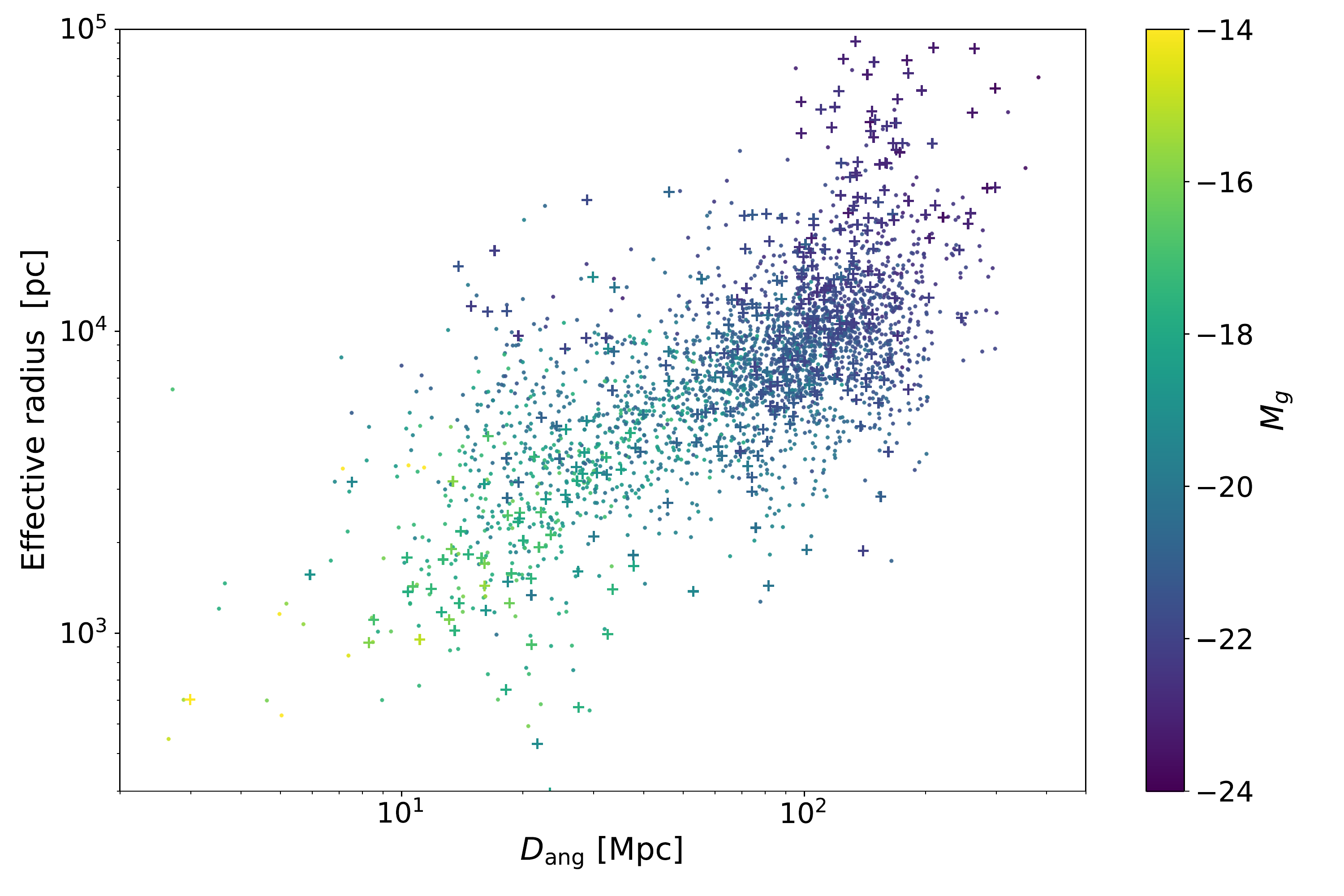}
\includegraphics[width=\columnwidth]{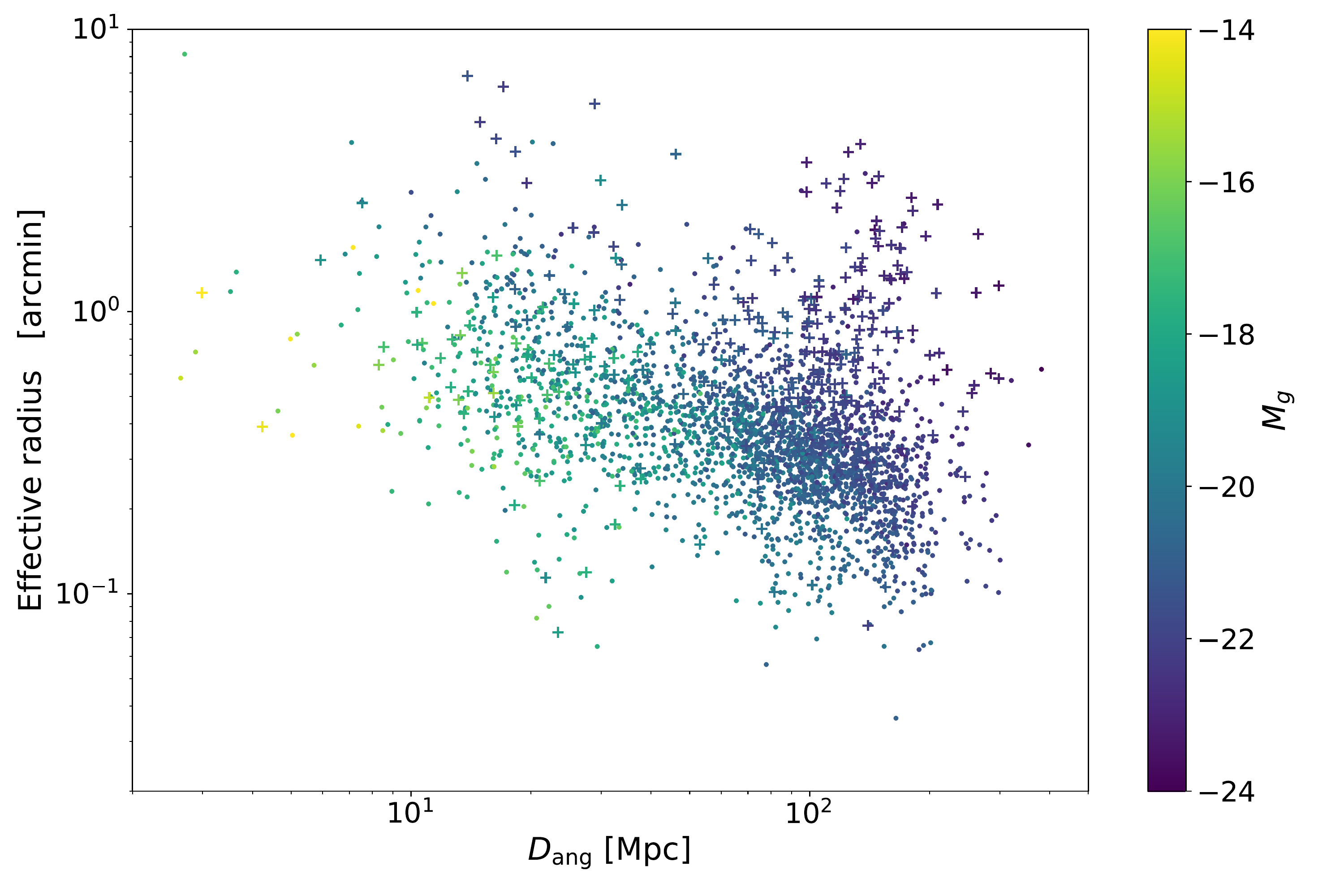}
\caption{Distribution of the effective radii of E, cD, dE, cE and Im galaxies modeled as a single S\'ersic profile (crosses), and the disk effective radii for spiral and lenticular EFIGI galaxies (dots), as a function of the angular diameter distance $D_{ang}$. Depending on the Hubble type, either radius can be considered as an estimate of the galaxy size. Left panel shows the physical radii, while the right panel shows the angular radii. The points are also color-coded with the absolute magnitude in the $g$ band. This graph shows that selection effects affecting the EFIGI sample leads to larger/brighter galaxies of a given type being preferentially located at large distances, and having preferentially small angular radii.}
\label{h-vs-D-ang-cmap-abs-mag-g}
\end{figure*}

\begin{figure*}
\includegraphics[width=\textwidth]{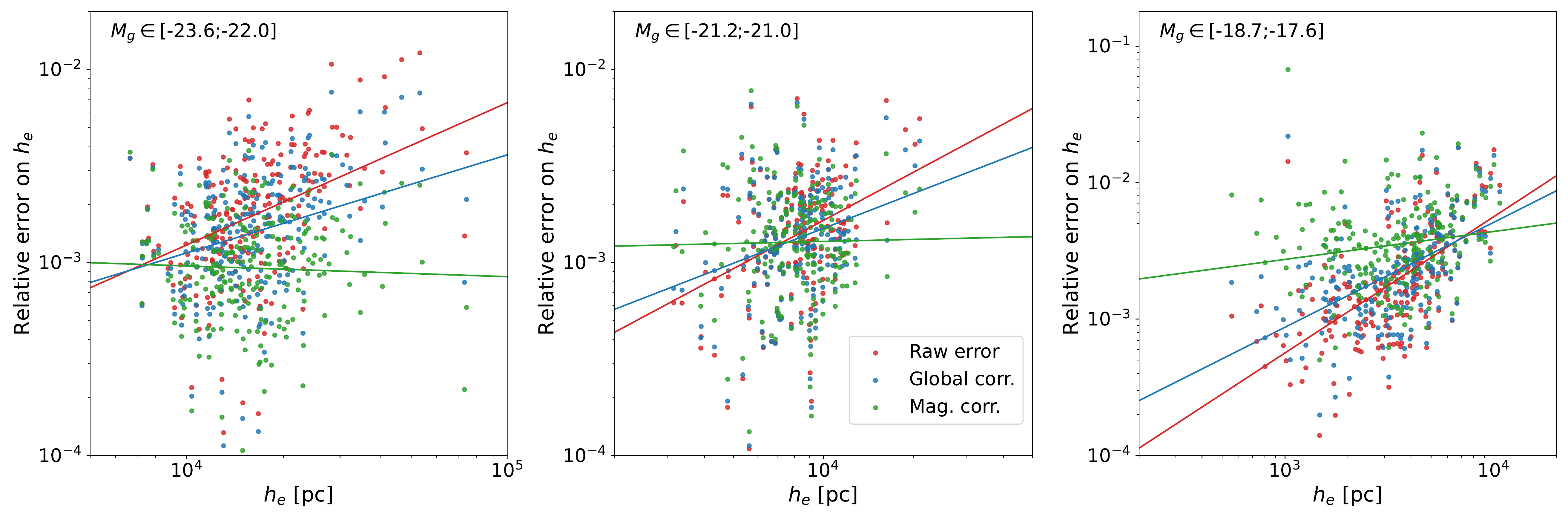}
\caption{Distribution of the relative error on the physical disk effective radii $\sigma(h_e)/h_e$ as a function of $h_e$ in bins of total galaxy absolute magnitude in the $g$ band. Only 3 bins are shown as examples of the general behavior. The color of the points represent the type of plotted error (see text for details) and the lines are the corresponding linear regression. There is a trend of relative errors increasing with $h_e$ for the raw errors (in red). The correction derived globally over all absolute magnitudes (in blue) reduces the bias, whereas the correction by minimization over 12 magnitude intervals (in green),  actually erases it.}
\label{error-corr}
\end{figure*} 

EFIGI is an incomplete sample (it is not volume limited). First, it is not limited in apparent magnitude as is usually the case for observed galaxy samples. It displays a fast decreasing incompleteness below an isophotal diameter of $D_{25}=1$ arcminute, but it is not either complete above this value  \citep{de-Lapparent-2011-EFIGI-stats}. Moreover,  the cone-shape of the sampled volume (resulting from the sky projected area selection) combined with the shape of the luminosity function of galaxies in which brighter galaxies are much rarer result in a cone-like volume, with small and faint galaxies detected in large numbers in the nearby tip of the cone, and large and bright galaxies detected predominantly at larger distances. Moreover, the depth of the EFIGI survey does not encompass a large number of walls and voids of the cosmic web, hence its redshift distribution is affected by these large-scale inhomogeneities \citep{Baillard-2011-EFIGI}. In particular, the survey has a nearby over-density at $z\sim0.005$ due to the Virgo cluster, in which reside most of the Im, cE and dE morphological types which add up to the previously mentioned over-selection of small and faint galaxies in the nearest part of the survey.

Combination of all the above mentioned selection effects yields the strong correlation of the physical effective radii of EFIGI galaxies with their distance from us. This can be seen in the left panel of \fg\ref{h-vs-D-ang-cmap-abs-mag-g}, showing the physical effective radii $h_e$ of disk galaxies (lenticular and spiral types, shown as dots), and the physical effective radii $R_e$ of E, cD, cE, dE and Im adjusted by a single S\'ersic profile (shown as crosses), which can all be considered as estimates of the various galaxy sizes (in the $g$ band), as a function of the angular diameter distance $D_{ang}$ of each object. One can see that the effective physical radii of the various plotted types of galaxies systematically increase when they are more distant (the color-map in absolute magnitude $M_g$ shows that more distant galaxies are also brighter). Ideally, one should have a sample in which galaxies of all physical sizes are equally sampled at all distances so that there is no such bias.

As a result, if one considers first EFIGI disk (lenticulars, spirals) galaxies, as well as Im galaxies, galaxies with large physical effective disk radii $h_e$ have smaller angular sizes, estimated by their angular disk effective radii $\mathfrak{h}_e$, as shown in the right panel of \fg\ref{h-vs-D-ang-cmap-abs-mag-g}. Therefore, the physically larger EFIGI disk galaxies are spread over a systematically smaller number of fixed size pixels of the SDSS imaging survey ($0.385$ arcsec) than smaller objects, causing a systematically larger relative uncertainty $\sigma(\mathfrak{h}_e)/\mathfrak{h}_e$ estimated by SourceXtractor++ for large $h_e$. This relative error is equal to the relative uncertainty in the physical disk effective radius $\sigma(h_e)/{h_e}$ (see \eq\ref{eq-radius}), therefore the latter is larger for larger $h_e$. This is indeed what we obtain for EFIGI disk and Im galaxies, as seen on \fg\ref{error-corr}: despite a large dispersion, there is a systematic increase in $\sigma({h_e})/{h_e}$ with $h_e$ (by $\sim1$ dex over the $\sim1$ dex interval of measured $h_e$), with slopes in the interval $[-0.2;0.2]$ (in log-log) across $12$ absolute $g$ band magnitude intervals of width $\Delta M_g$ in the $0.2-0.5$ range - except for extremal bins, adapted to contain a weakly varying number of galaxies (between $194$ and $272$).

Because we use the total least square estimation in all fits that are performed in the present article (see \sct\ref{sct-methodo-odr}), which takes into account the errors on both axes, any systematic trend in the errors as a function of any axis would bias the corresponding fit.
Therefore the biased distribution of relative errors on $h_e$ illustrated in \fg\ref{error-corr} would tend to give systematically more weight to intrinsically smaller galaxies, hence biasing the size-magnitude relations toward small radii. We therefore choose to eliminate this overall systematic increase in $\sigma(h_e)/h_e$ with $h_e$ by a linear regression, while keeping unchanged the dispersion around the linear fits. As seen in \fg\ref{error-corr}, the slopes are positive, and should be flattened. Nevertheless, using the factor that flattens the slope of the fit to galaxies of all $M_g$ together (shown in blue in \fg\ref{error-corr}, labeled as ``Global corr.'') is insufficient to correct for the bias. We therefore iteratively find the common factor to apply to the slopes of the fits per $M_g$ interval in order to minimize the sum of the squares of the slopes over the 12 defined $M_g$ intervals. This sum of squares behaves as a parabola without noise, and the minimum yields a factor $3.18$ that is applied to  $\sigma(h_e)/h_e$ to correct for its systematic and biased increase with $h_e$. These corrected slopes are shown in green in \fg\ref{error-corr} (labeled as ``Mag. corr.'').

When considering the effective angular and physical radii, $r_e$ and $R_e$ respectively, of the single-S\'ersic profile fits to E, cD, cE and dE galaxies, the gradients in $R_e$ and $r_e$ with $D_{ang}$ in both panels of \fg\ref{h-vs-D-ang-cmap-abs-mag-g} are less visible than for disk and Im galaxies, as these spheroid types populate different and narrow ranges of $R_e$ (see left panel of \fg\ref{binggeli-cmap-BT}): E and cD are among the largest galaxies with 80\% of objects having their $R_e$ in the interval $4-42$ kpc, whereas cE and dE are among the smallest with 80\% of both types of objects having their $R_e$ in the interval $0.9-5$ kpc. There is however also an overall systematic increase in $\sigma(R_e)/R_e$ with $R_e$ as in \fg\ref{error-corr} (although smaller, $\sim0.5$ dex for $\sim1$ dex in $R_e$), that we correct in the same iterative approach as for disks and Im galaxies, using the three [-23.7;-22.0], [-22.0;-21.0] and [-21.0;-13.3] $M_g$ magnitude intervals containing 102, 115, and 112 galaxies respectively. 

We emphasize that these corrections leave intact the fact that 2 galaxies with identical values of $h_e$ (or $R_e$) may have their $\sigma(h_e)/h_e$ (or $\sigma(R_e)/R_e$) differ by a factor as large as $10$. We also checked that this flattening correction preserves the decreasing trends of $\sigma(\mathfrak{h}_e)/\mathfrak{h}_e=\sigma(h_e)/h_e$ versus the angular effective radius $\mathfrak{h}_e$, and $\sigma(r_e)/r_e=\sigma(R_e)/R_e$ versus $r_e$.

We also performed tests on synthetic images of galaxies generated with Stuff and SkyMaker \citep{Bertin-2009-skymaker} in order to check the uncertainties provided by SourceXtractor++. We measured that the relative errors on bulge and disk effective radii are underestimated by a varying factor increasing from 1 to $\sim10$ at the smallest relative uncertainties.  We initially tried to correct for this effect, but the correction is insufficient to eliminate the biases in $\sigma(h_e)/h_e$ versus $h_e$ and $\sigma(R_e)/R_e$ versus $R_e$, which the minimization procedure per magnitude interval described here succeeds in doing. 

At last, we measure a similar $\sim1$ dex dispersion in $\sigma(R_e)/R_e$ versus $R_e$ for the effective radii of the bulges of lenticular and spiral galaxies as in \fg\ref{error-corr}, but we do not detect any systematic trend with $R_e$.  We suspect this is due to the fact that the bulges are internal smaller regions of lenticular and disk galaxies, and are less affected by the biases in the total galaxy size distribution with distance. Therefore we do not apply any correction to $\sigma(R_e)/R_e$ versus $R_e$ for the bulges in the bulge and disk decomposition of lenticular and spiral galaxies. 

\subsection{Orthogonal distance regression\label{sct-methodo-odr}}

In this article, we derive multiple relations between the parameters of the bulge and disk components for the galaxies in the EFIGI catalog. Because all parameters estimated by the SourceXtractor++ modeling undergo uncertainties, we use the ODRPACK Version 2.01 Software for Weighted Orthogonal Distance Regression (ODR hereafter, \citealt{ODR-1992}) of the \textit{scipy} \textit{Python} library \citep{2020SciPy-NMeth}, which allows one to fit any functional form. Although this is not stated, we suppose that this method corresponds to the total least squares estimation, which is the generalization of the Deming regression\footnote{\url{https://en.wikipedia.org/wiki/Deming_regression}} for the linear case, which is itself a generalization of the orthogonal distance regression for identical variance along both axes. The advantage of these various estimates is that they take into account the errors on both axes when performing the fit (including the covariances, which we neglect here), contrarily to the linear regression approach, which considers the x-axis values as the truth. The minimization of the ODR package is done on the distances between the data points and the fit along both axis, which simplifies to the distance orthogonal to the fit when both axis have the same weight, and not only along the y-axis as this is done in a linear regression. For this reason, the ODR package leads to different functions than the regression along the y-axis when fitting a linear model, but we checked that these differences do not alter the qualitative conclusions of the current analysis.

Moreover, we have discarded points with anomalously low errors compared to the rest of the distribution (at least one order of magnitude below the median value) when performing the ODR fits. This is mandatory to avoid that the resulting model does not only go through these data points while ignoring the rest of the sample. The minimum error threshold value was found empirically for all parameters involved in such fits, and such filtering only reduces the sample size by a few percents. Finally, the adjustment made on the errors on $h_e$ described in the previous section (\sct\ref{sct-methodo-uncertainties}) is pivotal to obtain a realistic fit of the size-luminosity relation for disks (see \sct\ref{sct-results-disk-scaling-relations} hereafter) but would still be needed if we had opted for a linear regression, as the systematic trends in the errors occur along the y-axis.

\section{Results    \label{sct-results}}

\subsection{Revisiting the Kormendy relation for E and bulges \label{sct-results-kormendy}}

\begin{figure*}
\begin{center}
\includegraphics[width=0.70\textwidth]{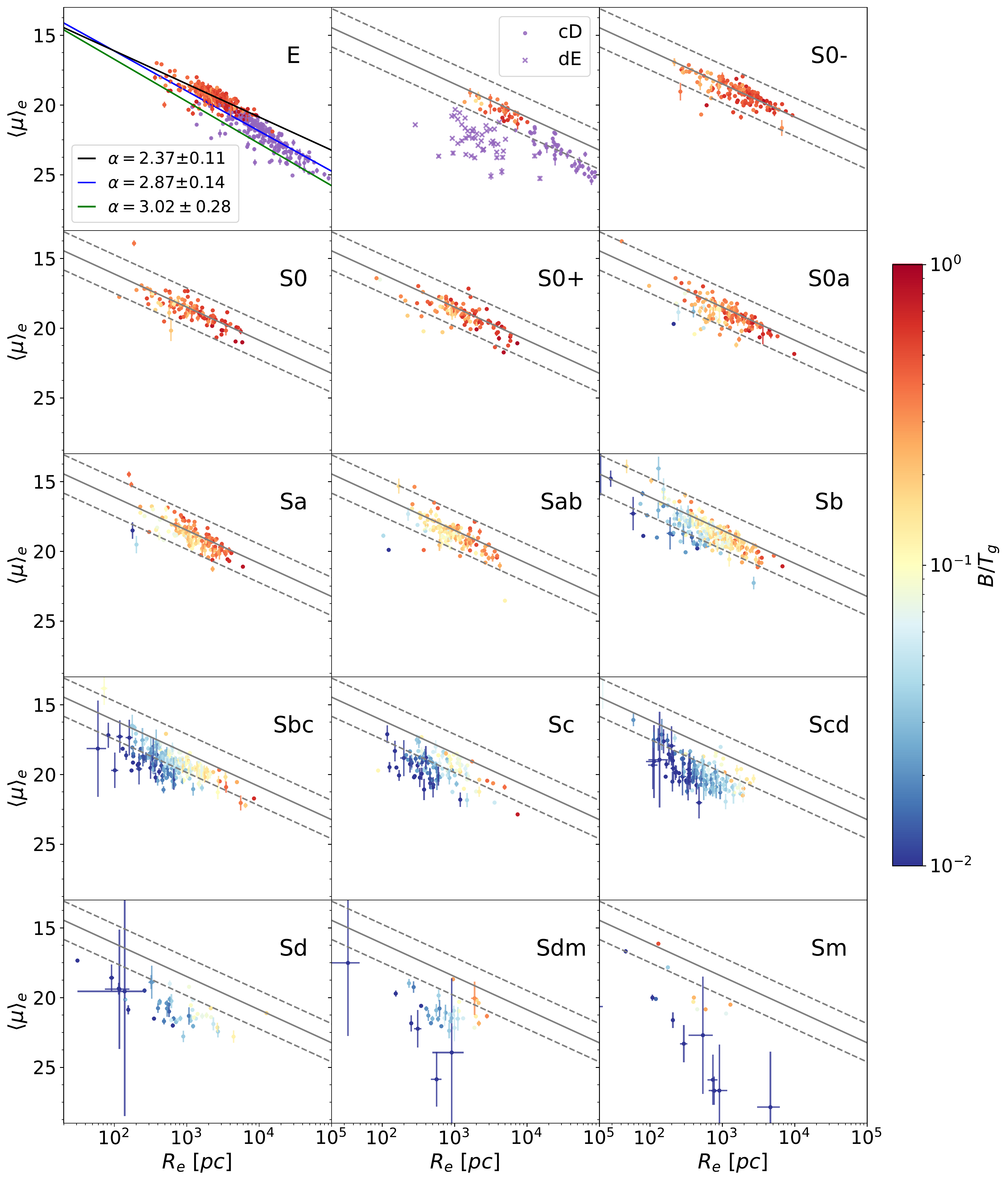}
\caption{Mean effective surface brightness $\langle\mu\rangle_e$ versus effective radius $R_e$ for the S\'ersic components of EFIGI E and cD galaxies, and for the bulges of lenticular and spiral types with {\tt Inclination} $\leq 2$, all derived from the S\'ersic bulge and exponential disk decompositions, in the $g$ band. The purple points in the 2 upper left panels represent the same relation for the E, cD and dE galaxies modeled as a single S\'ersic profile. In the upper-left panel are shown the linear fits of $\langle\mu\rangle_e$ as a function of $R_e$ for the S\'ersic components and single-S\'ersic fits to elliptical galaxies in black and blue respectively, as well as the \cite{Kormendy-1977-II-kormendy-relation} relation, in green. The fit to the E S\'ersic component (in black) is repeated in solid gray in the other panels, with the dashed lines showing the same line offset by $\pm 3$ times the \rms dispersion in $\langle\mu\rangle_e$ around the fit for the E types. The color of the points represent the bulge-to-total luminosity ratio in the $g$ band ${B/T}_g$. Almost all bulges of types S0$^-$ to Sb are within $3\sigma$ of the linear fit to the E S\'ersic components. Later types progressively shift to smaller effective radii and lower values of ${B/T}_g$, as well as dimmer effective surface brigthnesses than what would be expected from the Kormendy relation at these radii.}
\label{kormendy-g-per-type-cmap-BT}
\end{center}
\end{figure*}

\begin{figure*}
\includegraphics[width=\columnwidth]{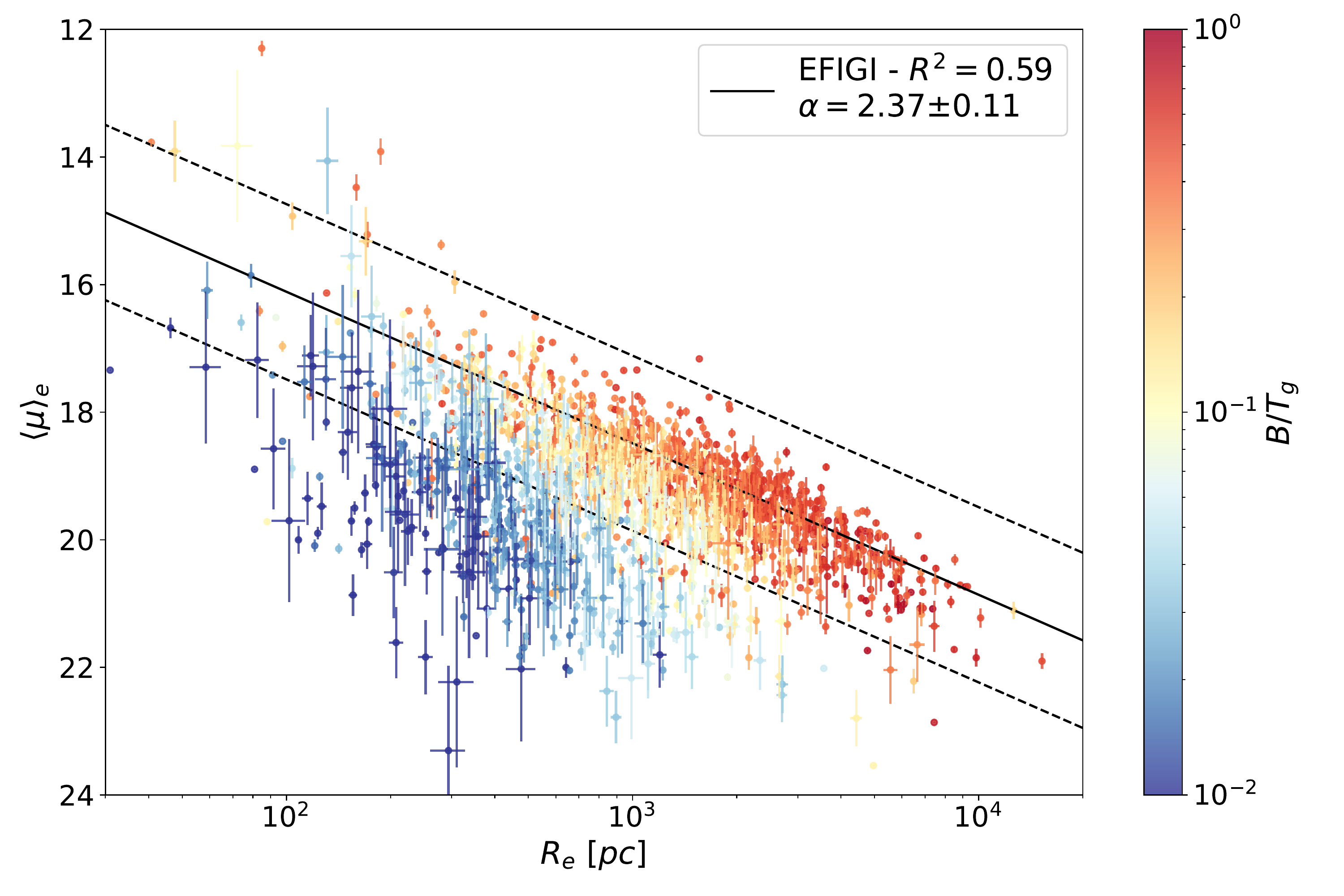}
\includegraphics[width=\columnwidth]{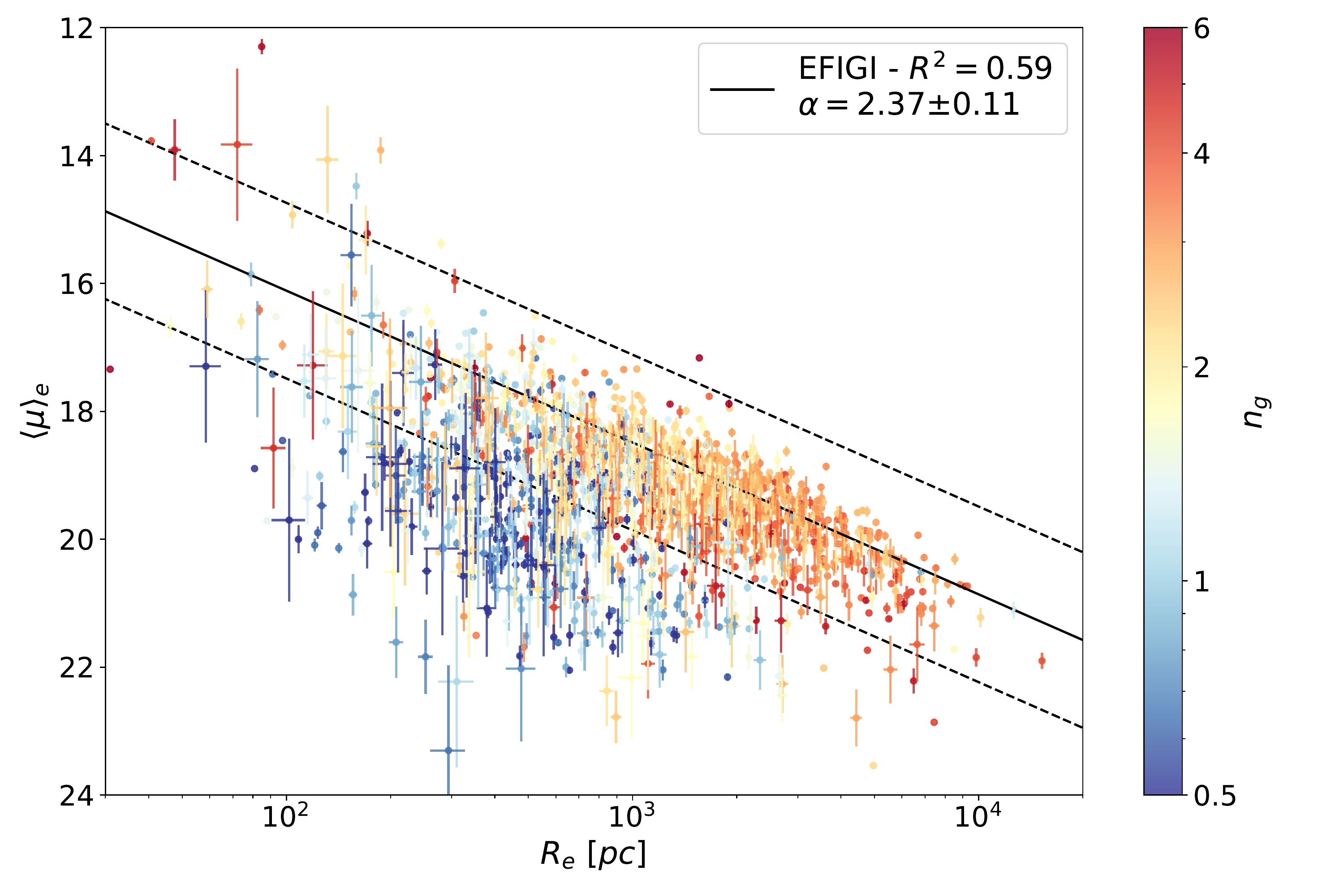}
\caption{Mean effective surface brightness $\langle\mu\rangle_e$ versus effective radius $R_e$ for the S\'ersic components of E types and the bulges of all Hubble types from S0$^-$ to Sm with {\tt Inclination} $\leq 2$, all in the $g$ band (dE types are excluded from this graph). The panels are color-coded by the bulge-to-total ratio ${B/T}_g$ (left) and the S\'ersic index $n_g$ (right) respectively. The black solid line in both panels is the linear fit for EFIGI E S\'ersic components, while the dashed lines have the same slope and are offset by $\pm 3$ times the \rms dispersion in $\langle\mu\rangle_e$ around that fit. The departure from the Kormendy relation occurs as both ${B/T}_g$ and $n_g$  decrease to the lowest possible values, while the highest ones are found for the highest radii along the Kormendy relation.}
\label{kormendy-cmap}
\end{figure*}

\cite{Kormendy-1977-II-kormendy-relation} was the first to show that elliptical galaxies showed a correlation between $\langle\mu\rangle_e$ and $R_e$, measured in a 4600-5400 $\AA$ band denoted ``$G$'', that corresponds to the red wavelength part of the SDSS $g$ band. One-dimensional surface brigthness profiles were obtained by  processing photographic plate images of the galaxies with a microdensitometer, and ``to minimize effects of the three-dimensionality of spheroids, the profile at $45^\circ$ to the major axis was used'' \citep{Kormendy-1977-II-kormendy-relation}. In contrast, the EFIGI effective radii derived by SourceXtractor++ are provided along the major axis. However, we calculated the effective radius at $45^\circ$ for our models on all EFIGI E galaxies and the ratio between the former and the latter has a median value of 0.89 and is below 0.8 (but higher than 0.67) for only 11.5\% of objects. More importantly, this limited difference between the two sets of values leads to a $10^{-3}$ difference in the slope of the $\langle\mu\rangle_e$ versus $R_e$ relation, so we directly compare below the Kormendy relation with our results based on the semi-major axes of the fitted profiles. We also investigated the effect of elongation for elliptical galaxies, and detected no effect on the $\langle\mu\rangle_e$ versus $R_e$ distribution for the three subsamples with {\tt Incl-Elong} attribute equal to 0, 1 and 2.

The upper left panel of \fg\ref{kormendy-g-per-type-cmap-BT} shows the relation between $\langle \mu \rangle_e$ obtained using \eq\ref{eq-mu-def}, and $R_e$ obtained using \eq\ref{eq-radius}, with $m$ and $r_e$ provided by the single-S\'ersic profile fits with SourceXtractor++ to all 226 EFIGI elliptical galaxies (in purple). The points are color-coded as a function of ${B/T}_g$, the bulge-to-total flux ratio in the $g$ band. An ODR linear fit (see \sct\ref{sct-methodo-odr}) of $\langle \mu \rangle_e$ as a function of $R_e$ (in blue) yields
\begin{equation}
    \langle \mu \rangle_e = 2.87^{\pm0.14}\log R_e + 19.00^{\pm0.53} 
    \label{eq-kormendy}
\end{equation}
with $R_e$ in kiloparsecs. The derived slope is compatible with the $3.02$ value measured by \citet{Kormendy-1977-II-kormendy-relation}, also plotted in green, 
but with a 0.74 magnitude per arcsec$^2$ brighter $\langle \mu \rangle_e$ for EFIGI E galaxies in $g$, which could be due to the different ($G$) photometric band used in the original measurement. No uncertainty on the slope is provided by the author for the original relation, but by performing a linear regression on the tabulated values in his article, we derived a 0.28 uncertainty in the slope, so the difference between the slope in \eq\ref{eq-kormendy} and that measured by \citet{Kormendy-1977-II-kormendy-relation} is at the $0.5\sigma$ level. We also obtain a nearly identical result to \eq\ref{eq-kormendy} if we use a linear regression, with a slope of $2.86 \pm 0.15$ and an intercept of $19.08\pm 0.58$ for EFIGI ellipticals.

In the upper left panel, the relation obtained for the S\'ersic components of elliptical galaxies is different from that obtained from the single-S\'ersic fits, with $R_e$ smaller by $\sim 0.7$ dex and $\langle \mu \rangle_e$ brighter by $\sim 3$ magnitudes, which highlights the strong impact of the modeling method. Indeed, the ODR linear fit for E S\'ersic components (in black in the graph) yields:
\begin{equation}
    \langle \mu \rangle_e = 2.37^{\pm0.11}\log R_e + 18.49^{\pm0.35} 
    \label{eq-kormendy-B}
\end{equation}
with $R_e$ in kpc.
There is a $2.8\sigma$ difference with the slope of the relation fitted to the E single S\'ersic profiles given in \eq\ref{eq-kormendy}, and a $0.8\sigma$ difference between the intercepts (these differences are not of concern as different modeling methods have been used). In the other panels of \fg\ref{kormendy-g-per-type-cmap-BT} showing the bulge components of the fits from cD to Sm types, the linear fit to the E S\'ersic components from the upper left panel (\eq\ref{eq-kormendy-B}) is also indicated as a reference, and the dashed lines above and below correspond to the value of the intercept offset by $\pm 3 \sigma$, where $\sigma$ is the \rms dispersion in $\langle \mu \rangle_e$ around the fit. 

The upper central panel of \fg\ref{kormendy-g-per-type-cmap-BT} shows that the S\'ersic components of cD galaxies follow the Kormendy relation for the S\'ersic components (in black), but only populate the larger values of effective radius for E elliptical galaxies. A similar effect is seen for the single S\'ersic profiles of cD versus E galaxies (purple circles). In contrast, the single-S\'ersic fits to the dE types shown in the same panel (purple crosses), are shifted to smaller $R_e$ than the single-S\'ersic fits to both cD and E types. Nevertheless, dE and E types exhibit a similar interval of $\mu_e$ between 20 and 25 $g$ magnitude, despite the intuitive expectation that the effective surface brightness of the centrally very dense elliptical galaxies should be significantly brighter than that of dE galaxies, as they are fitted by S\'ersic profiles with indices in the intervals $n=3.5-7$ and $n=1-3$ respectively (see \sct\ref{sct-results-binggeli}). This is due to the fact that the S\'ersic profile has a significant flux out to very large distances (in particular when $n>3$), and the effective radii of both E and dE type are therefore much larger than the central parts of the galaxies, which have markedly different appearances for these types and are used in particular to determine the visual morphological type. However, the flux accounted for in the calculation of the effective surface brightness, makes $\mu_e$ an average quantity dominated by the low level wings of the profiles.

\fg\ref{kormendy-g-per-type-cmap-BT} also shows separately and for each EFIGI morphological type $\langle \mu \rangle_e$ versus $R_e$ derived from a bulge and disk modeling, also color-coded with ${B/T}_g$, and compared to the fit to the S\'ersic component of E types shown in the upper left panel (\eq\ref{eq-binggeli-E-bulges}), as a reference. For all lenticular and spiral types, EFIGI galaxies display a systematically decreasing interval of $R_e$. Also, for a given surface brightness, bulges of later Hubble types have, on average, smaller $R_e$ than earlier types, or equivalently that they are fainter at a given $R_e$. For instance, the bulges of Sbc are $\sim6$ times smaller or $\SI{2.4}{mag.\; arcsec^{-2}}$ fainter than what is predicted by the fit for E galaxies, from their surface brightness or effective radius respectively. Size variations are further explored in \sct\ref{sct-results-size-evol} and \fg\ref{B-and-D-radius-distrib}. Moreover, for a given Hubble type, the variations in $R_e$ and $\langle \mu \rangle_e$ are linked to the value of the $B/T$ ratio. 

\fg\ref{kormendy-g-per-type-cmap-BT} also shows that the Kormendy relation remains valid for the bulges of lenticulars and early-type spirals up to Sab type. However, for Sb and later types, the relation between $\langle \mu \rangle_e$ and $R_e$ departs from the Kormendy relation, as ${B/T}_g$ decreases: these bulges have lower values of mean effective surface brightness $\langle \mu \rangle_e$ than what would be predicted from their effective radius $R_e$ using the Kormendy relation. To evaluate this difference in surface brightness, we compute linear fits for each type and note that the main change is that the intercept of the fit shifts toward fainter magnitudes, but the slope of the relation remains rather stable. For instance, Scd galaxies are fitted by a slope which differs by less than $1 \sigma$ from that for E galaxies, but the intercept is 2 magnitudes below that for the fit to E galaxies. There is also a systematic decrease of $R_e$ for later types, by 1 to 2 orders of magnitude between E and Scd types.

In \cite{Quilley-2022-bimodality}, we showed that bulges of EFIGI late-type spirals  not only have smaller $B/T$ values but also smaller S\'ersic indices than bulges of early-type spirals and lenticulars. In \fg\ref{kormendy-cmap}, we therefore plot $\langle \mu \rangle_e$ versus $R_e$ for the S\'ersic component of E types, and the bulges of lenticular and spiral galaxies, color-coded by ${B/T}_g$ (left panel), and by the bulge S\'ersic index (right panel). The three black lines are again the linear fit (solid line) of $\langle \mu \rangle_e$ as a function of $R_e$ for E galaxies (see upper left panel of \fg\ref{kormendy-g-per-type-cmap-BT} and  \eq\ref{eq-kormendy-B}), and $\pm 3$ times the \rms dispersion in $\langle \mu \rangle_e$ around that fit (dashed lines).   

One can see that the departure at $R_e \lesssim 2$ kpc from the Kormendy relation followed by S\'ersic components of E types occurs for bulges with smaller radii and fainter effective surface brightness, as well as with decreasing values of both ${B/T}_g$ and the S\'ersic index, but with more dispersion in the latter which is affected by larger relative uncertainties, likely due to the stronger degeneracies in this parameter when performing the luminosity profile fitting. This effect corresponds to the progressive shift below the Kormendy relation for later Hubble types, seen in \fg\ref{kormendy-cmap}. We also note that the left panel of \fg\ref{kormendy-cmap} is in agreement with \fg 8 of \cite{Kim-2016-bulges-SDSS}, who also showed a larger deviation from the Kormendy relation for bulges with lower $B/T$, also using disk and bulge decomposition on SDSS data.

We agree with the proposition of \cite{Gadotti-2009-bulge-structure-SDSS} that galaxies whose bulge deviate from the Kormendy relation are likely to be such bulges, and \fg\ref{kormendy-cmap} also shows that they have smaller ${B/T}_g$ and S\'ersic indexes. In contrast, classical bulges are probably those that fall along the Kormendy relation: these bulges have ${B/T}_g \gtrsim 0.1 $ and $n \gtrsim 2 $. This is also consistent with the S\'ersic index limit of 2 inferred by \citet{Fisher-Drory-2008-bulges-n-sersic}. We further discuss these interdependent trends in terms of bulge structure along the Hubble sequence, as well as the lack of kinematics to identify pseudo-bulges (as rotationally supported components), in \sct\ref{sct-discussion-bulge-types}.

\begin{figure}
\includegraphics[width=\columnwidth]{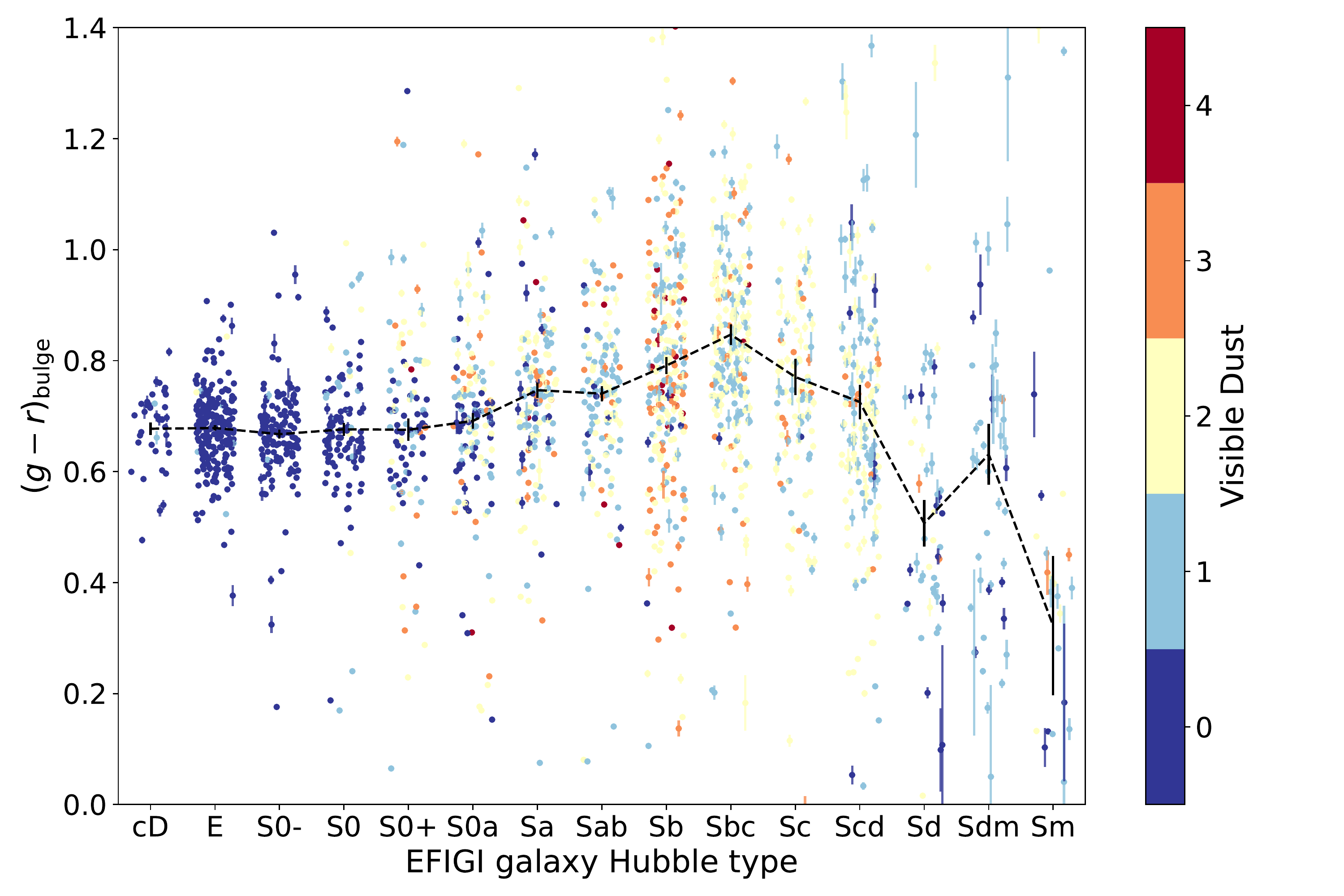}
\caption{Distribution of the $g-r$ absolute color of the EFIGI bulges (or S\'ersic component) for each Hubble type (with {\tt Inclination} $\leq 2$). The black dashed line represents the mean color by type and its associated error. Bulge color is overall stable with most bulges in the 0.5-0.9 range. There is a 0.17 mag reddening between lenticulars and intermediate spirals (Sbc), which could be due to dust reddening, as it is more frequently present in large amounts in these galaxies. This effect decreases for Scd types, with some bluing possibly being present for the bulges of Sd and later types, compared to the lenticular and early spirals.}
\label{bulge-color}
\end{figure}

\begin{figure*}
\begin{center}
\includegraphics[width=0.8\textwidth]{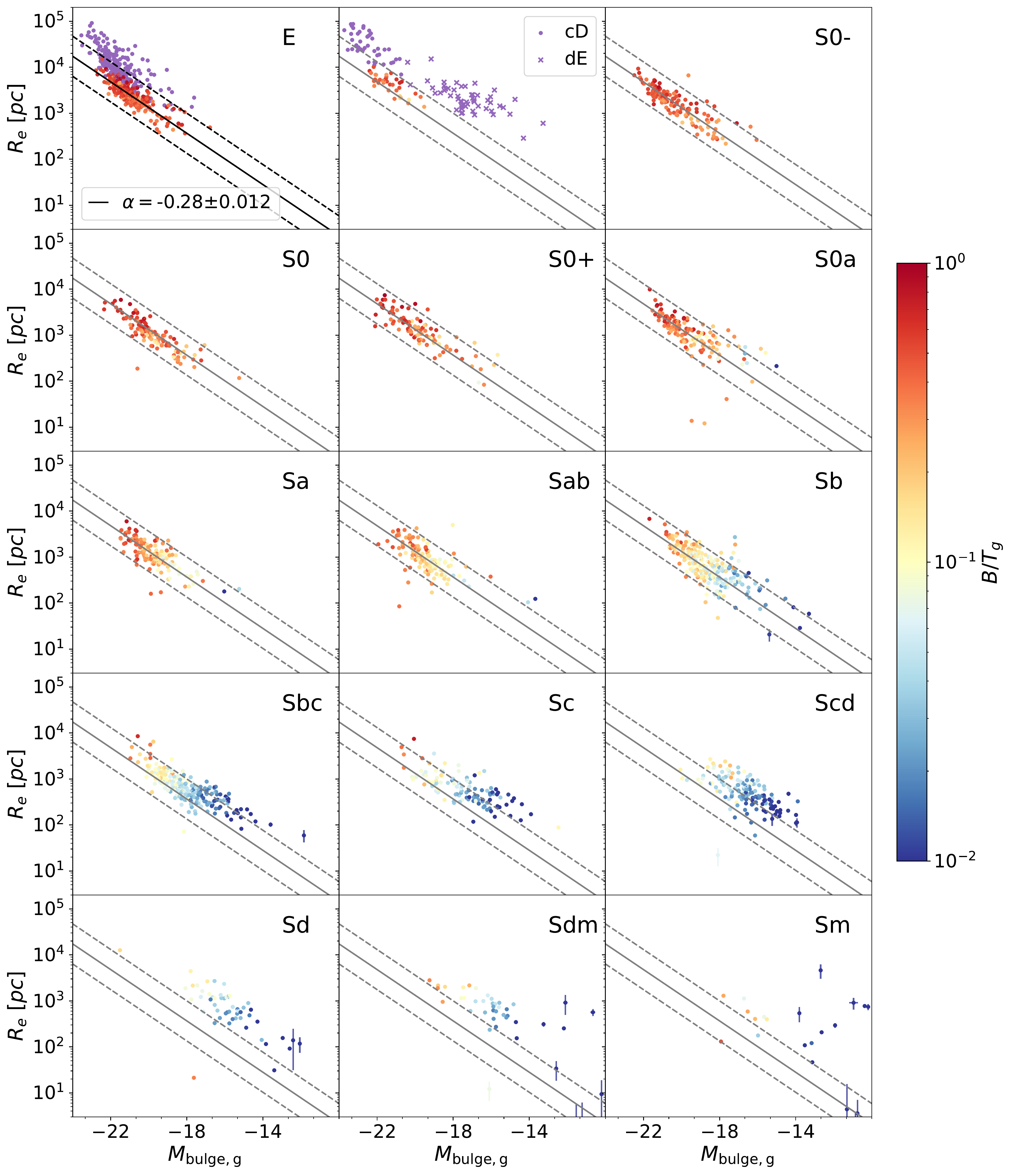}
\caption{Size-luminosity relation for the S\'ersic components of EFIGI E and cD galaxies, and for the bulges of lenticular and spiral types with {\tt Inclination} $\leq 2$, all in the $g$ band. The purple points in the 2 upper left panels represent the same relation for the E, cD and dE galaxies modeled as a single S\'ersic profile. The solid line in the upper-left panel shows the linear fit of $\log(R_e)$ as a function of $M_g$ for all EFIGI E S\'ersic components, and the dashed lines are offset by $\pm 3$ times the \rms dispersion in $\log(R_e)$ around the fit. These solid and dashed lines are repeated in gray in all other panels. The color of the points represents the bulge-to-total luminosity ratio in the $g$ band, ${B/T}_g$. Both effective radii and luminosities of bulges get smaller while spanning the Hubble sequence.}
\label{binggeli-per-type}
\end{center}
\end{figure*}

Another feature regarding the departure from the Kormendy relation is suggested by \cite{Allen-2006-MGC-BD-decomp} based on the Millenium Galaxy Catalogue: from their bulge and disk decompositions, the authors claim that bluer bulges tend to be below the relation compared to redder ones. However their \fg 18 depicting this effect shows a very small deviation between the two clouds of points corresponding to bluer and redder bulges in $u-r$ color, whose overlap is moreover not quantified. An agreement with \cite{Allen-2006-MGC-BD-decomp} would require that EFIGI bulges of later types be bluer. We do measure systematic variations of the $g-r$ bulge color along the Hubble sequence, but there is no clear color shift of bulges across the surface brightness versus effective radius plane. \fg\ref{bulge-color} shows that intermediate spirals host redder bulges than lenticulars in $g-r$, then the trend inverts itself. Indeed, there is a reddening of bulges between the lenticular types with a mean $g-r$ color of $0.67-0.69$ up to Sbc types with a $0.85$ mean ($8 \sigma$ difference between the S0 and Sbc types). This reddening could result from the presence of larger amounts of dust in intermediate spirals, as shown by the color-coding of the points with the value of the {\tt VisibleDust} attribute. The offset between the largest amounts of {\tt VisibleDust} in Sb types and the peak of the mean reddening trend in $g-r$ bulge color for Sbc types could result from the fact that this attribute estimates the presence of dust in the whole galaxy which is often dominated by disk dust, and does not necessarily represent well the dust impacting the bulge light. For later Hubble types, $g-r$ decreases down to $0.73$ for Scd types (with a $3.3 \sigma$ difference between the Sbc and Scd types), similar to the $0.74$ mean color for both Sa and Sab types. The bulges embedded in  later types of spirals (Sd to Sm) exhibit even bluer colors, but the low statistics and the larger individual  uncertainties, due to the difficulty to measure their faint bulges, do not allow for conclusive results. We note that this bulge reddening in $g-r$ is also detected in $g-i$. However, $NUV-r$ would be a better choice than optical colors to differentiate stellar populations \citep{Quilley-2022-bimodality}, but bulge and disk decomposition in the near ultraviolet has not yet been performed.

\begin{figure*}
\includegraphics[width=0.9\columnwidth]{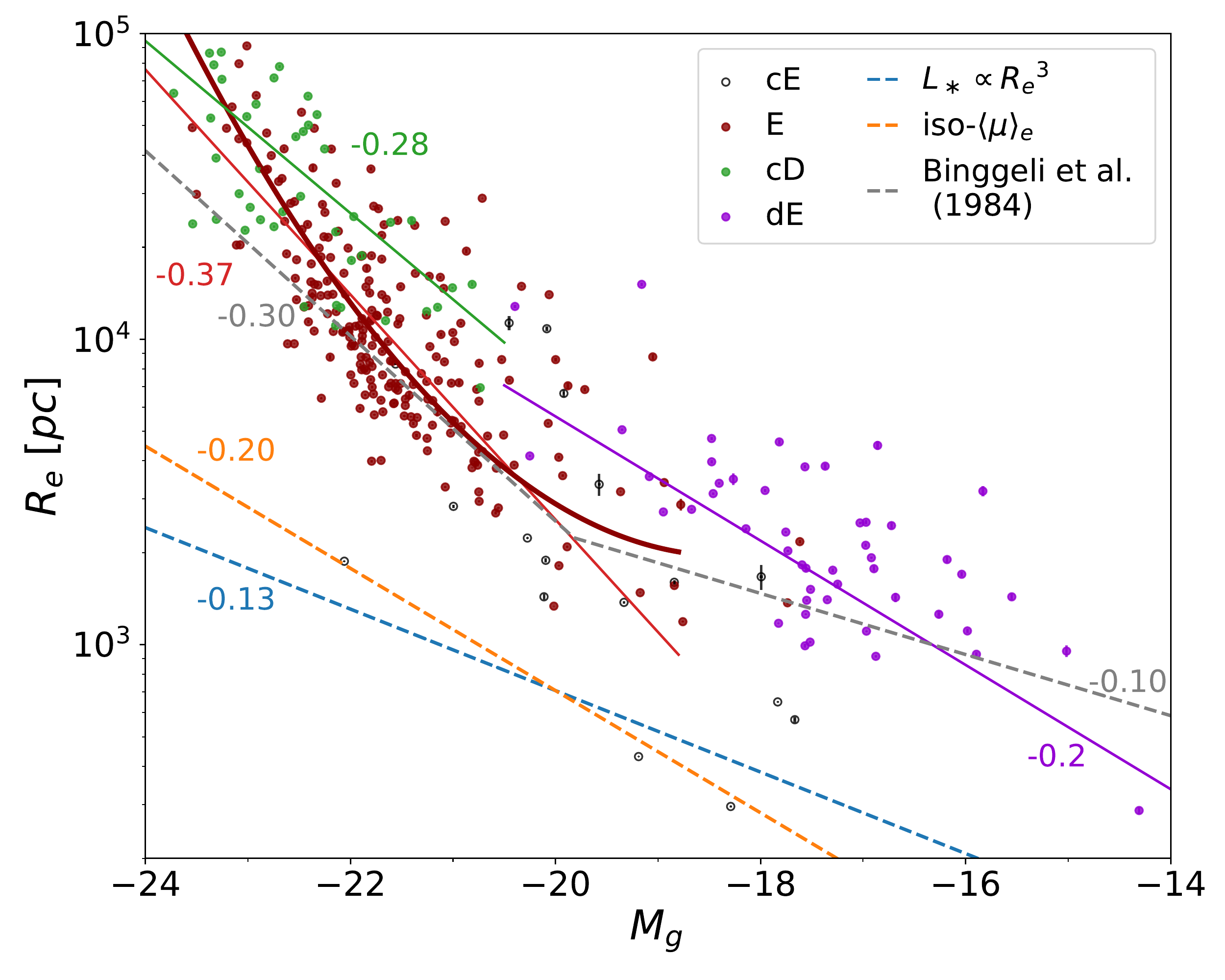}
\includegraphics[width=1.1\columnwidth]{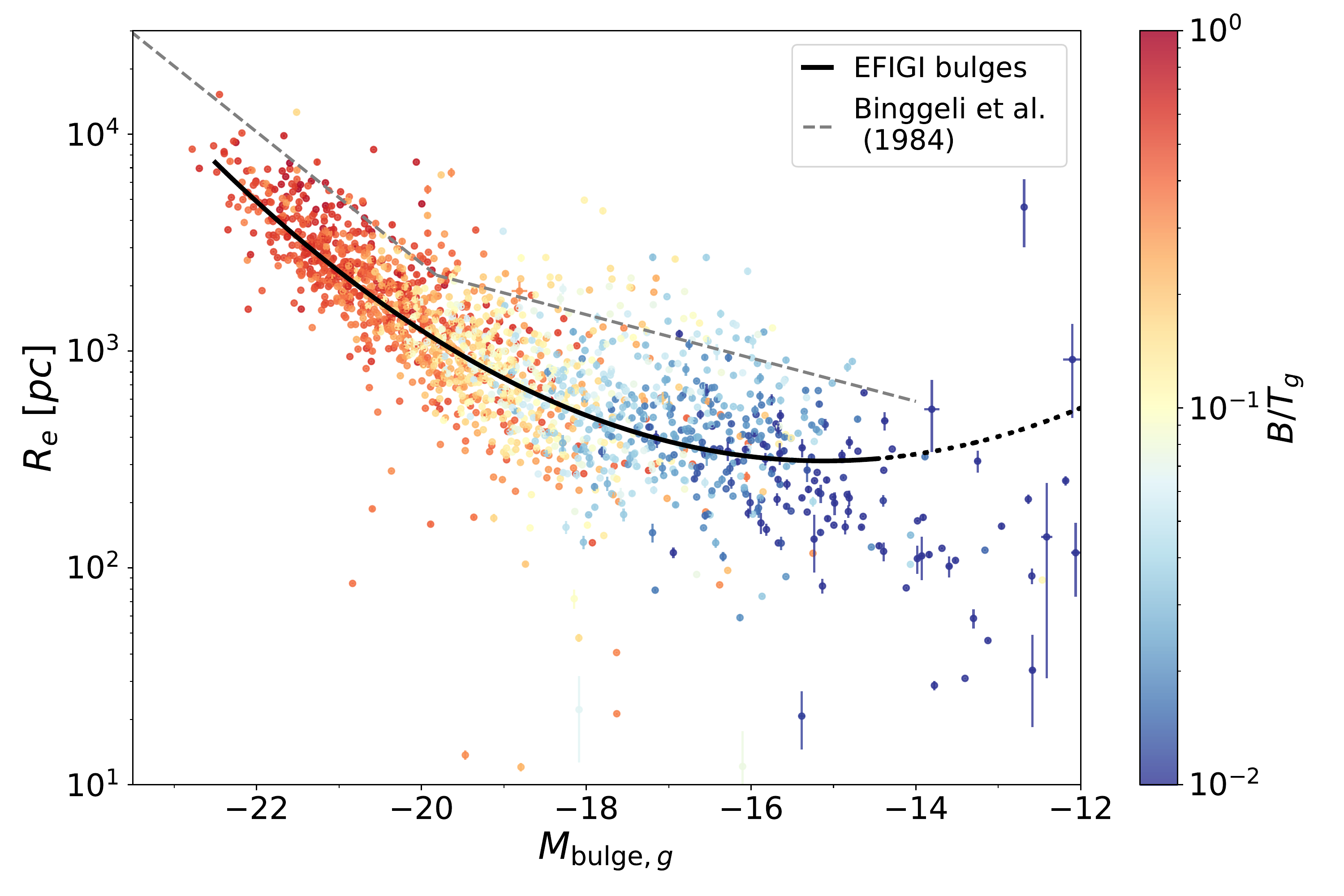}
\caption{Size-luminosity relations for elliptical, cD, cE and dE galaxies, and bulges of lenticulars and spirals. \textit{Left:} Size-luminosity relation for the single-S\'ersic fits to the E, cD, cE and dE galaxies color-coded by type (with effective radius and absolute magnitude as measures of the size and luminosity respectively), and their corresponding linear fits. A second order fit for E galaxies (thick dark red line) is also plotted. \textit{Right:} Size-luminosity relation for the S\'ersic components of E and cD types and the bulges of lenticular and spiral types with {\tt Inclination} $\leq 2$. The color of the points represent ${B/T}_g$, the bulge-to-total luminosity ratio in the $g$ band. The solid black line is the second degree polynomial fit to all points. In both panels, the dashed gray lines are the historical fits from \cite{Binggeli-1984}, using the intercepts defined in the text.}
\label{binggeli-cmap-BT}
\end{figure*}

\subsection{Revisiting the size-luminosity relation for pure spheroids     \label{sct-results-binggeli}}

In complement to the Kormendy relation between $\langle\mu\rangle_e$ and $R_e$, \cite{Binggeli-1984} brought to light a correlation between $R_e$ and $M$ for a sample of nearby E and dE galaxies (which is further described below), and is called the ``size-luminosity'' relation. The upper left panel of \fg\ref{binggeli-per-type} shows the relation between the effective radii $R_e$ in logarithmic scale and the $g$-band absolute magnitude $M_g$ for the S\'ersic components fitted with SourceXtractor++ to EFIGI E galaxies. The points are again color-coded as a function of  ${B/T}_g$, the bulge-to-total flux ratio in the $g$ band. A linear fit to these E components using the ODR package yields
\begin{equation}
    \log R_e = -0.279 ^{\pm0.012} M_g - 2.457 ^{\pm0.242}
    \label{eq-binggeli-E-bulges}
\end{equation}
also shown in the graph (solid line), along with the $\pm 3$ times the \rms dispersion in $\log R_e$ around that fit (dashed lines). In this first panel, we also show the distribution of $R_e$ versus $M$ for E galaxies modeled with a single S\'ersic profile (in purple). The upper central panel of \fg\ref{binggeli-per-type} similarly shows the distribution of $R_e$ and $M$ for the S\'ersic components and the single S\'ersic profiles of both the cD and dE types (in purple).

In the other panels of \fg\ref{binggeli-per-type}, we show  the relation between $R_e$ and $M_{\mathrm{bulge},g}$ both derived from the bulge and disk SourceXtractor++ profile modeling, separately for each morphological type from S0$^-$ to Sm (also color-coded with ${B/T}_g$), compared to the fit of the S\'ersic components for E types (upper left panel and \eq\ref{eq-binggeli-E-bulges}). One can see that there are similar size-luminosity relations for the bulges of disk galaxies (that is lenticulars and spirals). As morphological types advance along the Hubble sequence, bulges have smaller $R_e$ Moreover, while bulges of types until Sb follow the E relation, there is a progressive departure toward fainter magnitudes at a given $R_e$ for later types, similarly to what is observed for the Kormendy relation (see \sct\ref{sct-results-kormendy}). The color-coding of the points in \fg \ref{binggeli-per-type} by the bulge-to-total flux ratio in the $g$ band ${B/T}_g$ shows that for each lenticular and spiral type, there is a ${B/T}_g$ positive gradient for larger and brighter bulges. This is further explored in \sct\ref{sct-results-size-evol} and \fg\ref{B-and-D-radius-distrib}.

In the left panel of \fg\ref{binggeli-cmap-BT}, we gather on the same graph the variation of $R_e$ versus $M_g$ for EFIGI E (in dark red), cD (in green), dE (in purple) and cE types (as black open circles), derived from the single-S\'ersic profile fits to these types. We also include the ODR linear fits for E galaxies (in red)
\begin{equation}
    \log R_e = -0.368 ^{\pm0.017} M_g -3.955 ^{\pm0.368}
    \label{eq-binggeli}
\end{equation}
for cD galaxies (in green)
\begin{equation}
    \log R_e = -0.282^{\pm 0.046} M_g - 1.791^{\pm 1.032}
    \label{eq-binggeli-cD}
\end{equation}
and dE galaxies (in purple)
\begin{equation}
    \log R_e = -0.203^{\pm 0.027} M_g + -0.319^{\pm 0.488}
    \label{eq-binggeli-dE}
\end{equation}
No size-magnitude relation is fitted to the cE galaxies as they are too few and too dispersed for a fit to be meaningful. We note that the slope of the fit to cD types (\eq\ref{eq-binggeli-cD}) is flatter than that of the linear fit to E galaxies (\eq\ref{eq-binggeli}), but not at a significant level ($1.8 \sigma$). cD galaxies are located among the brightest and largest E galaxies, but are limited by the poor statistics of this rare type, hence are not further discussed in this study.

It is interesting to compare our derived size-magnitude relations to those obtained by linear regression by \cite{Binggeli-1984} for a sample of E and dE from the Virgo Cluster, E and dE from the local group, and dwarf spheroidal satellites of the Milky Way. \cite{Binggeli-1984} measured slopes of -0.3 and -0.1 by fitting $\log R_e$ as a function of absolute magnitude in the $B$ band for their sample galaxies brighter and fainter than $\sim -20$ respectively (with $H_0=\SI{50}{km.s^{-1}.Mpc^{-1}}$), with no distinction of type, leading to $M_g = -19.82$ (with $H_0=\SI{70}{km.s^{-1}.Mpc^{-1}}$ and using the $B$ to $g$ color correction from \citealp{Fukugita-1995-colors}). The dashed gray lines in \fg\ref{binggeli-cmap-BT} show both fits from \cite{Binggeli-1984}, while the intercept values (not provided in the article) were chosen to match the default parameters of the Stuff software for generating synthetic galaxies \citep{Bertin-2009-skymaker}, with an $R_e$ break value between the 2 regimes scaled to 3.35 kpc using $H_0=\SI{70}{km.s^{-1}.Mpc^{-1}}$, as used in the present article (see \sct\ref{sct-intro}).

\begin{figure*}
\includegraphics[width=\columnwidth]{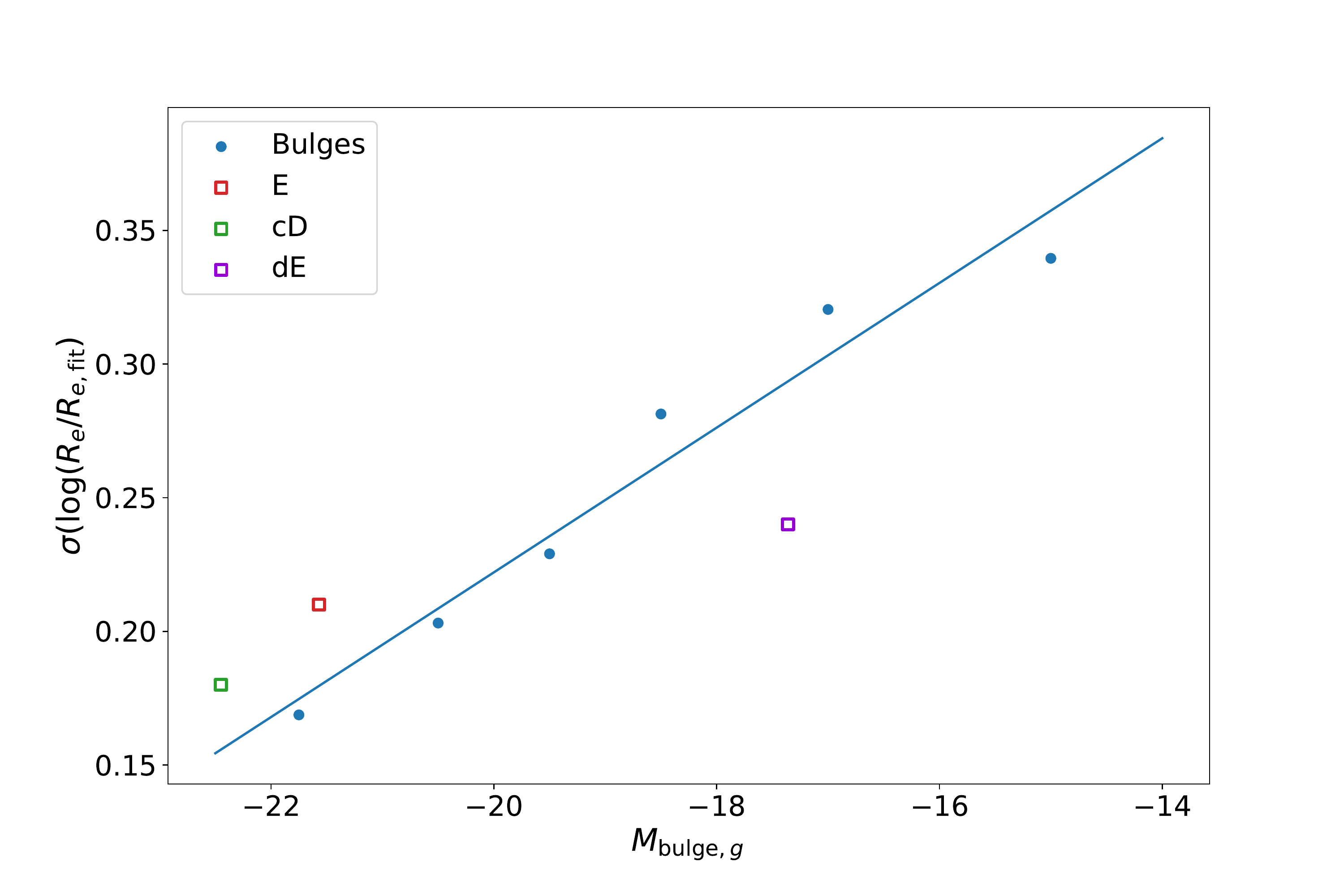}
\includegraphics[width=\columnwidth]{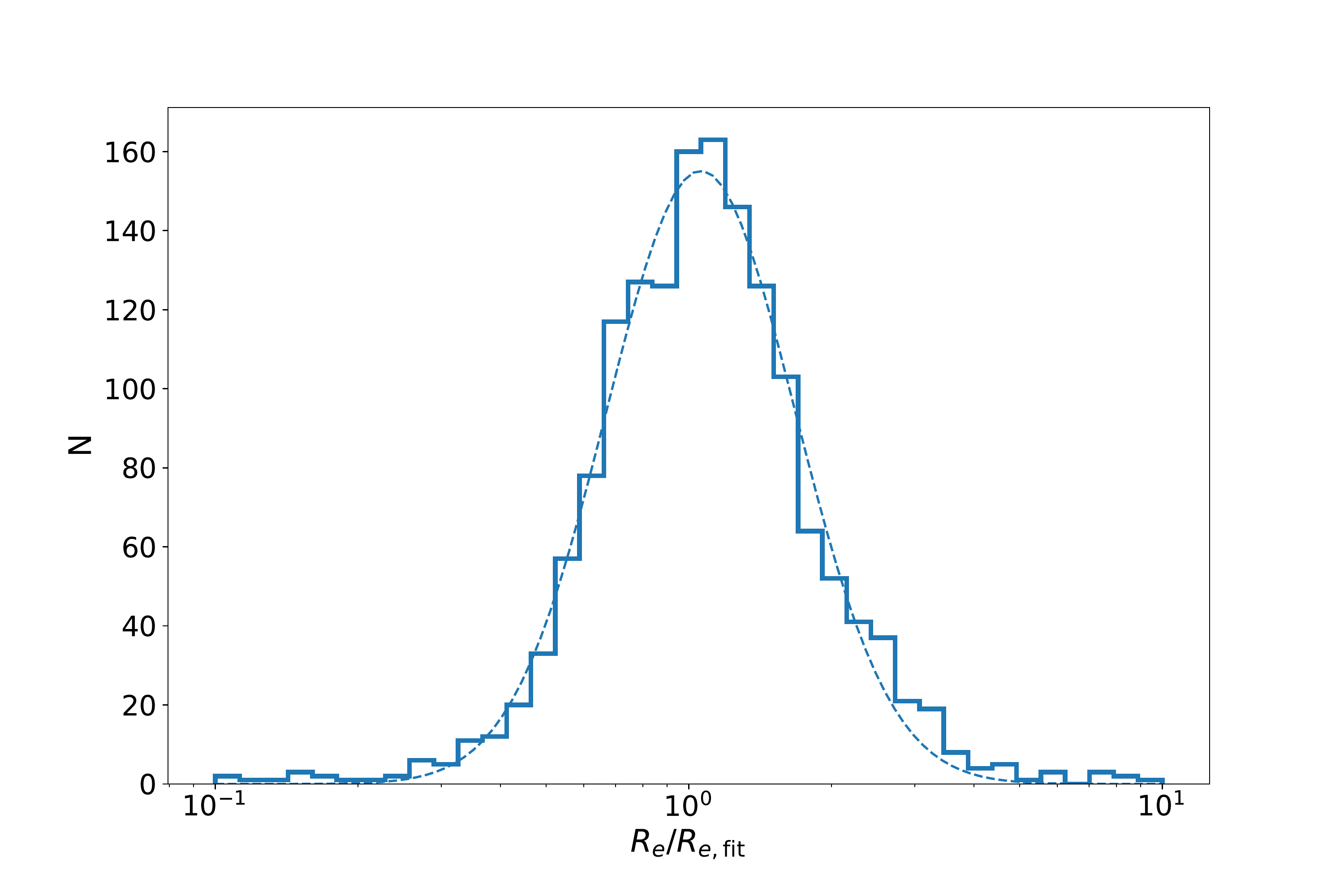}
\caption{Dispersion around the size-luminosity relations for the EFIGI S\'ersic and bulge components from \fg\ref{binggeli-cmap-BT}. \textit{Left}: Variation of the \rms dispersion in the logarithm of the ratio between the actual bulge effective radius $R_e$ and the fitted value $R_{e,\mathrm{ fit}}$ as a function of bulge magnitude $M_{{\rm bulge}, g}$ around the size-luminosity relation (\eq\ref{eq-binggeli-E-2nd-deg}) for the S\'ersic components and bulges of all types of EFIGI galaxies plotted in the right panel of \fg\ref{binggeli-cmap-BT}. The estimated \rms dispersion in $R_e/R_{e, \mathrm{fit}}$ increases for fainter bulges, and can be approximated by a linear regression. For comparison, the dispersion around the second-degree fit for the single-S\'ersic profile of E galaxies (\eq\ref{eq-binggeli-E-bulges}), and the linear fits to cD and dE galaxies (\eqs\ref{eq-binggeli-cD} and \ref{eq-binggeli-dE}) at the mean magnitude of galaxies of each type are shown with different symbols and colors. \textit{Right}: Histogram of $R_e/R_{e, \mathrm{fit}}$ for the S\'ersic components and bulges of all types of EFIGI galaxies. In order to account for the increasing dispersion around the fit seen in the left panel, the values of $\log (R_e/R_{e, \mathrm{fit}})$ are divided by the dispersion in the magnitude bin in which they lie, then renormalized to the average over the values for all magnitude intervals.}
\label{dispersion-Re}
\end{figure*}

Given the various linear size-magnitude relations plotted in the left panel of \fg\ref{binggeli-cmap-BT}, we first note a steeper slope $\alpha=-0.368$ for the size-magnitude relation of EFIGI E galaxies (\eq\ref{eq-binggeli}) compared to the $-0.3$ value measured by \citet{Binggeli-1984} for galaxies brighter than $-20$ in the $B$ band, which we estimate as a $2.8\sigma$ difference. Here again, as the authors do not provide errors on the derived slope, which may be larger than the one in \eq\ref{eq-binggeli} ($0.017$) due to the smaller statistics of their sample compared to EFIGI, we use this latter underestimated error for this fit by \citet{Binggeli-1984}, and we proceed similarly below for all comparisons with their results. For the dE galaxies, which dominate below $M_g > -19$, we compare the EFIGI dE slope in \eq\ref{eq-binggeli-dE} to the one obtained by \cite{Binggeli-1984} for E and mostly dE galaxies fainter than $M_B = -20$, that is $-0.10$, which is half the slope we measure, and differs from it by $3.4\sigma$. It is likely that the slope differences for E and dE types between EFIGI and \citet{Binggeli-1984} are due to the nonlinearity of the photographic plates that they used, as well as their profile extraction based on growth curve calculated from the two-dimensional galaxy surface brightnesses, and extracted from photographic plates using a microphotometer. The difference in photometric bands with EFIGI, and their limited sample of 109 E and dE in total, compared to 171 E and 48 dE used for the EFIGI fits, may also play a role in the discrepancies. 

We also compare the size-luminosity relation fitted to the S\'ersic components of EFIGI E galaxies in \fg\ref{binggeli-cmap-BT} (\eq\ref{eq-binggeli-E-bulges}) to both the size-luminosity relations for EFIGI E galaxies modeled by a single S\'ersic profile (\eq\ref{eq-binggeli}), and that of \cite{Binggeli-1984}, with the limitation of a varying fraction of the galaxy light taken into account. The slope obtained for the S\'ersic components is compatible with \cite{Binggeli-1984} results with a $1.75\sigma$ difference, but differs more strongly from the slope obtained for EFIGI single-S\'ersic fits to E galaxies, with a $4.3\sigma$ difference.

Nevertheless, we obtain for EFIGI galaxies the same qualitative result as \citet{Binggeli-1984}, namely that the slope for the size-magnitude relation of dE galaxies (\eq\ref{eq-binggeli-dE}) is flatter than for E galaxies (\eq\ref{eq-binggeli}), with a $5.7\sigma$ difference. We also note that EFIGI E and dE types dominate at $M_g$ magnitudes fainter than $-19$ and brighter than $-20$, respectively, which is not discordant with the interpretation by \cite{Binggeli-1984} that the slope break near absolute magnitude -20 is due to a change in surface brightness of elliptical galaxies. Indeed, the break in the size-magnitude relations of E and dE types may result from the markedly different profiles of the E and dE galaxies : for EFIGI E galaxies we measure a single-S\'ersic profile index in the interval $n=3.5$-$7$ with a peak at $n\simeq5.5$, whereas it is in the interval $n=1$-$3$ for dE galaxies with a peak at $n\simeq1.5$.

The dashed orange line in the left panel of \fg\ref{binggeli-cmap-BT} with a slope of -0.2 corresponds to a fixed surface brightness (see \eq\ref{eq-mu-scaling}), which is nearly identical to the slope for dE (\eq\ref{eq-binggeli-dE}). It is not the case for the E types with a steeper $-0.37\pm0.04$ slope (\eq\ref{eq-binggeli}), which indicates a varying mean $\mu_e$ within the population, as expected from the Kormendy relation (see \sct\ref{sct-results-kormendy}, \fgs\ref{kormendy-g-per-type-cmap-BT} and \ref{kormendy-cmap}). The dashed blue line with a slope of $-1/7.5 = -0.13$ in the left panel of \fg\ref{binggeli-cmap-BT} corresponds to the case of a scale-invariant spheroid for which the total luminosity grows as the cube of the radius. All slopes for the E, cD, dE types in the left panel of \fg\ref{binggeli-cmap-BT} are steeper than this ideal case of a scale-invariant spheroid. The implications are further discussed in \scts\ref{sct-discussion-diffuse-E} and \ref{sct-discussion-diffuse-B}.

At last, and because the linear fit to the E galaxies of \eq\ref{eq-binggeli}, shown as a red line in the left panel of \fg\ref{binggeli-cmap-BT}, would underestimate the effective radius of galaxies with $M_g<-22.7$, we also add in this graph an ODR second degree fit to the single-S\'ersic fit of EFIGI E galaxies
\begin{equation}
    \log R_e = 0.062^{\pm 0.010} M_g^{-2} + 2.268^{\pm 0.420} M_g - 24.093^{\pm 4.463} 
    \label{eq-binggeli-E-2nd-deg}
\end{equation}
which is steeper and better matches the E types at the bright ($M_g \lesssim-22.5$) and faint ends ($M_g\gtrsim20.$) than the linear fit in \eq\ref{eq-binggeli}. We quantify this improved fit by calculating the residuals of the $R_e/R_{e,\mathrm{fit}}$ ratios for  $R_{e,\mathrm{fit}}$ given by either \eq\ref{eq-binggeli} or \eq\ref{eq-binggeli-E-2nd-deg}, for the $M_g$ values of the considered sample. In both cases, the distribution of $\log(R_e/R_{e,\mathrm{fit}})$ in bins of 0.1 dex can be fitted by Gaussian distributions centered at -0.045 and -0.032, with standard deviations  0.207 and 0.186, and reduced $\chi_2$ of 3.3 and 1.9 for the linear and second degree fits,  respectively (with some skewness beyond $\pm0.3$ dex for both).

\subsection{The size-luminosity relation for bulges     \label{sct-results-binggeli-bulge}}

\begin{figure*}
\includegraphics[width=\columnwidth]{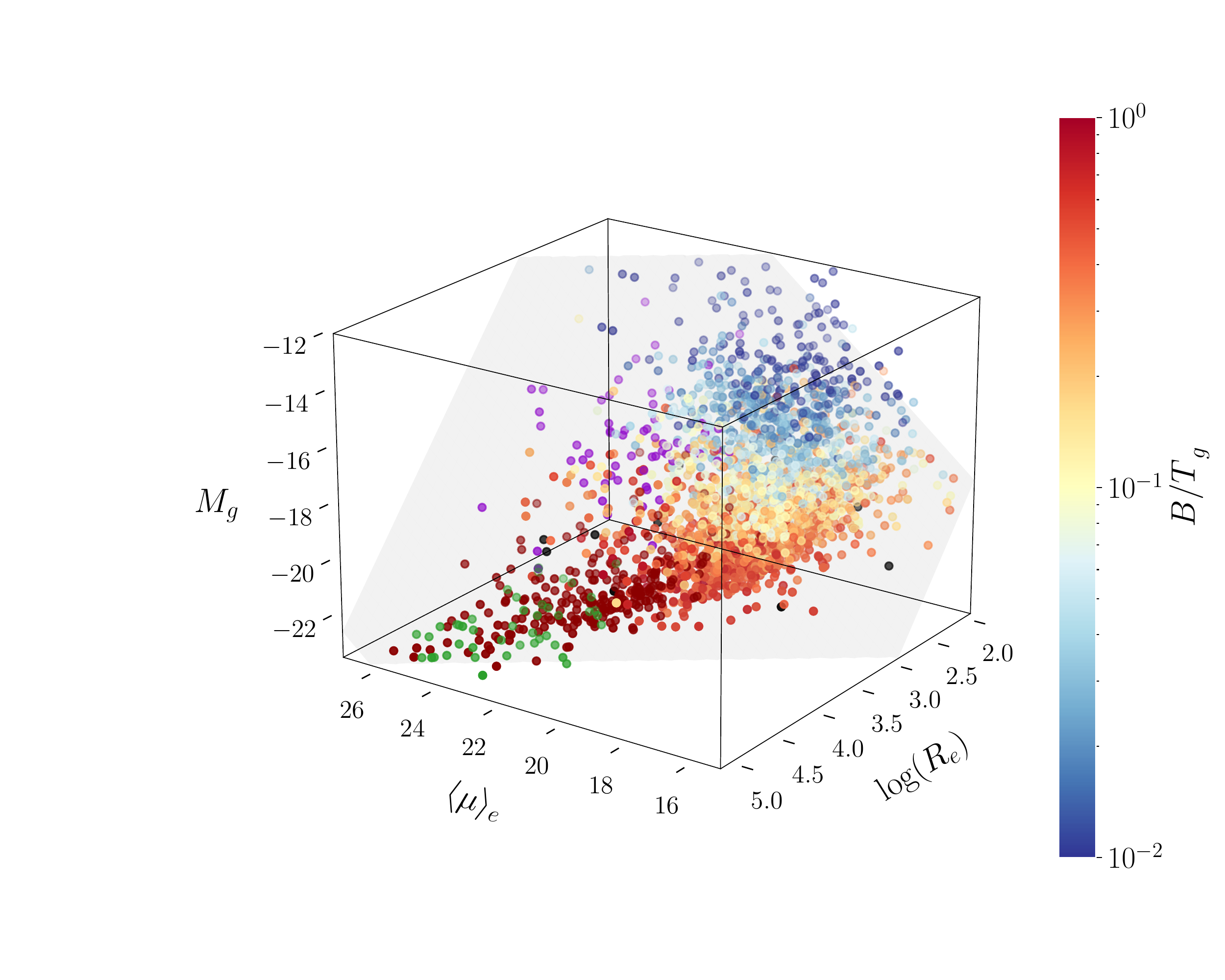}
\includegraphics[width=\columnwidth]{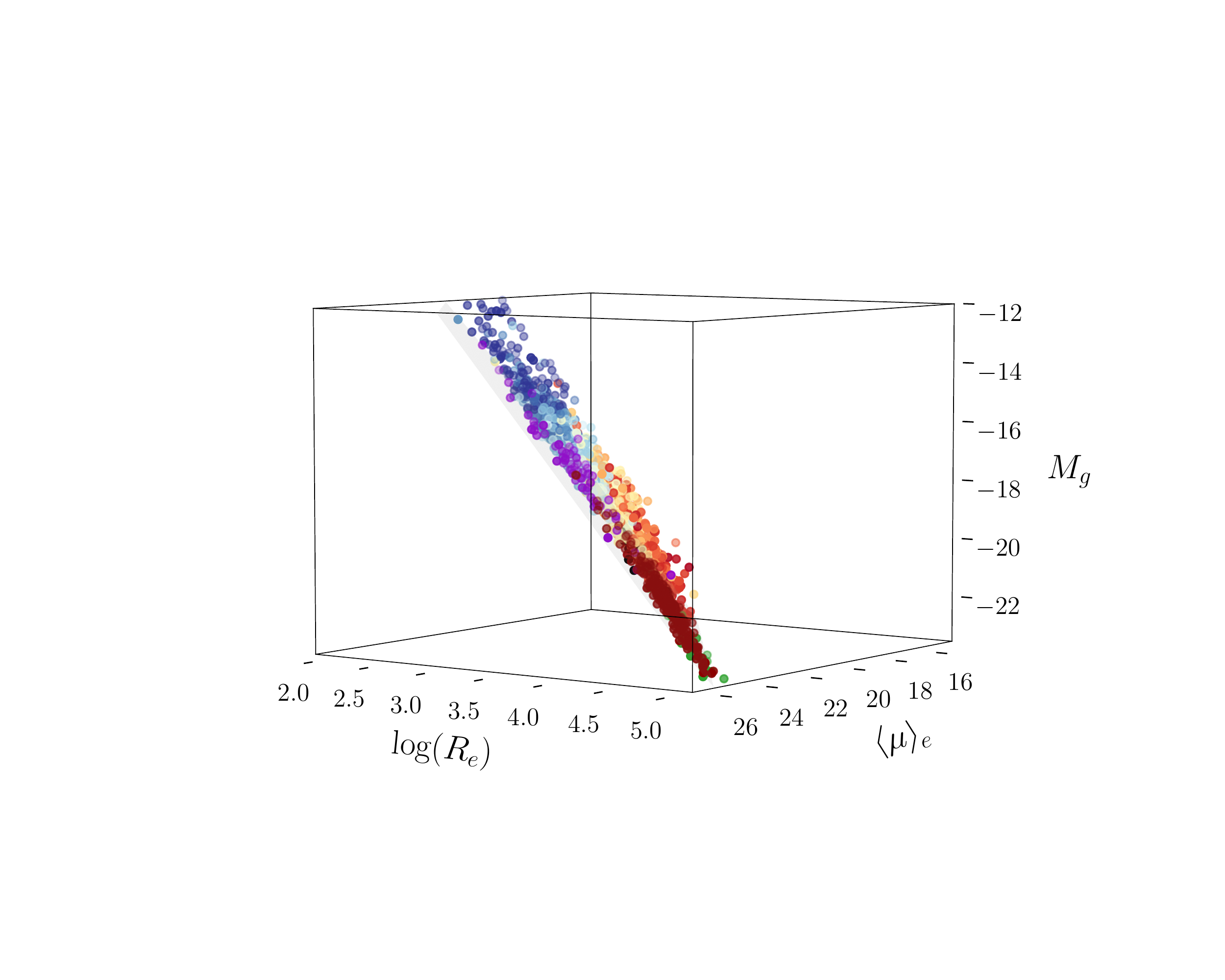}
\caption{Distribution in the $\langle\mu\rangle_e$, $R_e$, $M_g$ tridimensional space of E (dark red), cD (purple), dE (green), cE (black) galaxies using their single-S\'ersic fits, and of the bulges of all lenticular and spiral types from bulge and disk decompositions color-coded by ${B/T}_g$, the bulge-to-total luminosity ratio in the $g$ band. Two different views are shown on the left and right, in order to highlight the fact that galaxies are distributed along a plane, as predicted from \eq\ref{eq-mu-scaling}, with a small dispersion around it due to redshift: the gray plane is drawn for $z=0$ and is best seen in the left panel. The Kormendy and size-luminosity relations are projections against the corresponding faces of the cube.}
\label{3D-figure}
\end{figure*}

In the right panel of \fg\ref{binggeli-cmap-BT}, we plot the effective radii versus the magnitudes of the bulges from the bulge and disk decomposition for all galaxies. The bulge data points are again color-coded by ${B/T}_g$, the bulge-to-total luminosity ratio of each galaxy in the $g$ band. As already seen in \fg\ref{binggeli-per-type}, ${B/T}_g$ determines the position of bulges in this 2D plane. The right panel of \fg\ref{binggeli-cmap-BT} shows that both the luminosity and radii of the bulges continuously decrease as ${B/T}_g$ decreases from $\lesssim1$ to $10^{-2}$, down to $M_{\mathrm{bulge},g} > -17$. At lower luminosities, $R_e$ decreases less steeply as the luminosity decreases. This bending of the trend justifies the use of a second degree polynomial fit rather than a linear model for the size-luminosity relation of EFIGI bulges. The result of this fit appears as a black solid line and has the following equation:
\begin{align}
\begin{split}
    \log R_e &= 0.025^{\pm 0.0014}\; M_{\mathrm{bulge},g}^2\\
    &+ 0.762^{\pm 0.053}\; M_{\mathrm{bulge},g} + 8.253^{\pm 0.498}
    \label{eq-size-lum-bulge}
\end{split}
\end{align}

We now examine the dispersion around the size-luminosity relation of the EFIGI bulges presented in the right panel of \fg\ref{binggeli-cmap-BT}. We first compute for all bulges the ratio of $R_e$ to the value $R_{e,\mathrm{fit}}$ given by \eq\ref{eq-size-lum-bulge} for the $M_{\mathrm{bulge},g}$ bulge magnitude of each EFIGI galaxy. We then calculate the \rms dispersion around the value of 1 in log-scale, which is the quadratic mean of $\log(R_e/R_{e,\mathrm{fit}})$, in the six following intervals of $M_{\mathrm{bulge},g}$: $[-22.5,-21]$, $[-21,-20]$, $[-20,-19]$, $[-19,-18]$, $[-18,-16]$ and $[-16,-14]$. Left panel of \fg\ref{dispersion-Re} shows the variation in these dispersions as a function of the mean $M_{\mathrm{bulge},g}$ for each interval. For bright bulges, the dispersion is the lowest and is also similar to those  measured around the single-S\'ersic size-luminosity relations for cD, E and dE galaxies, also plotted in the graph. There is a systematic increase in the dispersion for fainter $M_{\mathrm{bulge},g}$, which we fit using a linear regression (as a blue line in the graph), whose coefficients are given in Table~\ref{tab-fits-size-lum}. In the right panel of \fg\ref{dispersion-Re}, we show the histogram of the values of the log-ratios $\log(R_e/R_{e,\mathrm{fit}})$ for all EFIGI galaxies in the sample, divided by the dispersion in the bulge magnitude interval in which they lie, and renormalized by the mean dispersion ($0.27$) over the six $M_{\mathrm{bulge},g}$ intervals. This histogram can be fitted by a Gaussian with a central offset of only $0.025$ dex in $R_e/R_{e,\mathrm{fit}}$, an \rms dispersion of $0.20$ dex, and a reduced $\chi^2 = 1.508$, hence validating the Gaussian shape of the residual distribution.

\subsection{Understanding the surface brightness, effective radius and absolute magnitude relationships for E galaxies        \label{sct-results-3D-space}}

Both the Kormendy and Binggeli relation are actually 2D projections of a 3D relation in the parameter space $\langle\mu\rangle_e$, $R_e$, $M$, where galaxies are distributed along a plane. This is illustrated by the approximate relation \eq\ref{eq-mu-scaling}, which is the equation of a plane.
\fg\ref{3D-figure} shows this plane from two different angles: face-on (left panel) and edge-on (right panel). There is a small dispersion perpendicular to the plane which is due to the redshift surface brightness dimming, the K-correction and the profile elongation, that we all neglect when deriving \eq\ref{eq-mu-scaling} from \eq\ref{eq-full} (see \sct\ref{sct-methodo-math}). This plane is not homogeneously populated: most E galaxies (in dark red) have $M_g$ in the range $[-23;-20]$, they span 2 dex in $R_e$ but are mostly within $\log R_e \in [3.3;4.3]$ while the range of surface brightness $\langle \mu \rangle_e$ is large and encompass $\sim 6$ magnitude. The cD galaxies are among the most massive and largest E galaxies. On the other side, prominent bulges, mostly found in lenticulars, are mixed with the smallest and faintest E galaxies (here we consider the parameters from the single-S\'ersic fits to E, cD, dE et cD types, and from the bulge and disk decomposition for lenticulars and spirals). As the ${B/T}_g$ ratio decreases, bulges get smaller and fainter (in magnitude), but their effective surface brightness $\langle \mu \rangle_e$ has a more complex behavior as seen with the Kormendy relation in \fg\ref{kormendy-cmap}: it brightens for decreasing ${B/T}_g$ with ${B/T}_g \gtrsim 0.1$, then dims for ${B/T}_g \lesssim 0.1$.

\begin{figure*}
\centering
\includegraphics[width=0.95\textwidth]{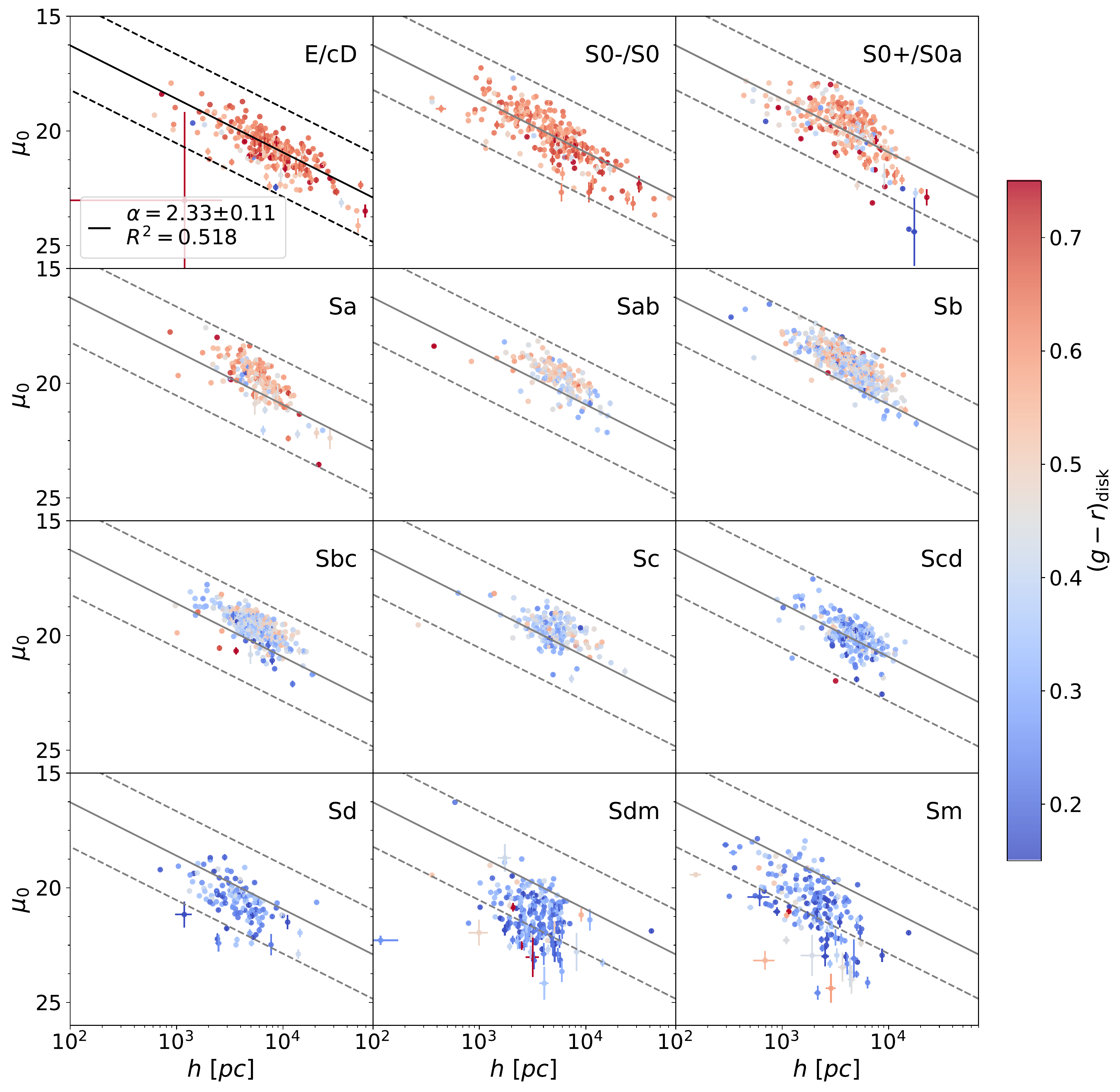}
\caption{Central surface brightness $\mu_0$ versus scale-length $h$ for the disk (or exponential) component of the different Hubble Types. The linear fit obtained for E and cD types (upper left panel) is shown in all panels as solid lines, along with the parallel dashed lines at $\pm 3$ times the \rms dispersion in $\mu_0$ around the fit. Points are color-coded with the $g-r$ color of the disk (or exponential) component. For all lenticular types as well as early and intermediate spirals up to Scd, $h$ is correlated with $\mu_0$ with different intercept values, but a common slope would be an acceptable approximation to each distribution. For Sdm and Sm late-type spirals, the distribution is more dispersed.}
\label{de-jong-g-per-type}
\end{figure*}

\subsection{Novelty of disk scaling relations \label{sct-results-disk-scaling-relations}}

\cite{Freeman-1970} modeled the luminosity profile of galactic disks using an exponential profile and found an approximately constant central surface brightness $\mu_0$ of $21.65\pm 0.3$ mag arcsec$^{-2}$ for 28 of the 36 disks considered, even though they cover 5 magnitudes and span the Hubble sequence from S0 to Im types. Such a nearly constant surface brightness for very different disks would strongly constrain their formation scenario based on angular momentum considerations. 
However, \cite{De-Jong-1996} disproved this result by examining, for nearly face-on disks, the distributions of their $\mu_0$ and scale-lengths $h$, that fully parameterizes the variation of the mean surface brightness in an exponential disk (see \eq\ref{eq-exp-h}).
The low statistics of \cite{De-Jong-1996} did not allow him to perform any fit but both panels of his \fg 5 showing $\mu_0$ versus $h$ in the $B$ and $K$ bands respectively, seem to show that disks with a larger $h$ are dimmer. 

Using the EFIGI large statistical samples of all morphological types with improved profile modeling, we show in \fg\ref{de-jong-g-per-type} the relations between $\mu_0$ and $h$ in the $g$ band for the exponential profile of all EFIGI elliptical, spiral and lenticular galaxies decomposed into the sum of a S\'ersic bulge and an exponential disk. We perform a linear fit using the ODR package (see \sct\ref{sct-methodo-odr}) for E and cD types taken together and obtain the relation
\begin{equation}
    \mu_0 = 2.33^{\pm 0.11} \log h + 18.60^{\pm 0.42} 
\end{equation}
with $h$ in kpc, shown as a black solid line in the top left panel of \fg\ref{de-jong-g-per-type}. The fit is repeated in gray in the other panels, showing the variations of $\mu_0$ versus $h$ for the disks of all lenticular and spiral types, in order to guide the eye for comparisons between Hubble types. The joint E and cD fit (shown in the upper left panel) could match the S0$^-$, S0, S0$^+$, S0a and Scd types, whereas the disk of all other types have a different behavior: early and intermediate spirals (Sa to Sc) follow a similar slope but with a brighter zero-point than for E-cD (and S0$^-$-S0), while disks of types Sd and later have a  weaker correlation between $\mu_0$ and $h$, and a fainter zero-point than for E-cD (and S0$^-$-S0).

We have examined the $\mu_0$ versus $\log h$ relations in the $r$ and $i$ filter and note that these shifts are filter-dependent for spiral types: the Scd fall mostly below the joint S0$^-$-S0 fit in the $r$ and $i$ bands, whereas the Sc types match this fit in both bands, and Sbc match it in $i$ only. There is nevertheless no visual change in zero-point for the S0$^+$ -S0a compared to the joint E and cD fit in the $r$ and $i$ bands. This is due to the fact that the elliptical and lenticular types (including S0a) have similar colors, as shown by the color coding of the points by $g-r$ disk color in \fg\ref{de-jong-g-per-type}, whereas the disks of spiral types become bluer and bluer for later and later types along the Hubble sequence.

\begin{figure}
\includegraphics[width=\columnwidth]{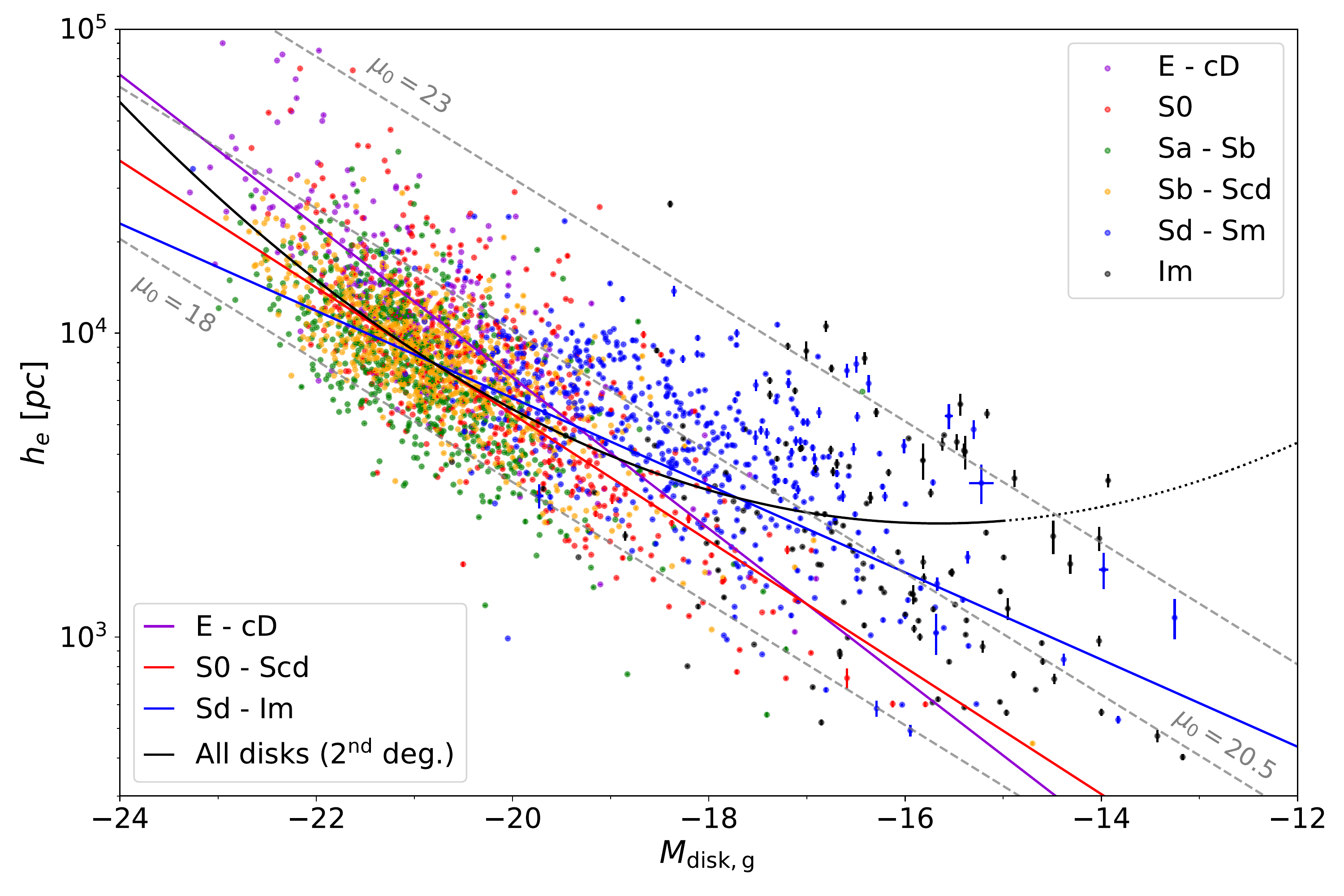}
\caption{Effective disk radius versus absolute $g$ magnitude for the EFIGI disk (or exponential) components. The color of the points correspond to groups of Hubble type. Dashed gray lines are iso-$\mu_0$ lines, of slope -0.2, which for an exponential profile corresponds to a disk luminosity scaling as ${h_e}^2$. They allow to see that disks span $\sim 6$ dex in central surface brightness at all magnitudes. The solid lines are ODR fits: three linear models for disks of types E-cD, S0-Scd and Sd-Im, in purple, red and blue respectively, whereas the black line is the second degree polynomial fit of $\log h_e$ as a function of $M_{\mathrm{disk},g}$ for all disks. The size-luminosity relation is close to an iso-surface brightness growth for lenticulars as well as early and intermediate spirals, but there is a tail of faint disks for (dimmer) late-type spirals and irregulars with a larger size than earlier spirals at these faint magnitudes, for $M>-19$.}
\label{binggeli-disk}
\end{figure}

\begin{figure*}
\includegraphics[width=\columnwidth]{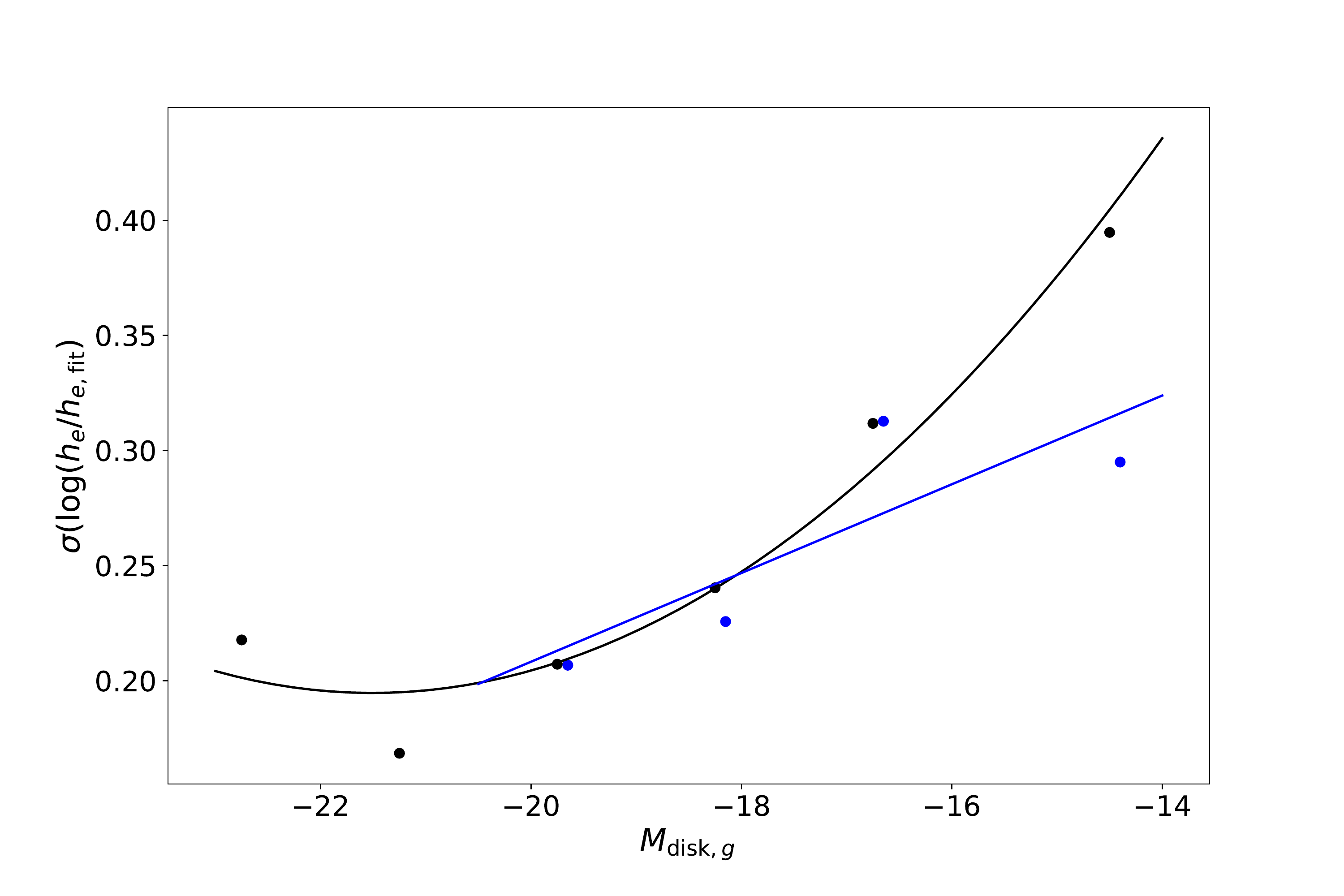}
\includegraphics[width=\columnwidth]{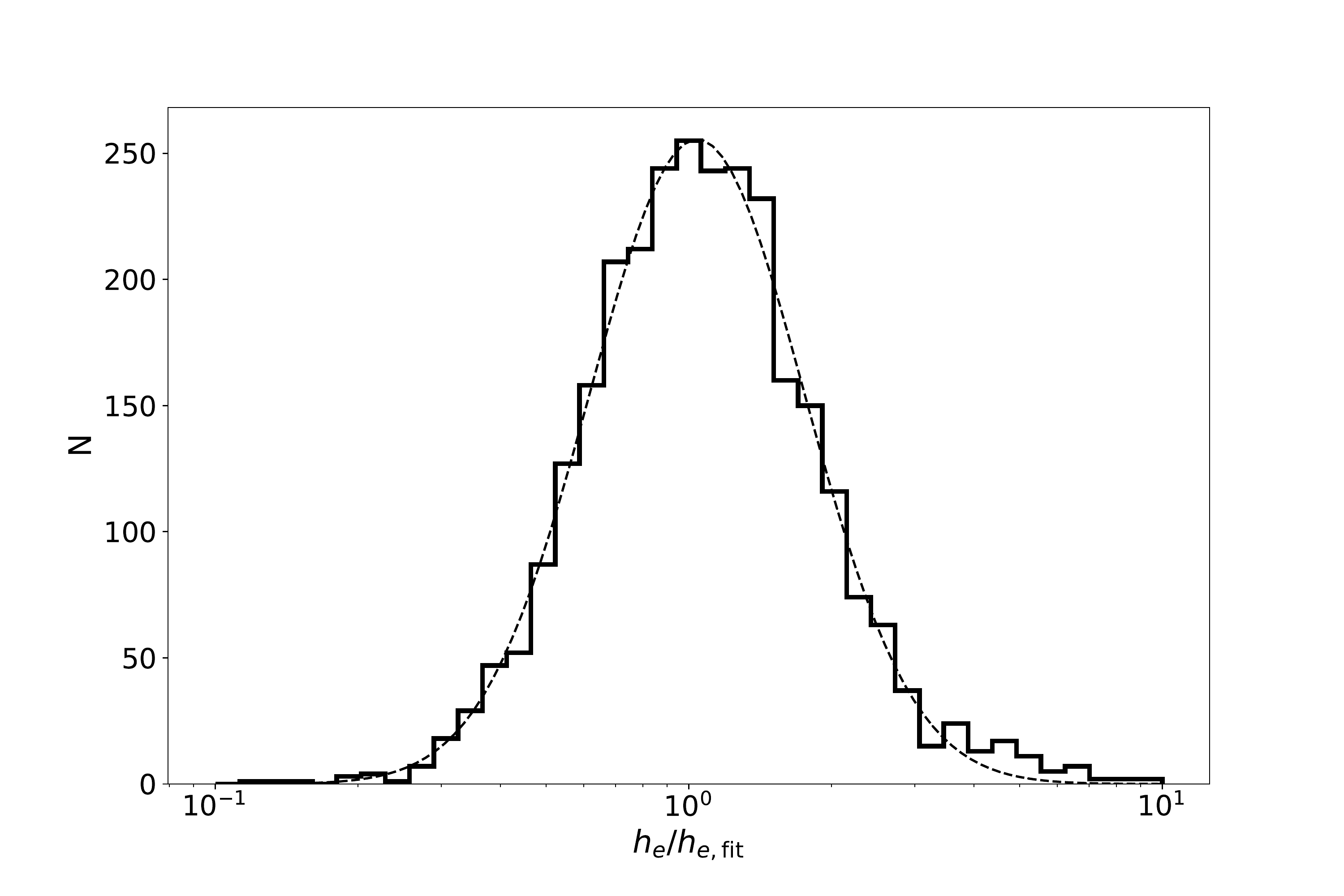}
\caption{Dispersion around the size-luminosity relations for the EFIGI exponential and disk components from \fg\ref{binggeli-disk}. \textit{Left}: Variation of the \rms dispersion in the logarithm of the ratio between the actual disk effective radius $h_e$ and the fitted value $h_{e,\mathrm{ fit}}$ as a function of disk magnitude $M_{{\rm disk}, g}$ around the size-luminosity relations plotted in \fg\ref{binggeli-disk} for EFIGI galaxies, as a function of the mean disk magnitude in 1.5 or 3 mag{.} intervals. The dispersion around the second degree fit is shown as black points, and that around the linear fit to Sd-Im types only as blue points (and calculated only for $M_{\mathrm{disk},g}>-20.5$). The resulting dispersion values are fitted with a second degree polynomial (black), and a linear regression (blue) respectively. In both cases, the estimated dispersion increases for fainter disks. \textit{Right}: Histogram of $h_e/h_{e, \mathrm{fit}}$ for the size-luminosity relation for all disks of EFIGI lenticulars and spirals as well as irregulars. In order to account for the increasing dispersion around the fit seen in the left panel, the values of $\log(h_e/h_{e, \mathrm{fit}})$ are divided by the dispersion in the magnitude bin in which they lie, then renormalized to the average over the values for all magnitude intervals.}
\label{dispersion-he}
\end{figure*}

We also measure the size-luminosity relation of the disks of EFIGI galaxies by examining in \fg\ref{binggeli-disk} the distribution of their disk effective radii $h_e$ versus absolute magnitude $M_{\mathrm{disk},g}$. Despite the large range of central disk surface brightness from $17.5$ to $23.5$ encompassed by these disks and illustrated with gray dashed lines of constant surface brightness (with a slope of $-0.2$, see \eq\ref{eq-mu-scaling}), EFIGI disks exhibit a correlation between the disk effective radii and their absolute magnitudes.

On one hand, a linear fit (using the ODR package) to the size-magnitude relation of disks of lenticulars and spirals of Scd type and earlier yields
\begin{equation}
    \log h_e = -0.208^{\pm 0.004} M_{\mathrm{disk},g} - 0.434^{\pm 0.084} 
    \label{eq-size-lum-disk1}
\end{equation}
shown in red in \fg\ref{binggeli-disk}. This fit is close to the iso-surface brightness trend characterized by a slope of -0.2, thus indicating that the luminosities of these disks scale as $h^2$: $M_{\mathrm{disk}}\sim5\log h_e$ (see \eq\ref{eq-mu-scaling} and \sct\ref{sct-discussion-diffuse}). The dispersion around this fit can be parameterized by a large range of $\mu_0$ values ($2.8$ magnitude/arcsec$^2$ for $90\%$ of the galaxies). 

On the other hand, disks of Sd and later spiral types, that are fainter and bluer (see \fg\ref{de-jong-g-per-type}, deviate at $M_{\mathrm{disk},g}>-19$ from the extrapolation of the relation for earlier types (\eq\ref{eq-size-lum-disk1}), with systematically larger disk radii and fainter surface brightness. We also include the Im galaxies (in black) modeled as a single S\'ersic in \fg\ref{binggeli-disk}, as they appear to extend the size-magnitude relation of late-type disks. The ODR linear fit to Sd, Sdm, Sm and Im types altogether is 
\begin{equation}
    \log h_e = -0.140^{\pm 0.007} M_{\mathrm{disk},g} + 0.983^{\pm 0.145}
    \label{eq-size-lum-disk2}
\end{equation}
shown in blue in \fg\ref{binggeli-disk}. The dispersion around this fit can be parameterized by an even larger range of $\mu_0$ than for earlier type spirals ($4.4$ magnitude/arcsec$^2$ for $90\%$ of the galaxies). Moreover, there is a $7.3\sigma$ difference between the slopes in \eqs\ref{eq-size-lum-disk1} and \ref{eq-size-lum-disk2}, validating the two different trends. 

We also show in \fg\ref{binggeli-disk} the measured effective radii versus absolute $g$-band magnitude for the exponential component of the E and cD types (in purple). A linear fit (also shown in purple) to these data points using the ODR package yields
\begin{equation}
    \log h_e = -0.249^{\pm 0.012} M_{\mathrm{disk},g} - 1.123^{\pm 0.266} 
    \label{eq-size-lum-E-cD}
\end{equation}
The E and cD components exhibit larger sizes than the disks of lenticulars and early spirals at the bright-end of their size-luminosity relation. Therefore, its slope is steeper than in the fits for the early disks (\eq\ref{eq-size-lum-disk1}). These components consequently have fainter surface brightnesses than lenticulars and early spirals, with values in the $[19;23]$ mag arcsec$^{-2}$ interval, similarly to the disks of the latest spiral types, but with a much smaller contribution to the total galaxy light.

The decreasing slopes of the size-magnitude relations with morphological types in \ref{eq-size-lum-disk1}, \ref{eq-size-lum-disk2} and \eq\ref{eq-size-lum-E-cD} justify that we perform a second degree polynomial fit of $\log h_e$ as a function of magnitude for the exponential or disk or single S\'ersic component of all types altogether (E, cD, all lenticulars, all spirals, Im\footnote{Because Im types are fitted with single S\'ersic profile, it is their total absolute magnitude that is plotted along the x-axis of \fg\ref{binggeli-disk}, despite its labeling as $M_{\mathrm{disk},g}$.}), which yields the following relation: 
\begin{align}
\begin{split}
    \log h_e = &0.020^{\pm 0.0019}\; M_{\mathrm{disk},g}^2 \\
    &+ 0.623^{\pm 0.078}\; M_{\mathrm{disk},g} + 8.25^{\pm 0.808}
    \label{eq-size-lum-disk}
\end{split}
\end{align}
The dispersion around this fit can be parameterized by an even larger range of $\mu_0$ values than for earlier type spirals ($\sim3$ magnitude/arcsec$^2$). 

To quantify the increasing spread in surface brightness of EFIGI disks at fainter magnitudes seen in \fg\ref{binggeli-disk}, we examine the dispersion in $h_e$ with disk magnitude (as in \sct\ref{sct-results-binggeli-bulge}). As in \eq\ref{eq-size-lum-disk}, we include in this calculation the Im single-S\'ersic profile magnitudes. We first compute for all disks as well as the E and cD exponential component, the ratio of $h_e$ to the value $h_{e,\mathrm{fit}}$ given by \eq\ref{eq-size-lum-disk}. We then calculate the \rms dispersion around the value of 1 in log-scale, which is the quadratic mean of $\log(h_e/h_{e,\mathrm{fit}})$, in the six following intervals of $M_{\mathrm{disk},g}$ (or magnitude of the E-cD exponential component, or Im single-S\'ersic total magnitude): $[-23.5,-22]$, $[-22,-20.5]$, $[-20.5,-19]$, $[-19,-17.5]$, $[-17.5,-16]$ and $[-16,-13]$. Left panel of \fg\ref{dispersion-he} shows the variation in these dispersions as a function of the mean magnitude for each interval. There is a systematic increase in the dispersion for fainter disks, which we fit using a second degree regression (shown in the graph as the black solid line), and whose coefficients are given in Table~\ref{tab-fits-size-lum}. 

We also calculate similarly the dispersion around the linear fit of \eq\ref{eq-size-lum-disk2} restricted to the disks of Sd, Sdm, Sm types and the Im single S\'ersic profiles that dominate the faint-end of the size-luminosity relation in \fg\ref{binggeli-disk}. A linear model to the dispersion in $\log(h_e/h_{e,\mathrm{fit}})$ around the fit in \eq\ref{eq-size-lum-disk2} computed in the four faintest magnitude intervals, and plotted in blue, appears sufficient at magnitudes fainter than $-20.5$ for these late types, as shown in \fg\ref{dispersion-he}. The dispersion around this linear fit does not increase as  steeply as for the second degree polynomial fitted to the dispersion for all disk types. In the right panel of \fg\ref{dispersion-he}, we show the histogram of the $h_e/h_{e,\mathrm{fit}}$ ratios around the second degree size-luminosity relation of \eq\ref{eq-size-lum-disk}, for all EFIGI galaxies in the sample. The $\log(h_e/h_{e,\mathrm{fit}})$ are divided by the dispersion in the disk magnitude bin they lie in, and renormalized by the mean dispersion ($0.211$) over the six $M_{\mathrm{disk},g}$ intervals. This histogram can be fitted by a Gaussian with a central offset of only $0.015$ dex in $h_e/h_{e,\mathrm{fit}}$, an \rms dispersion of $0.20$ dex, and a reduced $\chi^2 = 1.864$, hence validating the Gaussian shape of the residual distribution.

\subsection{Bi-variate luminosity-radii distribution function for disks\label{sct-results-disk-bivar}}

\begin{figure}
\includegraphics[width=\columnwidth]{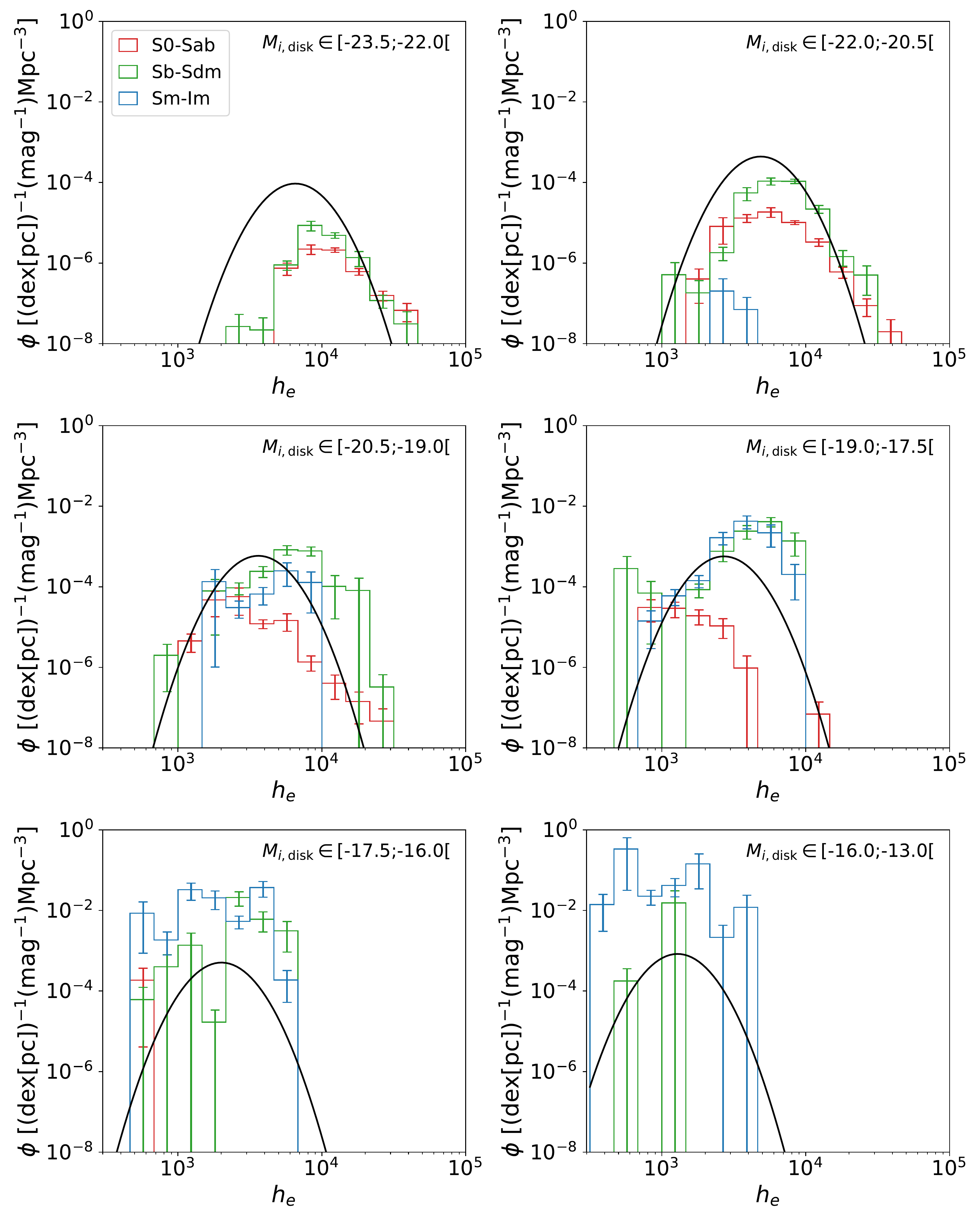}
\caption{Spatial density of disk effective radius for EFIGI morphological types grouped as S0$^-$-Sab, Sb-Sdm, Sm-Im (the effective radii of single-S\'ersic fits are used for the Im), compared to the expected density from the bi-variate luminosity-disk radius function proposed by \cite{de-Jong-Lacey-2000-spiral-galaxies-functions} for a sample of $\sim 1007$ Sb-Sdm galaxies. For each magnitude interval, its mean magnitude is used to derive the plotted analytical curve.}
\label{histograms-he-mag}
\end{figure}

We now compare the EFIGI disk sizes with those obtained by \citet{de-Jong-Lacey-2000-spiral-galaxies-functions} for a sample of 1007 Sb-Sdm spirals with $z < 0.025$ (widely distributed on the sky). We display  in \fg\ref{histograms-he-mag} the distribution of EFIGI galaxies as a function of $h_e$, per interval of disk absolute magnitude $M_{\mathrm{disk},i}$, and per group of morphological types. Here we use EFIGI magnitudes in the $i$ band, in order to compare with the \citet{de-Jong-Lacey-2000-spiral-galaxies-functions} analysis performed in the Cousins $I_C$ band. All galaxies fainter than $m_i=15.5$ are excluded, and those remaining are weighted by $w/V_{\mathrm{max}}$, where $w$ is the incompleteness correction, calculated as the ratio of galaxies, per bin of 0.5 apparent $i$ magnitude, between a power law fitted in the $8.5\le i\le14$ interval on the number counts of the complete magnitude limited MorCat sample (see \sct\ref{sct-data}) and EFIGI number counts. The volume correction $V_{\mathrm{max}} = \Omega \frac{4\pi}{3}D_{\mathrm{lum, max}}^3$ is obtained using $\Omega=6670$ deg$^2$, the solid angle of sky covered by EFIGI \citep{Baillard-2011-EFIGI}, and $D_{\mathrm{lum, max}}$ the luminosity distance to a galaxy with absolute magnitude $M_i$ if its apparent $i$ magnitude was equal to the survey magnitude limit $m_{i, {\rm lim}}=15.5$, chosen for the present calculation (we thus use \eq\ref{eq-mag-def} and the k-correction of each object). The $V_{\mathrm{max}}$ weighting allows one to correct for the fact that galaxies of fainter absolute magnitudes (hence later types among the star-forming galaxies) are visible out to shorter distances (hence a smaller volume) than brighter galaxies, therefore providing a fair comparison of volume densities of the different galaxy types. \fg\ref{histograms-he-mag} shows the resulting density distributions of $h_e$ measured in the SDSS $i$ band for EFIGI S0$^-$ to Sab types (in red), Sb to Sbm types (in green) and Sm to Im types (in blue): the Sb-Sdm grouping corresponds to the types studied by \citet{de-Jong-Lacey-2000-spiral-galaxies-functions}, while the S0$^-$-Sab and Sm-Im types group the earlier and later lenticular and spiral types, respectively, along the Hubble sequence in the EFIGI sample (the Im are fitted as single S\'ersic profiles). We also chose the SDSS $i$ magnitude intervals derived from those of \fg 5 of \citet{de-Jong-Lacey-2000-spiral-galaxies-functions} using a 0.51 color correction between the $i_C$ band that  they use, and the SDSS $i$ band for EFIGI \citep{Fukugita-1995-colors}.

For comparison with the results of \citet{de-Jong-Lacey-2000-spiral-galaxies-functions}, we show as black curves in \fg\ref{histograms-he-mag} the density distributions per absolute $i$ magnitude interval of the bi-variate function that they propose to model the distribution of disk effective radii for their Sb-Sdm sample. 
Given the error bars, one can see marked differences with the EFIGI distribution of disk effective radii of the corresponding type group, in the magnitude intervals of all disks (or single-S\'ersic for Im). First, in the brightest intervals $[-23.5;-22.0[$ and $[-22.0;-20.5[$, the curve over predicts the density of small disk radii compared to the data but there is agreement on the decreasing trends at large $h_e$ in both panels. Then, in the three magnitude intervals $[-20.5;-19.0[$, $[-19.0;-17.5[$ and $[-17.5;-16.0[$ the under prediction of galaxies by the curves shifts to large $h_e$. The quasi absence of EFIGI Sb-Sdm galaxies in the faintest $[-16.0;-13.0]$ interval, in which \citet{de-Jong-Lacey-2000-spiral-galaxies-functions} predict a significant volume density of objects, whereas the EFIGI density of Sm-Im galaxies is above the curve, may result from classification errors of the latest spirals in the Sb-Sdm sample of \citet{de-Jong-Lacey-2000-spiral-galaxies-functions}. As this magnitude interval does not appear in their \fg 5 (probably due to an absence of data), it could be beyond the range of validity of their proposed function. 

Moreover, we do not show in \fg\ref{histograms-he-mag} the $[-25.0;-23.5[$ $M_i$ interval of \citet{de-Jong-Lacey-2000-spiral-galaxies-functions} (corresponding to their brightest magnitude interval), because it only contains 10 EFIGI galaxies (4 S0, and 6 Sb-Sdm spirals) that are all located near the faint edge of the interval (with $M_{\mathrm{disk},i}\ge-23.83$). These bright disks all have $h_e > 10^4$ pc and no $h_e$ bin contains more than 2 galaxies. Altogether, this prevents any meaningful analysis. One may wonder whether the sample of \citet{de-Jong-Lacey-2000-spiral-galaxies-functions} contains disks which are clustered near the faint edge of the interval, and in that case the bi-variate functional form cannot be validated for $M_i\lesssim-24$, or it does include a significant number of disks brigther than $M_i=-24$. In the latter case, we suspect that a bulge and disk decomposition that underestimates the bulge contribution, hence overestimates the disk component could explain why the EFIGI sample does not contain such bright disks.

Another bias that could affect the comparison of EFIGI disk sizes with the model of \citet{de-Jong-Lacey-2000-spiral-galaxies-functions}, shown in \fg\ref{histograms-he-mag}, could be a higher threshold in surface brightness detection in their data, as these observations are based on photographic plates: at a fixed radius $h_e$, a faint absolute magnitude implies a fainter central surface brightness $\mu_0$ (see \fg\ref{binggeli-disk}), and such objects could fail to be detected. The joint bias resulting from the fact that at a fixed absolute magnitude, a larger $h_e$ radius implies a fainter central surface brightness $\mu_0$ (see \fg\ref{binggeli-disk}) could also explain the large $h_e$ under prediction of the galaxy densities by the curves derived from \citet{de-Jong-Lacey-2000-spiral-galaxies-functions} compared to EFIGI for magnitudes intervals $[-20.5;-19.0[$, $[-19.0;-17.5[$, and $[-17.5;-16.0[$. 

At last, the distribution of the other EFIGI type groupings, namely S0-Sab types and Sm-Im types show different density distributions from the Sb-Sdm types, with smaller $h_e$ for S0-Sab for $M_{{\rm disk}, i}\ge -20.5$, and a similar interval of $h_e$ for the Sm-Im, except for the two brightest magnitude intervals in which there are no or very few EFIGI galaxies of Sm-Im types.  In the 2 faintest magnitude intervals, the Sm-Im types have a similar density distribution to that of the model, but with a 1 dex higher density. The model function proposed by \citet{de-Jong-Lacey-2000-spiral-galaxies-functions} therefore does not appear appropriate to describe any of the broad groups of Hubble types considered in the EFIGI sample. We intend to derive an updated functional form in the $g$ band based on the magnitude limited MorCat catalog to $g\le15.5$, hence with higher statistics and smaller type groupings. The large incompleteness corrections performed near the $15.5$ apparent $i$ magnitude limit used for producing here \fg\ref{histograms-he-mag} will then be circumvented.

\subsection{How bulge and disk bulge prominence and size vary among Hubble types\label{sct-results-size-evol}}

In \cite{Quilley-2022-bimodality}, we showed that the Hubble sequence is an inverse evolutionary sequence, characterized by disk reddening and an increase in the bulge-to-total mass and light ratio (denoted $B/T$). Here we examine the changes in effective radii of bulge and disk that accompany these changes in color and luminosity. In all of this subsection, even though the Hubble types in all graphs are ordered from left to right along the historical sequence, we present and discuss all variations of EFIGI galaxies from right to left, that is across types from late to earlier types, as this is the main direction of galaxy evolution. 

\subsubsection{Bulge-to-total ratio growth with Hubble type\label{sct-results-size-evol-bt}}

\begin{figure}
\includegraphics[width=\columnwidth]{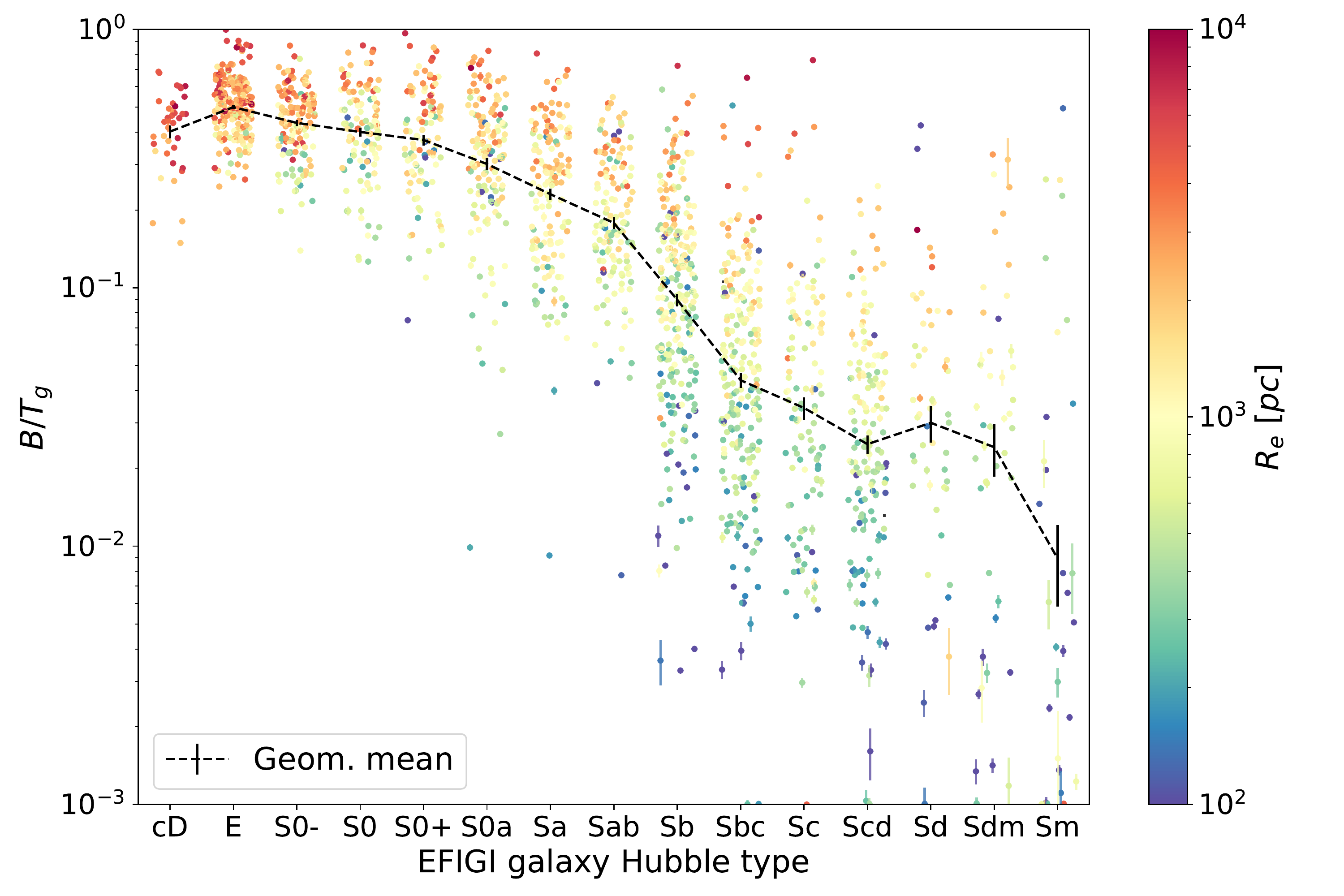}
\caption{Bulge-to-total luminosity ratio in the $g$ band ${B/T}_g$ from the bulge and disk decomposition of all EFIGI galaxies with {\tt Inclination} $\leq 2$ as a function of Hubble Type. The color of the points represent the effective radius $R_e$ of the bulge (or S\'ersic component for E and cD types). The black dashed line shows the geometric mean value per type, and the vertical bars the estimated uncertainties in this mean (see text for details). There is a correlation between Hubble type and ${B/T}_g$ with the latter increasing sharply along the sequence toward earlier types. However, there is also a significant dispersion of ${B/T}_g$ within each type, ranging from $\sim 0.25$ dex for lenticulars to almost 1 dex for late-type spirals, with the other trend that larger ${B/T}_g$ correspond to larger bulge $R_e$ overall, as well as within each type.}
\label{B-T-g-per-type-cmap-Re}
\end{figure}

The strong increase in $B/T$ toward earlier Hubble types that we highlighted in \cite{Quilley-2022-bimodality} can be seen in \fg\ref{B-T-g-per-type-cmap-Re}, showing the distribution of the bulge-to-total luminosity ratio in the $g$ band as a function of Hubble type. There is moreover a significant dispersion in ${B/T}_g$ within each type. The black dashed line shows the geometric mean\footnote{We use the geometric mean as it is less sensitive to outliers in logarithmic scale.} value per type and the associated error, which is estimated as the \rms deviation in $\log B/T_g$ divided by $\sqrt{N}$ with $N$ the number of galaxies in the type bin (we verified that it is larger than the error derived from the quadratic mean of errors on individual points). 

The frequently failed bulge fits for types Sd, Sdm and Sm, as these are very faint, are discarded (see \sct\ref{sct-results-kormendy}), leading to very low statistics and a large dispersion in ${B/T}_g$ (describing nearly the whole plotted interval $10^{-3}-1$ in \fg\ref{B-T-g-per-type-cmap-Re}) for these late spiral types. The graph then shows that for earlier types, that is from Scd to Sb galaxies, each Hubble type displays an interval of $\sim1.5$ dex in ${B/T}_g$, and a strong systematic increase of the mean ${B/T}_g$ per type from the Scd late-type spirals to earlier spiral types and lenticulars. Indeed Scd, Sc and Sbc types have geometric mean values of $B/T_g\sim 0.025-0.045$, while the mean reaches $0.09$ for Sb galaxies. The increase persists but not as steeply for earlier types, with a geometrical mean ${B/T}_g$ of $0.18$ for Sab galaxies, and reaches $0.37$, $0.40$ and $0.43$ for S0$^+$, S0 and S0$^-$ lenticular types, respectively. At last, E galaxies exhibit the highest average ${B/T}_g$ value at $0.50$. 

\begin{figure*}
\includegraphics[width=\columnwidth]{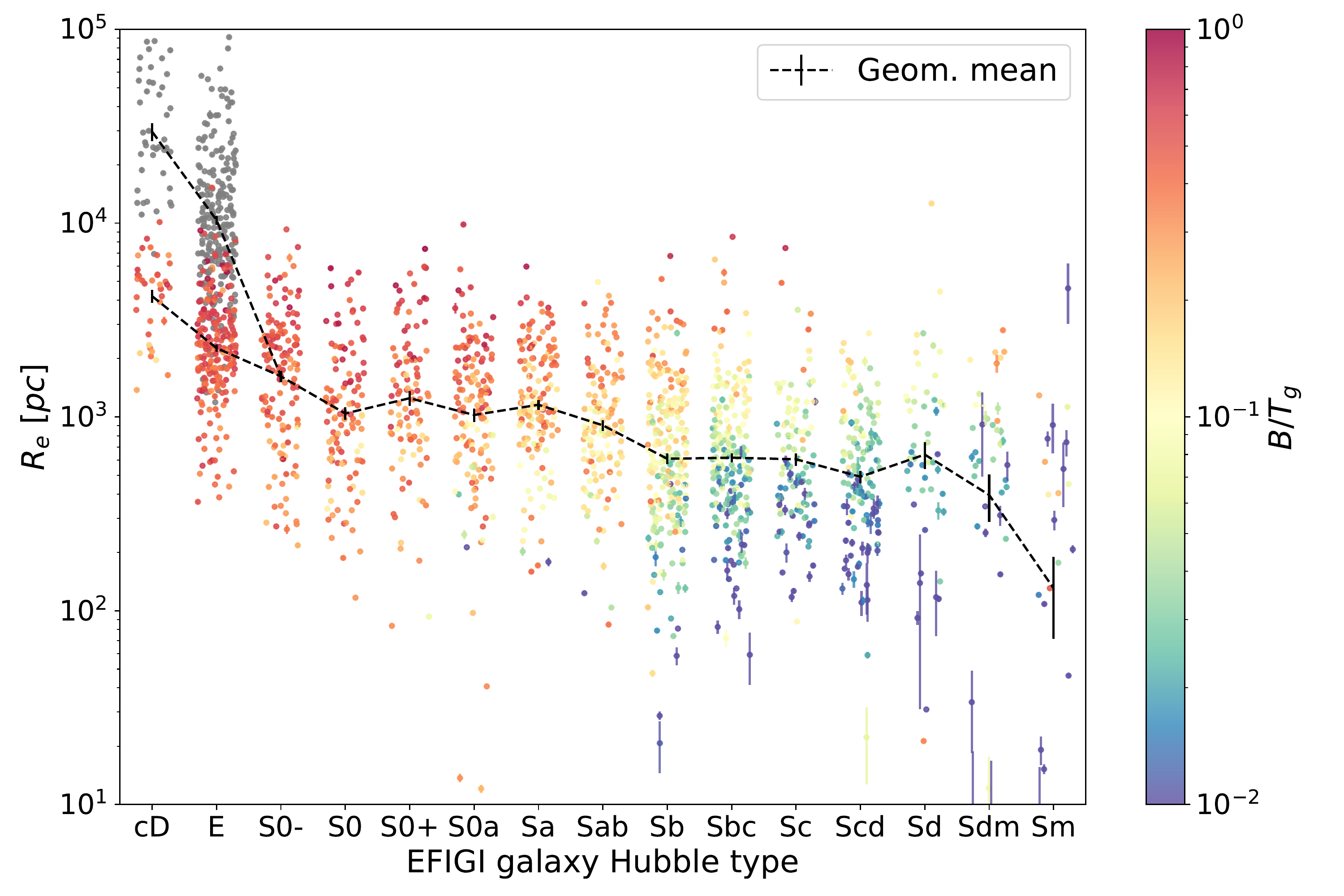}
\includegraphics[width=\columnwidth]{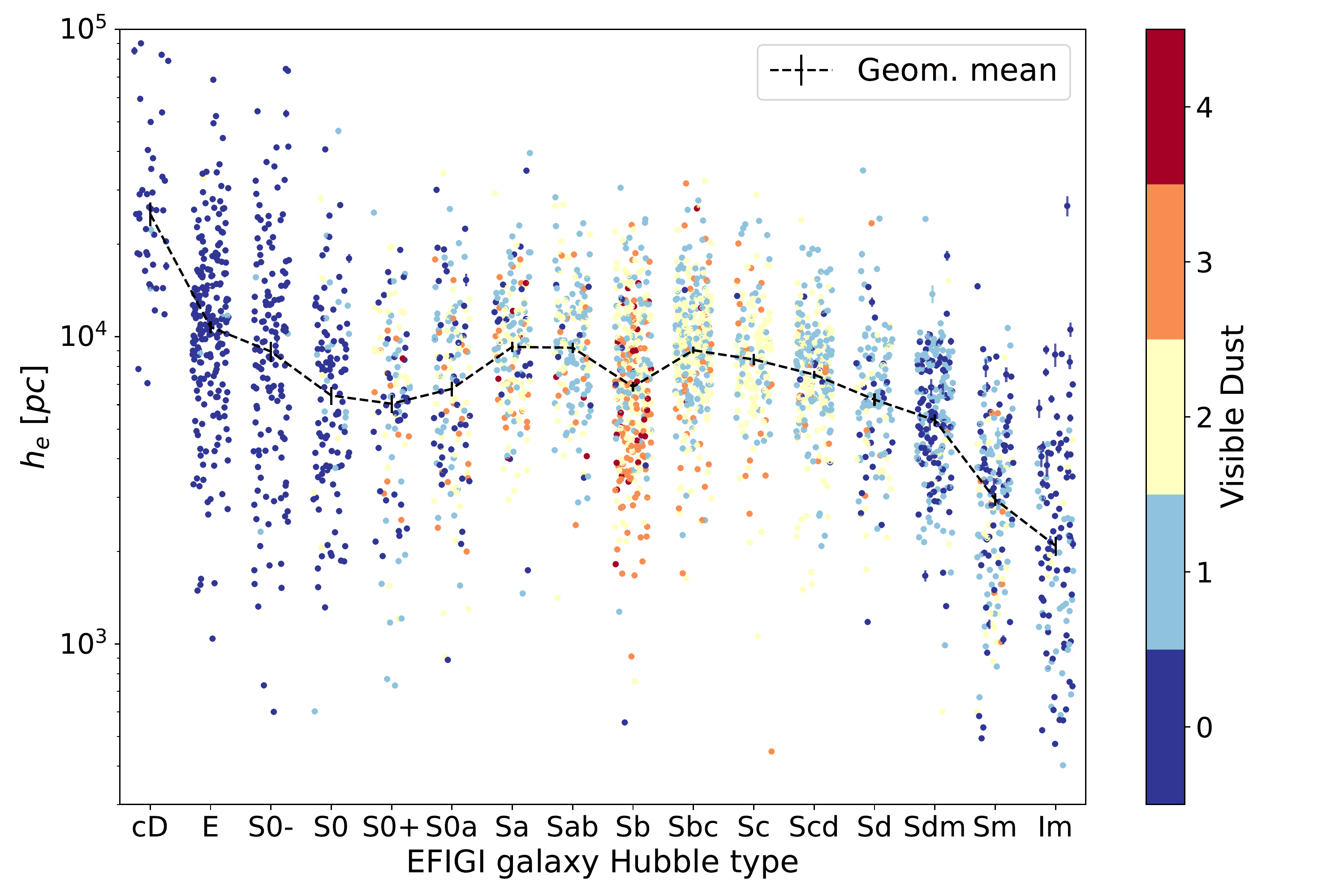}
\caption{Distributions of the effective radius $R_e$ of the bulge (or S\'ersic) component (left) and $h_e$ of the disk (or exponential) component (right) from the bulge and disk decomposition of all EFIGI galaxies with {\tt Inclination}$\leq 2$, as a function of Hubble Type. The black dashed lines represent the geometric mean value for each Hubble type, and the vertical bars the estimated uncertainties on this mean (see \sct\ref{sct-results-size-evol-bt} for details). In the left panel points are color-coded with the bulge-to-total ratio in the $g$ band ${B/T}_g$, except for the gray dots, corresponding to the single-S\'ersic profile modeling. In the right panel, Im sizes are derived from their single-S\'ersic profile modeling, and the points are color-coded with the EFIGI {\tt Visible Dust} attribute, which suggests that the lower radii measured for Sb galaxies compared to adjacent types are likely due to strong dust extinction.}
\label{B-and-D-radius-distrib}
\end{figure*}

The increase in ${B/T}_g$ for earlier types along the Hubble sequence seen in \fg\ref{B-T-g-per-type-cmap-Re} is a direct consequence of the Hubble sequence classification system which includes the visually perceived ${B/T}_g$ as one of the criteria to differentiate among the spiral types. It is also physically meaningful because the bulge growth in both ${B/T}_g$ and absolute bulge luminosity is a key factor in the evolution of galaxies, and is related to the fading of their star formation \citep{Bluck-2014-bulge-mass, Lang-2014-bulge-growth-quenching-CANDELS, Bremer-2018-GAMA-survey-morph-transf-GV, Dimauro-2022-bulge-growth, Quilley-2022-bimodality}.
\fg\ref{B-T-g-per-type-cmap-Re} also shows that the dispersion in ${B/T}_g$ abruptly decreases from an \rms dispersion in $\log B/T_g$ included in the 0.4-0.5 interval for types Scd to Sb, to the 0.2-0.3 interval for types Sab to S0a, and to the 0.1-0.2 interval for types S0$^+$ to E. At last, each point representing a galaxy in \fg\ref{B-T-g-per-type-cmap-Re} is color-coded according to the effective radius $R_e$ of the bulge (or S\'ersic component for E and cD types), and this shows a systematic trend that within a given Hubble type, as well as across types, galaxies with larger ${B/T}_g$ also have larger bulge $R_e$. These large dispersions in ${B/T}_g$ and bulge $R_e$ for each Hubble type means that any given set of values of these parameters cannot solely define a galaxy morphological type, at least the disk characteristics also need to be defined.

\subsubsection{Variation of bulge and disk radius with Hubble type\label{sct-results-size-evol-rad-vs-type}}

Here we further examine the implications of the observed trends in bulge predominance between morphological types in terms of the effective radii of both the bulge and disk components. The scaling relations per Hubble type of the left panel of \fg\ref{kormendy-cmap} and the right panel of \fg\ref{binggeli-cmap-BT} show that in addition to a systematically decreasing interval of $R_e$ for later Hubble types, $R_e$ also varies within each type with effective surface brightness and absolute magnitude respectively, and that this variation is linked to the variations in the $B/T$ ratio. The left panel of \fg\ref{B-and-D-radius-distrib}, directly shows the distribution of the effective radius $R_e$ of the bulge (or S\'ersic component for E and cD types) as a function of Hubble type, color-coded with $B/T$ for EFIGI galaxies, as well as the geometric means and the estimated errors per type (calculated as in \fg\ref{B-T-g-per-type-cmap-Re}, see \sct\ref{sct-results-size-evol-bt}). Again, the very faint bulges of EFIGI galaxy types Sm, Sdm and Sd (see \sct \ref{sct-results-kormendy}) lead to low statistics for these spiral types, but the range of effective radii for the few successfully modeled bulges (with an uncertainty in $\log R_e\lesssim0.2$ dex) nevertheless lies within the same interval as the earlier spiral types (Sb to Scd types). For the earlier Scd to Sb types, there is a ``plateau'' in effective radius with a mean value of $R_e = 0.60, 0.62, 0.61$ kpc for Sc, Sbc, Sb respectively. 

The mean $R_e$ then increases between late and early spirals, with 0.9 kpc for Sab galaxies to 1.2 kpc for Sa galaxies. The rms dispersion in $\log R_e$ for any given type of lenticulars and spirals from Scd to S0$^{-}$ in the left panel of \fg\ref{B-and-D-radius-distrib} is between $0.31$ and $0.43$ dex. When weighted by the square root of the number of galaxies and adopted as an estimate of the uncertainty in the geometric mean, the doubling of the mean $R_e$ from Sb to Sa types corresponds to a $7.5\sigma$ increase. This significant step in mean $B/T$ is obtained thanks to the large statistical size of the EFIGI sample per Hubble type. 

There is then another ``plateau'' in bulge effective radii for types between Sa up to S0. We note that the Sb to Sa increase in mean $R_e$ corresponds to the entry of galaxies into the Green Valley, which we characterized in \cite{Quilley-2022-bimodality} by a stronger bulge-to-total ratio in both luminosity and mass. What the left panel of \fg\ref{B-and-D-radius-distrib} provides here is the additional information that this stronger bulge prominence is also detected by larger effective radii by a factor of 2 on average. 

For the lenticulars and ellipticals, the significant increase in $R_e$ from a geometric mean of 1.0 kpc for S0 types to a mean of 1.6 kpc for S0$^-$ types could be the result of misclassifications of E galaxies into S0$^-$, driving the mean value for S0$^-$ types higher than it should be. Indeed, the bulge and disk decompositions applied to E galaxies (with high $B/T$, hence mostly red colored points) lead to larger values of $R_e$ with a geometric mean at $\SI{2.3}{kpc}$, and a dispersion of 0.27 dex.

The left panel of \fg\ref{B-and-D-radius-distrib} also shows the effective radii derived from the single-S\'ersic profile modeling of E and cD galaxies as gray dots (see \sct\ref{sct-methodo-srx}), and indicates a shift in $R_e$ from $1.1$ kpc for the mean over S0 to S0a types (if one excludes S0$^-$ as they may be contaminated by E galaxies), to a mean of $10.4$ kpc and $29.6$ kpc for E and cD types respectively. 

We now examine the corresponding effective radii $h_e$ of the disk (or exponential component for E and cD types) for EFIGI galaxies in the right panel of \fg\ref{B-and-D-radius-distrib}, which shows the variations of $h_e$ for each Hubble Type, as well as the geometric means and the estimated errors per type (determined as for \fg\ref{B-T-g-per-type-cmap-Re}, see \sct\ref{sct-results-size-evol-bt}). One can see that there is a marked increase in the geometric mean of $h_e$ from $\SI{2.1}{kpc}$ to $\SI{5.3}{kpc}$ for Im and Sdm types respectively, that is by a factor of $\sim2.6$ and a $11.3\sigma$ increase. The increase in the mean $h_e$ is less abrupt between Sdm and Sbc types with a factor of $\sim1.7$ and a $10.5 \sigma$ increase between these two types.

Between Sbc and Sa types, the mean effective radii of the disks presents a ``plateau'' at $9.0-9.3$ kpc, with a systematic shift toward lower values for Sb types, at a geometric mean radius of $\SI{6.9}{kpc}$. As the distribution of $h_e$ as a function of Hubble type does not display any systematic trend with $B/T$, we color-code the points in the right panel of \fg\ref{B-and-D-radius-distrib} with the EFIGI {\tt VisibleDust} attribute. Interestingly, the more frequent presence of large amounts of dust in disks of Sb types compared to the other spiral types may cause the lower tail and lower mean $h_e$ for the Sb types: the high dust content of these galaxies would obscure their disks and make them apparently smaller (there is a similar effect with the isophotal diameter $D_{25}$, see \citealp{de-Lapparent-2011-EFIGI-stats}).

Moreover, in the right panel of \fg\ref{B-and-D-radius-distrib}, S0a and lenticulars display smaller mean disk radii than early-type spirals except for the S0$^-$ type, whose large $\SI{9.0}{kpc}$ radii could be due to a contamination by misclassified E galaxies, as already discussed for the bulge radii (left panel of \fg\ref{B-and-D-radius-distrib}). Taken together, S0a, S0$^+$ and S0 types have a mean $h_e=6.4$ kpc, which is $\sim42\%$ lower (and $4.9\sigma$) than the $9.1$ kpc ``plateau'' value for Sbc, Sab and Sa spirals. At last, the exponential components (of the bulge and disk decompositions) of E and cD galaxies have mean $h_e$ of 10.7 and 25.1 kpc respectively, confirming again that these types are the largest of the Hubble sequence. We can note that these values are nearly identical to and consistent with the single profile effective radii of $10.4$ kpc and $29.6$ kpc for E and cD types respectively (plotted in gray in the left panel of \fg\ref{B-and-D-radius-distrib}). This confirms the larger overall size of E compared to lenticular and spiral disks (hence the full galaxy sizes), and the fact the cD galaxies are giant galaxies built by the merger of galaxies in dense regions such as clusters of galaxies \citep{Edwards_2020_formation_BCG, Chu_2022_BCG_CFHTLS}.

\subsubsection{Variation of bulge and disk radius with $B/T$  \label{sct-results-size-evol-rad-vs-bt}}

\begin{figure*}
\includegraphics[width=\columnwidth]
{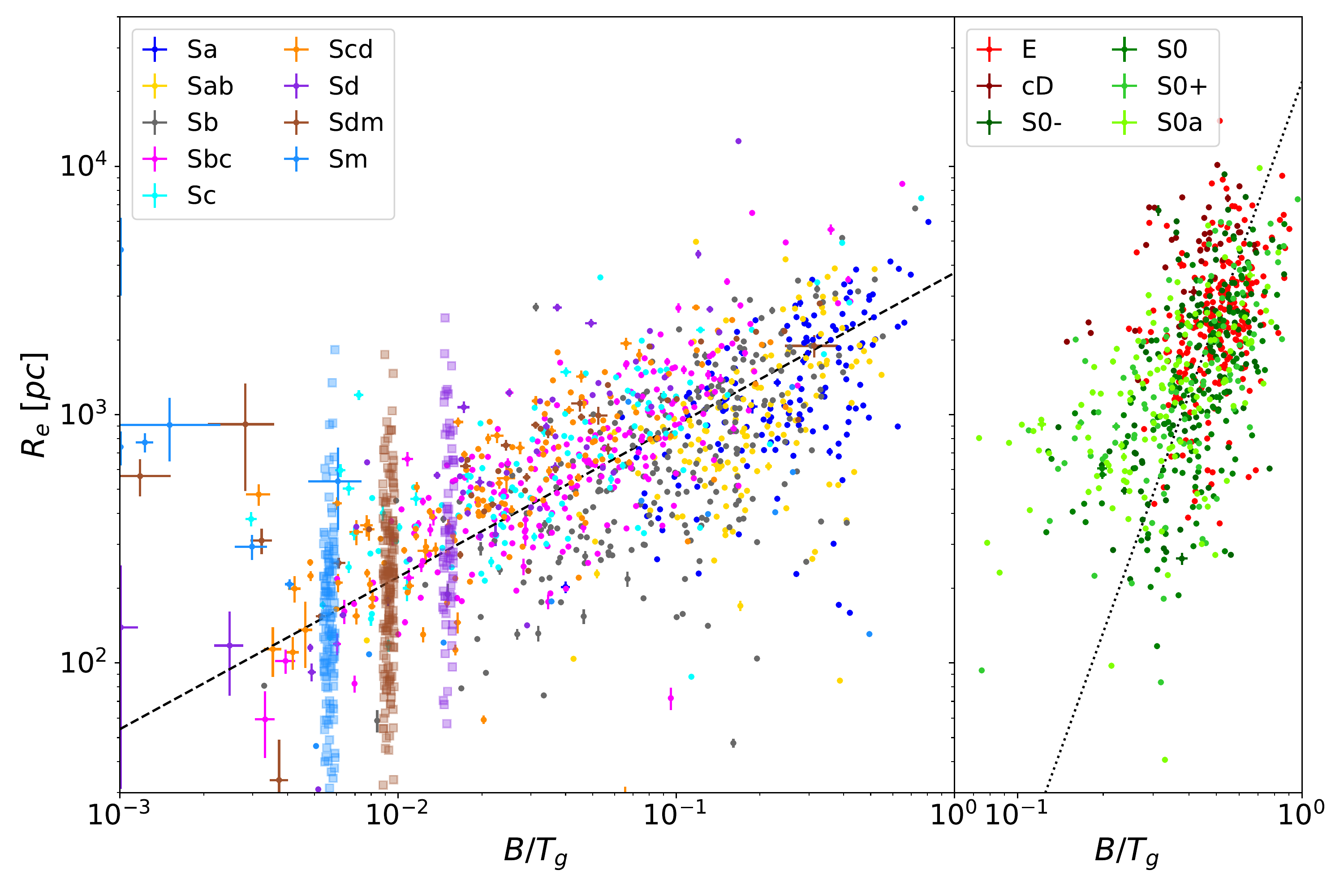}
\includegraphics[width=\columnwidth]
{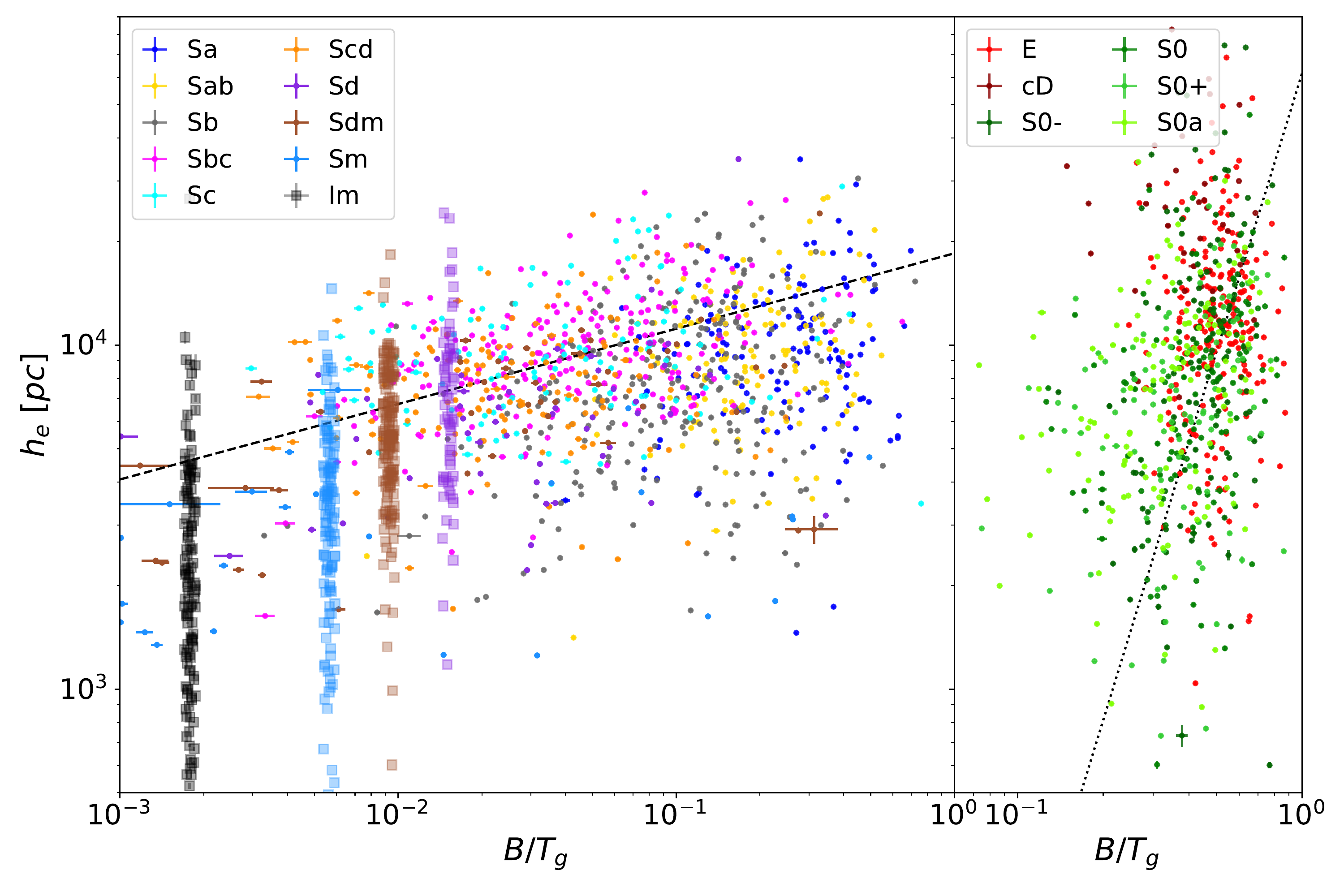}
\caption{Distributions of the effective radius $R_e$ of the bulge (or S\'ersic) component (left), and $h_e$ of the disk (or exponential) component (right) of all types of EFIGI galaxies with {\tt Inclination}$\leq 2$, as a function of the bulge-to-total ratio ${B/T}_g$ in the $g$ band, identified in \cite{Quilley-2022-bimodality} as a tracer of Hubble type. Both panels are color-coded with the Hubble type of the galaxy, with subplots corresponding late (spiral and Im) types on the left, and to early (E and S0) types on the right. The corresponding ODR fits are plotted in each panel. The indicative positions in ${B/T}_g$, $R_e$ and $h_e$ for the Sd, Sdm, and Sm types are added in both panels (see text for details). For the Im types (added in the right panel), the effective radii from the single-S\'ersic modeling are added at an arbitrary small ${B/T}_g$ value (see text). These graphs show that for spiral types, $R_e$ increase strongly with ${B/T}_g$, whereas there is only a weak increase in $h_e$ with ${B/T}_g$. S0 and E galaxies have the highest ${B/T}_g$ values, but only some of them actually have the highest $R_e$ or $h_e$, with a tail of S0 toward lower values encompassing $\sim1$ dex for both their $R_e$ and $h_e$. The smallest S0 $R_e$ correspond to the smallest $R_e$ of early spirals (left panel), and the smallest S0 $h_e$ correspond to the intermediate $h_e$ between those of the smallest disks (Sm types) and of the Im galaxies (right panel).}
\label{B-and-D-radius-fits}
\end{figure*}

Another major feature seen in the left panel of \fg\ref{B-and-D-radius-distrib} is that there is a strong $B/T$ gradient with $R_e$ within each Hubble type, as well as from type to type, suggesting that $B/T$ may actually be a useful quantity to parameterize the variations in $R_e$ among galaxies. This can be seen directly in the left panel of \fg\ref{B-and-D-radius-fits}, which shows the distribution of the effective radius $R_e$ of the bulge (or S\'ersic component for E and cD types) as a function of $B/T$ in the $g$ band for EFIGI galaxies separated into spirals (left subpanel) and early Hubble types from E to S0a (right subpanel). Despite a large dispersion, there is a continuous linear increase in the mean $\log R_e$ with $\log B/T_g$, independently of the galaxy type, with earlier Hubble types having on average larger ${B/T}_g$ and $R_e$ in both subpanels.

More specifically, one can see in the left panel of \fg\ref{B-and-D-radius-fits} an increase in $R_e$ with ${B/T}_g$ for all spiral types from Sa to Sm in the $B/T\sim0.002-0.6$ interval (left subpanel). For the E to S0a types, there is a steeper increase in the $B/T\sim0.2-0.7$ interval. We therefore linearly model the $\log R_e$ variations with $\log B/T_g$ for both samples using the ODR package (see \sct\ref{sct-methodo-odr}), and obtain 
\begin{equation}
    \log R_e = 0.612^{\pm0.018} \log B/T + 3.571^{\pm0.021}   
    \label{eq-re-bt-sp}
\end{equation}
for Sa to Sm spiral types, and  
\begin{equation}
    \log R_e = 3.174^{\pm0.166} \log B/T  + 4.341^{\pm0.063}  
    \label{eq-re-bt-s0}
\end{equation}
for E to S0a lenticular types. Indeed, \eq\ref{eq-re-bt-s0} indicates a steeper increase of $R_e$ with ${B/T}_g$ for E to S0a types compared to spirals. The \rms dispersion in $\log R_e$ around both fits is higher for E to S0a types, with $0.55$ dex compared to $0.31$ dex for spirals. We also calculate the residuals of the $R_e/R_{e,\mathrm{fit}}$ ratios for  $R_{e,\mathrm{fit}}$ given by \eq\ref{eq-re-bt-sp}, for the ${B/T}_g$ values of the considered sample. The distribution of $\log(R_e/R_{e,\mathrm{fit}})$ in bins of 0.1 dex can be fitted by a Gaussian distribution centered at 0.021, and with reduced $\chi_2$ of 3.3 (with some skewness beyond $\pm0.4$ dex). All parameters, including the \rms dispersion, about both fits are listed in Table~\ref{tab-fits-BT} in \sct\ref{sct-discussion-mock}. 

We now examine in the right panel of \fg\ref{B-and-D-radius-fits} the distributions of the effective radii $h_e$ of the disk (or exponential component for E and cD type) as a function of ${B/T}_g$. E to S0a types show again a steeper variation than for spiral types. We therefore perform ODR linear fits to the separate early and late type subsamples, and obtain 
\begin{equation}
    \log h_e = 0.219^{\pm0.013} \log B/T + 4.267^{\pm0.019}  
    \label{eq-he-bt-sp}
\end{equation}
for Sa to Sm spiral types, and 
\begin{equation}
    \log h_e = 2.691^{\pm0.133} \log B/T + 4.790^{\pm0.042}   
    \label{eq-he-bt-s0}
\end{equation}
for E to S0a types. As for $R_e$, the \rms dispersion in $\log h_e$ around both fits is higher for E to S0a types, with $0.59$ dex compared to $0.22$ dex for spirals (all parameters are also listed in Table~\ref{tab-fits-BT} in \sct\ref{sct-discussion-mock}). There is therefore a similar and steep increase in the effective radii of the disk (or exponential component) of E to S0a types with ${B/T}_g$ as for their bulges (see \eqs\ref{eq-re-bt-s0} and \ref{eq-he-bt-s0}), whereas the disk radii increase with ${B/T}_g$ for spiral types is lower by a factor of $\sim3$ in log-log compared to the bulge radii variations of these types (see \eqs\ref{eq-re-bt-sp} and \ref{eq-he-bt-sp}). 

We see in both panels of \fg\ref{B-and-D-radius-fits} that the Sa types show a larger dispersion in ${B/T}_g$, as well as in $R_e$ and $h_e$, compared to the other types, encompassing both the spiral and lenticular sequences. Excluding them from the spiral ODR fits increases the slopes by small amounts ($0.025$ and $0.024$ for \eq\ref{eq-re-bt-sp} and \ref{eq-he-bt-sp} respectively), and would lead to the same conclusions. We also calculate the residuals of the $h_e/h_{e,\mathrm{fit}}$ ratios for $h_{e,\mathrm{fit}}$ given by \eq\ref{eq-he-bt-sp}, for the ${B/T}_g$ values of the considered sample. The distribution of $\log(h_e/h_{e,\mathrm{fit}})$ in bins of 0.1 dex can be fitted by a Gaussian distribution centered at -0.064, and with reduced $\chi_2$ of 3.3 (with some skewness beyond $\pm0.4$ dex).

\begin{figure*}
\includegraphics[width=\columnwidth]{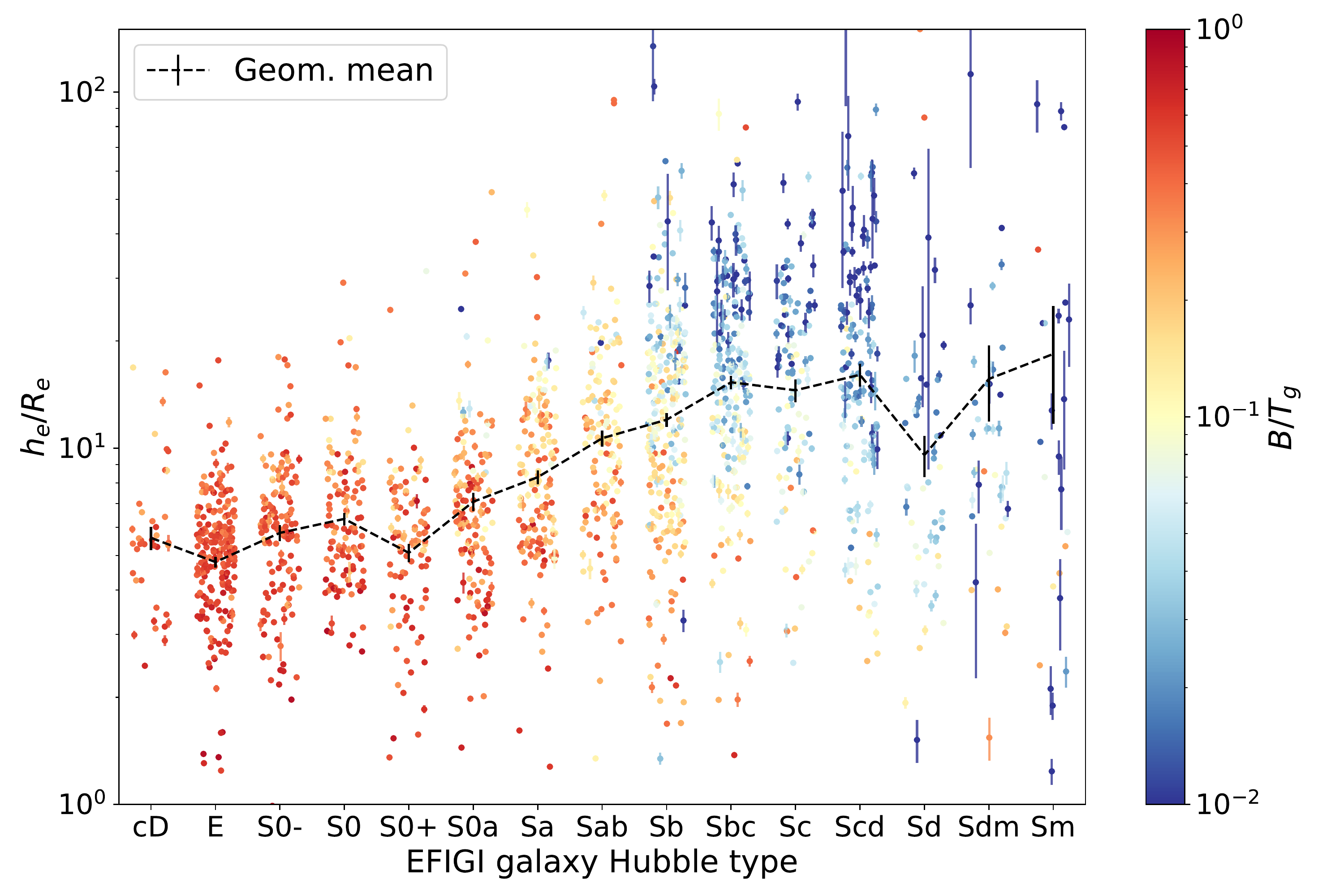}
\includegraphics[width=\columnwidth]
{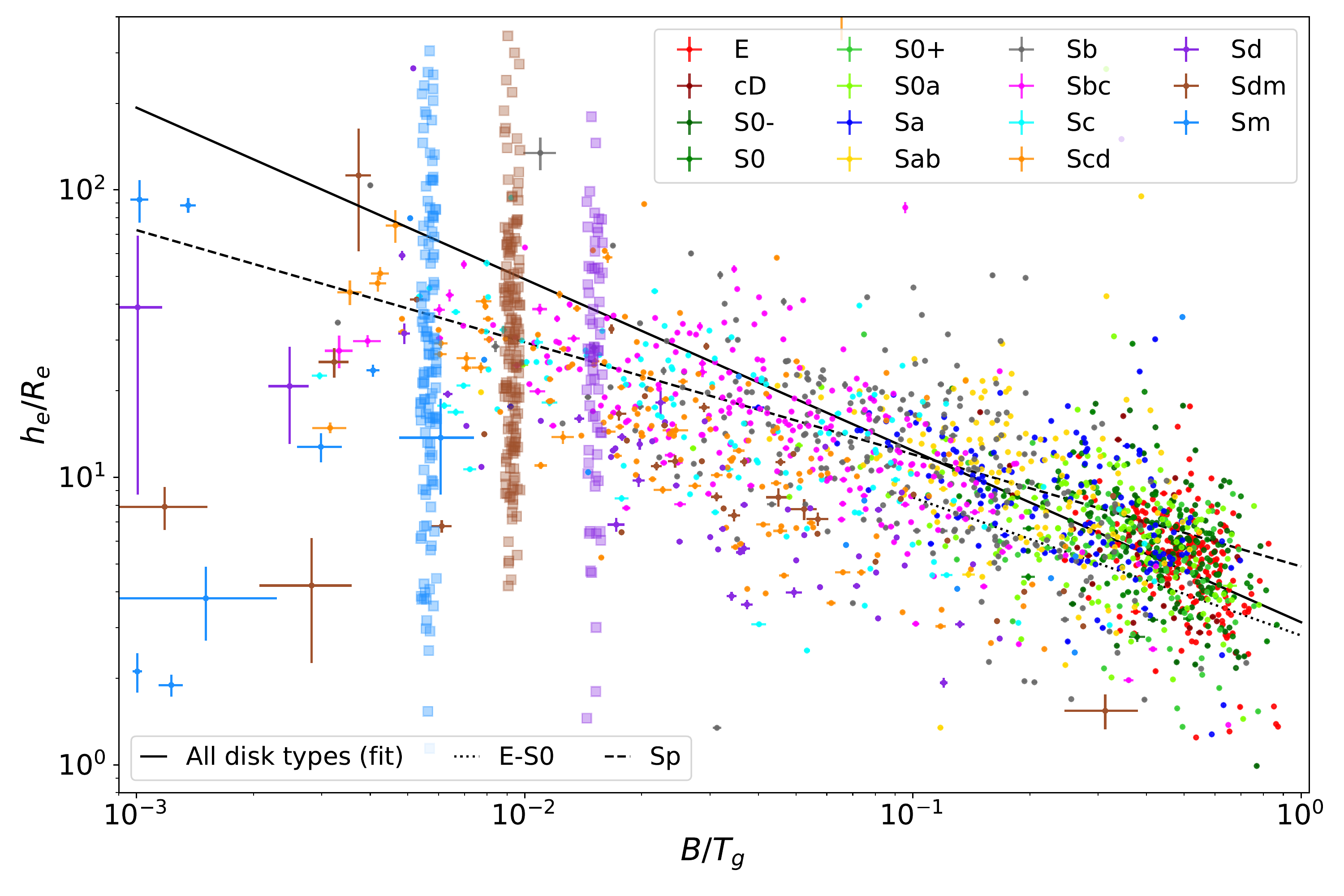}
\caption{Ratio between $h_e$ the effective radius of the disk (or exponential) component and $R_e$ the effective radius of the bulge (or S\'ersic) component as a function of Hubble type (left) and ${B/T}_g$ (right). The black dashed line in the left panel is the geometric mean value of this $h_e/R_e$ ratio, and the vertical bars the estimated uncertainties on this mean (see \sct\ref{sct-results-size-evol-bt} for details). Disk radii are on average 6 times larger than their bulge radii for lenticular galaxies. The size ratio increases along the Hubble sequence toward later and later spiral types, reaching $\sim 15$ for Sbc, Sc and Scd types. This ratio is linked to the ${B/T}_g$ luminosity bulge-to-total ratio, with higher values corresponding to relatively larger bulges, as highlighted by the linear fit (solid line, right panel). The dashed and dotted lines are the linear relations (in log-log scale) of $h_e/R_e$ versus ${B/T}_g$ derived from the two separate relations for $h_e$ or $R_e$ versus ${B/T}_g$ obtained in \fg\ref{B-and-D-radius-fits} for Sa to Sm, and E to S0a types, respectively.}
\label{D-B-radius-ratio}
\end{figure*}

We position, in the right panel of \fg\ref{B-and-D-radius-fits}, the effective radii measured for the single-profile S\'ersic modeling of Im galaxies near the lower left limit of the graph of $B/T_g\simeq10^{-3}$, namely at $B/T_g=1.7\;10^{-3}$ (in black). We also display the missing late spiral galaxies in both panels of \fg\ref{B-and-D-radius-fits}. As shown in the left panel of \fg\ref{B-and-D-radius-distrib}, only 52, 43 and 24 galaxies of Sd, Sdm, and Sm types respectively have their bulge successfully modeled by the bulge and disk decomposition (see \sct\ref{sct-results-kormendy}). These galaxies nevertheless do have small or very small bulges. In order to show indicative positions for the bulge and disk effective radii of these types in \fg\ref{B-and-D-radius-fits}, we first estimate their mean ${B/T}_g$ by linearly interpolating (in logarithmic scale) in \fg\ref{B-T-g-per-type-cmap-Re}: we use the geometric mean values of ${B/T}_g$ for Sab and Scd types and extend the line between these points to later types, leading to extrapolated values of $B/T_g = 0.015, 0.009, 0.006 $ for Sd, Sdm and Sm types, respectively. We also linearly interpolate (in logarithmic scale) in the left panel of \fg\ref{B-and-D-radius-distrib} between the geometric mean value of $R_e$ for Scd types ($0.49$ kpc), and that for Sm types ($0.13$ kpc), and obtain mean estimates of $0.32$ kpc and $0.20$ kpc respectively for Sd and Sdm types. We then introduce a dispersion in these $R_e$ values using a normal distribution in logarithmic scale with an \rms dispersion equal to that measured for Scd types (the most populated of the late types) in \fg\ref{B-and-D-radius-distrib}, and plot the resulting points in the left panel of \fg\ref{B-and-D-radius-fits} as dodger blue, brown and purple squares for Sm, Sdm and Sd types respectively, in order to differentiate them from the data points from the successful fits (shown as dots). To position these galaxies in the right panel of \fg\ref{B-and-D-radius-fits}, we use the actual geometric mean $h_e$ of the disk component provided by the bulge and disk decomposition (if the bulge fit fails for these small bulges, the disk fits are robust and the parameters can be used). For all the added Sd-Sdm-Sm and Im types in both panels of \fg\ref{B-and-D-radius-fits}, an arbitrary uniform spread of $\pm 5\%$ in $\log B/T_g$ is used in order to visualize the density of added galaxies for each Hubble type. 

We emphasize that if the E to S0a types and spiral samples in \fg\ref{B-and-D-radius-fits} are fitted based on a visual split according to their observed distribution, this choice is comforted by the fact that they are located in the Blue Cloud and the Red Sequence, respectively, and we showed in \citet{Quilley-2022-bimodality} that they exhibit marked differences in their combination of ${B/T}_g$ and disk color. The intermediate location of the Sa galaxies in both panels of \fg\ref{B-and-D-radius-fits} is consistent with the fact that they populate the core of the Green Valley (the transition region between the Blue Cloud and the Red Sequence). 

Moreover, in \citet{Quilley-2022-bimodality}, we suggested that spiral galaxies increase in mass (under the effects of mergers and gas accretion) while becoming of earlier and earlier type, and that this can be quantified by the growth of their light and mass bulge-to-total ratio ${B/T}_g$, and the quenching of their disks. In this article, \fg\ref{B-and-D-radius-fits}, and \eqs\ref{eq-re-bt-sp}, \ref{eq-re-bt-s0}, \ref{eq-he-bt-sp}, and \ref{eq-he-bt-s0} provide more information on the changes in the bulge and disk parameters in this scenario. \fgs\ref{B-T-g-per-type-cmap-Re} and \ref{B-and-D-radius-fits} shows that the measurable range in ${B/T}_g$ from the EFIGI spiral galaxies using our bulge and disk decompositions (see \sct\ref{sct-methodo-srx}) is from $B/T\sim0.002$, but with low statistics (see \sct\ref{sct-results-kormendy}),\footnote{There are some galaxies with $B/T\sim0.001-0.002$ in both panel of \fg\ref{B-and-D-radius-fits}, but larger statistics are necessary to confirm them.} to $B/T\sim0.2$, that is a factor $\sim 100$. Over this ${B/T}_g$ interval, the mean increase in the galaxy bulge radii $R_e$ is a factor of $\sim100^{0.64}=19\pm6$ (derived using \eq\ref{eq-re-bt-sp} and the dispersion around this fit), and the mean increase in their disk radii $h_e$ is a factor of $\sim7\pm2$ (calculated from the mean $h_e$ of $\sim2.1$ kpc for Im in the right panel of \fg\ref{B-and-D-radius-distrib}, to $13$ kpc for $h_e$ at $B/T=0.2$ using \eq\ref{eq-he-bt-sp}, and the dispersion around the fit in \eq\ref{eq-he-bt-sp}).

When comparing the left and right panels of \fg\ref{B-and-D-radius-fits}, lenticular galaxies display different behaviors for their disks and bulges compared to spirals. The range of disk radii spanned by the S0$^-$ to S0a types reaches smaller values than for early spirals, and as low as the lowest values for the latest Sm spirals (dodger blue squares). Only the irregulars (black squares) have even smaller values of $h_e$ using single-S\'ersic profiles. These small disk radii of lenticulars coincide with the low luminosity and low mass end of the Red Sequence shown to be dominated by lenticulars in \citet{Quilley-2022-bimodality}. In contrast, the S0$^-$ to S0a types span a narrower range of bulge effective radius $R_e$ from the values for the smallest bulges of the intermediate Sc types of radii $\sim200$ pc, up to $\sim6$ kpc, which is beyond the values for the largest spiral bulges by a factor of $\sim3$. This suggests the possible existence of a direct evolutive channel from small spirals (with small bulge radii and light fraction ($B/T\sim0.0025-0.05$) to small lenticulars with more prominent bulges (with $B/T_g\sim0.25-0.5$). It however remains to be explained how the bulge could grow by a factor of $10$ to $100$. We also note that the effective radii of the S\'ersic and exponential components of the E and cD galaxies populate the high radii and high ${B/T}_g$ part of the full range encompassed by the lenticulars (in the right subpanels of both panels of \fg\ref{B-and-D-radius-fits}), and appear in the continuity of the spiral relations for both bulges and disks.

\subsubsection{Ratio of disk-to-bulge radius variation with $B/T$  \label{sct-results-size-evol-ratio}}

We now examine in the left panel of \fg\ref{D-B-radius-ratio} the ratio of the effective radii of the disk (or exponential component of E and cD types) and the bulge (or S\'ersic component of the E and cD types) as a function of Hubble morphological type (with color-coding of each point according to the ${B/T}_g$ luminosity ratio). Starting from the latest type, one can see again the low statistics for Sm to Sd types, because they rarely host a measurable bulge (see \sct\ref{sct-results-kormendy}). For Scd to Sbc types, the geometric mean of the radii ratio and its dispersion have stable values of $<h_e/R_e>\simeq15$ and $\sigma_{h_e/R_e}$ of 0.29, 0.35 and 0.40 dex for types Sbc, Sc, and Scd respectively. The mean radii ratio steadily decreases from Sbc types all the way to S0$^+$ types where it reaches  $<h_e/R_e>\simeq5$, and remains at this value for all S0, E and cD galaxies. We measure an \rms dispersion of $\sigma_{h_e/R_e}$ between 0.19 and 0.32 dex for all types between Sbc and S0$^-$, with no noticeable trend, and a slight increase in dispersion for later-types with Sc, Scd having $\sigma_{h_e/R_e}$ of 0.35 and 0.40 dex respectively. The decrease is therefore significant at a $3.7-4.0\sigma$ level between most individual subsequent types across the Sbc to S0$^+$ interval, except for the Sbc to Sb and Sab to Sa transitions at $1.7\sigma$ and $2.2\sigma$ respectively. The full shift between Sbc and S0$^+$ is significant at the $15\sigma$ level. 

Comparison of the left panel of \fg\ref{D-B-radius-ratio} with both panels of \fg\ref{B-and-D-radius-distrib} shows that the decrease of $h_e/R_e$ from Sb to S0$^+$ types results from 2 different regimes: the distribution of disk $h_e$ is stable from types Sc to Sa (except for the already mentioned low $h_e$ tail of Sb galaxies, likely due to dust extinction, see \sct\ref{sct-results-size-evol-rad-vs-type}), whereas the bulge $R_e$ increases for earlier types across this interval, and this is concomitant with the marked increase in the mean ${B/T}_g$ per $NUV-r$ color-mass bin from $\sim0.15$ up to $\sim0.45$ that we detected through the Green Valley \citep{Quilley-2022-bimodality}. The continuing decrease in $h_e/R_e$ into the lenticular types, namely for S0a and S0$^+$ types, is explained by the fact that the mean $h_e$ shifts to lower values, whereas the distribution of $R_e$ is stable (see \fg\ref{B-and-D-radius-distrib}). We emphasize that these marked variations of $h_e/R_e$ for EFIGI galaxies disprove the claim by \cite{Courteau-1996-late-spirals-secular-evolution} that the radii ratio is independent of Hubble type. This is probably due to the low statistics of their study, combined with the large dispersion among each Hubble type that the EFIGI sample reveals.

There is moreover a strong ${B/T}_g$ gradient with $h_e/R_e$ and with Hubble type in the left panel of \fg\ref{D-B-radius-ratio} (as seen for $R_e$ in the left panel of \fg\ref{B-and-D-radius-distrib}). For any Hubble type, the larger $h_e/R_e$ values correspond to smaller ${B/T}_g$. However, similar values of $h_e/R_e$ can also correspond to very different values of ${B/T}_g$. Indeed, in \fg\ref{D-B-radius-ratio}, galaxies with a radius ratio of $\sim 10$ may span more than a dex in ${B/T}_g$ between the lenticulars and the intermediate to late spirals. This is due to the strong difference in the surface brightness between the bulges of these different morphological types, that we quantified in \sct\ref{sct-results-kormendy}. 

The right panel of \fg\ref{D-B-radius-ratio}  directly shows the variation of $h_e/R_e$ with ${B/T}_g$. In log-log scale, there is a linear increase in $h_e/R_e$ with decreasing ${B/T}_g$ by $\sim0.7$ dex from $B/T\sim0.5$ to $\sim0.01$. In contrast with the variations of $R_e$ and $h_e$ with ${B/T}_g$ separately shown in both panels of \fg\ref{B-and-D-radius-fits}, the E, cD and lenticulars do not show a distinct behavior from the spiral types, and $h_e/R_e$ shows a smoothly decreasing trend with ${B/T}_g$ over $\sim2.5$ orders of magnitude and all morphological types. Therefore, despite the steeper variations of $h_e$ and $R_e$ with ${B/T}_g$ for lenticular galaxies, their ratio $h_e/R_e$ follows the same scaling law as the spiral galaxies, meaning that both tails of lower radii seen in the right subpanels of both panels of \fg\ref{B-and-D-radius-fits} correspond to the same objects.

Overall, the monotonous variation of the mean $h_e/R_e$ with ${B/T}_g$ indicates that the increase of the radii ratio occurs jointly with the increase in the luminosity ratio between the bulge and the disk. In other words, moving toward earlier types of the Hubble sequence, the share of flux within the bulge increases on average as its relative size compared to that of the disk (similar to the size of the whole galaxy) also increases. This testifies that on average, mass is transferred from the disk into the bulge as galaxies evolve backward along the Hubble sequence, therefore increasing the bulge radius, even when the disk radius remains constant, as it is the case for early type spirals (see left and right panel of \fg\ref{B-and-D-radius-distrib}). Of course, this average scenario does not exclude more complex paths of individual galaxies as they evolve from star-forming to quiescent.

We show in the right panel of \fg\ref{D-B-radius-ratio} as a black solid line the ODR linear fit of $h_e/R_e$ versus ${B/T}_g$ for all types from E to Sm:
\begin{equation}
    \log{h_e\over R_e} = -0.597^{\pm0.013} \log B/T + 0.496^{\pm0.010}  
    \label{eq-he-re-bt}
\end{equation}
We add in this graph the indicative position of the late spirals types Sd-Sdm-Sm (plotted in \fg\ref{B-and-D-radius-fits}), by using for $h_e/R_e$ the mere ratio of the true $h_e$ values by the indicative $R_e$ values, as a function of the interpolated values of ${B/T}_g$ for these types (see \sct\ref{sct-results-size-evol-rad-vs-bt}). These added late spirals, as well as earlier spirals of Sbc-Sc-Scd types suggest a possible curving down of $h_e/R_e$ below the linear fit at low ${B/T}_g$, but more statistics are required to clarify this issue.

We also show as dashed and dotted lines in the right panel of \fg\ref{D-B-radius-ratio} the relation of $h_e/R_e$ versus ${B/T}_g$ calculated as the ratio of the linear relations of $h_e$ versus ${B/T}_g$ (\eqs\ref{eq-he-bt-sp} and \ref{eq-he-bt-s0}), and $R_e$ versus ${B/T}_g$ (\eqs\ref{eq-re-bt-sp} and \ref{eq-re-bt-s0}) for spirals and for S0 and E galaxies respectively. This allows us to test whether the disk-to-bulge size ratios deducted from the size versus ${B/T}_g$ relation for each component are consistent with the observed ones. It appears that for, on one hand the E, cD and S0 types, and on the other hand the spiral types, both ratios of the linear relations (dotted and dashed lines) would yield an acceptable variation with ${B/T}_g$, the picture is more complex for E, cD and S0 : despite the large dispersion in both $h_e$ and $R_e$ and a strong covariance between $R_e$ and $h_e$, there is a much smaller dispersion in $h_e/R_e$ compared to those in the individual radii for these types. As a result, when assigning a bulge and disk effective radii to lenticular galaxies for a given ${B/T}_g$ value, small bulges should be matched with small disks and large bulges with large disks. We therefore advise to use only one among the bulge and disk versus ${B/T}_g$ relations (\eqs\ref{eq-re-bt-s0} and \ref{eq-he-bt-s0}) and complement it with the $h_e/R_e$ vs ${B/T}_g$ relation, rather than using both size versus ${B/T}_g$ relations separately, in order to generate physically meaningful bulge and disk sizes for the lenticular galaxies.

\section{Discussion \label{sct-discussion}}

\subsection{Galaxies evolve both in mass and size        \label{sct-discussion-B-D-sizes}}

In \cite{Quilley-2022-bimodality}, we showed that the Hubble sequence was an inverse evolutionary sequence, with all its types spanning continuously the color-mass diagram. Three consecutive phases of evolution were identified: (i) 3 orders of magnitude in luminosity and mass growth through mergers and consumption of the gas reservoir from irregulars to Sb spirals forming the Blue Cloud; (ii) star formation fading between Sab early spirals and lenticulars along with a marked growth by a factor of 2 in bulge-to-total ratio ($B/T$), with Sa and S0a types populating the Green Valley; (iii) another mass growth by 1 order of magnitude between the faintest lenticulars and the most massive ellipticals (a factor of $\sim 4$ above the upper mass limit of spirals).

In the current analysis, we show that these three evolution phases along the Hubble sequence not only determine the luminosity and stellar mass growth of galaxies along the Hubble sequence, but also their growth in size. This is a consequence of the size-luminosity scaling relations for bulges and disks measured in the present analysis. \fg\ref{B-and-D-radius-distrib} in \sct\ref{sct-results-size-evol} indicates that the above three evolution phases can be matched to the three following regimes of bulge and disk size variations: (i) a 0.45 dex increase in the mean effective radius of the disks $h_e$ from late-type (Sm) to intermediate-type (Sc) spirals; (ii) a stable mean disk $h_e$ of $\sim8$ kpc from intermediate (Sc) to early (Sa) spirals, whereas  the mean bulge effective radius $R_e$ doubles from $\sim0.6$ kpc for Sb types to $\sim1.2$ kpc for Sa types, which corresponds to the entry into the Green Valley; (iii) a smaller mean disk $h_e$ of lenticular galaxies (S0a to S0$^+$) in the interval $6-7$ kpc, whereas the mean bulge $R_e$ remain stable in the $1-1.5$ kpc range, and ultimately, reaches $2.3$ kpc and $4.2$ kpc for the S\'ersic components of E and cD types, respectively. The mean effective radii for the E and cD galaxies fitted as single S\'ersic profiles (or equivalently the $h_e$ of the exponential component) reach 10 and 30 kpc respectively (see \fg\ref{B-and-D-radius-distrib}), making these two types the largest among the Hubble sequence, in the same way that they are the most luminous and massive. This is consistent with them being built in part by major mergers of the lenticular and spiral galaxies. 

\subsection{Different types of bulges: Pseudo and classical \label{sct-discussion-bulge-types}}

Pseudo-bulges are central concentration of stars built through secular evolution processes within the disk plane, whereas classical bulges are spheroids built by violent relaxation in mergers \citep{Kormendy-Kennicutt-2004-bulges-disk-galaxies-review}. In order to properly distinguish between these two classes, one would need to know for each bulge its formation scenario, which could be deciphered observationally from the bulge stellar kinematics. In an attempt to have available a quantitative criterion for differentiating among bulge types based on their photometric properties, two criteria have been widely used, which are improperly identified with the physical definition of pseudo and classical bulges. \cite{Kormendy-Kennicutt-2004-bulges-disk-galaxies-review} and \cite{Fisher-Drory-2008-bulges-n-sersic} first proposed to use the S\'ersic index $n_{\text{S\'ersic}}$ of the bulge as a criterion, with a transition value of $n_{\text{S\'ersic}}=2$, pseudo-bulges and classical bulges having lower and higher indexes respectively. However, the S\'ersic index is the parameter of the model-fitting showing the largest uncertainties, which can hinder the reliability of this criterion. In addition, we emphasize that \cite{Fisher-Drory-2008-bulges-n-sersic} use HST-ACS high-resolution imaging to label bulges as either pseudo or classical based on the presence or absence of morphological features within the images (nuclear ring, spiral or bar) respectively, and then show that the value of the S\'ersic index correlates with this labeling. There is however no evidence in their analysis that bulges labeled as classical or pseudo went through the related formation processes, or have the corresponding dynamical signatures. Later on, \cite{Gadotti-2009-bulge-structure-SDSS} suggested that bulges following the Kormendy relation for ellipticals are likely to have a similar structure, hence to be classical bulges. The author therefore proposed to use the Kormendy relation as a criterion to differentiate classical bulges from pseudo-bulges, the latter being outliers of this relation, with larger values of $\langle\mu\rangle_e$, that is fainter surface brightnesses, than what the relation predicts based on their $R_e$. Some of these effects are also detected in EFIGI galaxies and are presented below.

In \cite{Quilley-2022-bimodality}, we identified the bulge growth through the Green Valley and demonstrated that it was concomitant to a change toward steeper light profiles of the bulges, which could correspond to the transition from pseudo to classical bulges. Indeed, \fgs 20 and 26 of \citet{Quilley-2022-bimodality} show that the EFIGI color-mass bins with $n_{\text{S\'ersic}}\gtrsim2$ are mainly populated by Sb types, compared to Sbc and Sc types for the bins with $n_{\text{S\'ersic}}\lesssim2$. That analysis also revealed how EFIGI galaxies display a continuous variation in their bulge S\'ersic index along the Hubble morphological sequence within the $NUV-r$ color versus stellar mass diagram. This is in agreement with the present result that the $n_{\text{S\'ersic}} \gtrsim 2$ bulges follow the Kormendy relation for ellipticals, whereas the $n_{\text{S\'ersic}}\lesssim 2$ bulges deviate from it, and this departure from the Kormendy relation for elliptical galaxies starts between Sb and Sbc types (see \sct\ref{sct-results-kormendy} and \fgs\ref{kormendy-g-per-type-cmap-BT} and \ref{kormendy-cmap}). As the latter morphological types deviate more from the relation, we conclude that they are the types more likely to be dominated by pseudo-bulges. Therefore, we confirm that both criteria of \cite{Fisher-Drory-2008-bulges-n-sersic} and \cite{Gadotti-2009-bulge-structure-SDSS} agree between each other for characterizing bulges when applied to the EFIGI sample. As both the departure from the Kormendy relation, and the $n_{\text{S\'ersic}}=2$ transition occur at $B/T \sim 0.1$, our analysis highlights the additional morphological information that the small $B/T$ bulges of late spiral type galaxies are predominantly pseudo-bulges, whereas the larger $B/T$ bulges of early spirals and lenticulars tend to be classical bulges. 

We have therefore identified three parameters that can be used to characterize the nature of bulges: S\'ersic index, distance from the Kormendy relation for ellipticals, and the bulge-to-total ratio $B/T$ (characterizing the whole galaxy). As the three parameters vary simultaneously in a plane within the 3D $\langle\mu_e\rangle$-$M$-$R_e$ space (see \fgs\ref{kormendy-cmap}, \ref{binggeli-cmap-BT} and \ref{3D-figure}), they could a priori be used in isolation. Using them jointly could be a way to check the reliability of the bulge modeling, by allowing to spot cases in which one of these parameters has an inconsistent value with the others. 

Taken altogether, these elements paint a picture of continuous bulge evolution from small, faint and low contrast central disk concentrations toward prominent spheroids with steeper profiles, rather than a dichotomy between two separate classes of objects (pseudo-bulges and classical bulges).
By analyzing spatially resolved star formation histories (based on population spectral synthesis models) of the bulges and disks of 135 late-type spirals from the CALIFA survey, \cite{Breda-2022-continuous-seq-bulges} reach the same conclusion that the two types of bulges are extremities of a continuous sequence rather than clearly distinguishable classes. This is further confirmed by kinematical studies such as those by \cite{Mendez-Abreu-2014-composite-bulges} and \cite{Erwin-2015-composite-bulges}. The two teams investigated the stellar dynamics inside the bulges of 10 face-on barred galaxies (5 spirals and 5 lenticulars) and 9 S0-Sb galaxies, respectively. For 7 out of 10 galaxies studied by \cite{Mendez-Abreu-2014-composite-bulges}, and for all 9 objects from \cite{Erwin-2015-composite-bulges}, the authors found that a disky pseudo-bulge and a classical one were actually coexisting, with one galaxy in each sample showing evidence for an additional boxy-peanut component to its bulge. These studies therefore indicate that bulges are often composite systems, so that the continuous transition in S\'ersic index and deviation from the Kormendy relation that we detect for EFIGI galaxies can be interpreted as a predominance of either one of the possible central components: a rotation-supported accumulation of stars from the thick and old disk at its center \citep{Di-Matteo-2019-thick-disk-MW}, a steep profile and dispersion-supported spheroid build by mergers. 

In the next subsection (\sct\ref{sct-discussion-diffuse-B}), we propose an empirical magnitude interval, as well as a corresponding $B/T$ interval, in which the bulges may transition between pseudo to classical structure.

\subsection{Spatial density of bulges and disks\label{sct-discussion-diffuse}}

\subsubsection{Larger ellipticals are more diffuse       
\label{sct-discussion-diffuse-E}}

The Kormendy and size-luminosity relations of \eq\ref{eq-kormendy} and \ref{eq-binggeli}, based on the single S\'ersic modeling of elliptical galaxies, can be interpreted in terms of their stellar density. If elliptical galaxies were spatially scale-invariant (that is all had the same 3D density profile, only scaled by their spatial effective radius\footnote{Or more generally semi-major axis of the isophote enclosing half of the total light.}, denoted $\mathcal{R}_e$), their flux would grow as ${\mathcal{R}_e}^3$. \citet{Young-1976-sersic-3D} showed that an angular $r^{1/4}$ profile in projection on the sky (\ie, a S\'ersic profile with $n=4$) can be deprojected into a function that is indistinguishable from a 3D $r^{1/4}$ profile (except in the very central region), and he measured $\mathcal{R}_e = 1.350 R_e$ (see their \eq 17). As a result, the variation of the absolute flux with ${\mathcal{R}_e}^3$ can be written as a variation in ${R_e}^3$ (the physical effective radius derived from the sky projected profile, used in the present analysis), which yields, in terms of the absolute magnitude:
\begin{equation}
    M \simeq -2.5\log {R_e}^3 + \kappa'
    \label{eq-scaling-M-Re3}
\end{equation}
with $\kappa'$ a constant.
This equation can be applied to elliptical galaxies as they can be well fitted by a S\'ersic profile with an index $n_{\text{S\'ersic}}$ index varying from $3.5$ to 7, with a peak at $n_{\text{S\'ersic}}\simeq5.5$. Although ellipticals tend to be oblate \citep{Costantin-2018-shape-bulges-CALIFA}, and this should also be taken into account when deprojecting their profile, we assume in the following that \eq\ref{eq-scaling-M-Re3} remains valid in the case of scale-invariant elliptical galaxies.

The size-luminosity relation for EFIGI ellipticals obtained in \eq\ref{eq-binggeli} (see \sct\ref{sct-results-binggeli} and left panel of \fg\ref{binggeli-per-type})
can be rewritten as
\begin{equation}
    M \simeq -2.5\log{R_e}^{1.09} + \kappa''
    \label{eq-scaling-diffuse}
\end{equation}
with $\kappa''$ a constant. 
We interpret the difference between \eqs\ref{eq-scaling-M-Re3} and \ref{eq-scaling-diffuse} as an indication that E galaxies get more diffuse as they grow in size, with a dilution factor of ${R_e}^{-1.91}$. 
This remains valid using the scaling relations for the S\'ersic component of E galaxies (\eq\ref{eq-binggeli-E-bulges}): the $R_e$ exponent in \eq\ref{eq-scaling-diffuse} becomes $1.43$ and the dilution factor decreases to ${R_e}^{-1.57}$. Given that that their magnitude is tightly anticorrelated to the stellar mass (see \citealt{Quilley-2022-bimodality}), we further suggest that these estimated dilution factors may also be valid for the stellar mass.

\subsubsection{Spatial densities of pseudo and classical bulges     
\label{sct-discussion-diffuse-B}}

We perform for bulges of lenticular and spiral galaxies a derivation similar to that for ellipticals (in \sct\ref{sct-discussion-diffuse-E}). Let us denote $\alpha$ the slope of a linear approximation of the size-luminosity relation
\begin{equation}
    \log R_e = \alpha M+\lambda
\end{equation}
where $\lambda$ is a constant. It can be rewritten as
\begin{align}
\begin{split}
    M &\simeq -2.5\log{R_e}^{\beta} + \kappa'''\\
    \mathrm{with}\, \beta &=-{1\over2.5\alpha}\\
    \label{eq-scaling-diffuse-bulge}
\end{split}
\end{align}    
and $\kappa'''$ a constant, so that $\beta$ measures the scaling of the luminosity $L$ with the effective radius.
\begin{equation}
    L \propto {R_e}^{\beta}
    \label{eq-scaling-luminosity}
\end{equation}
If we define the volume density $\rho$ of the bulges, and their surface density $\Sigma$, depending on whether they can be considered as spheroidal or disky (as can be the case for pseudo-bulges \citealt{Athanassoula-2005-nature-of-bulges-n-body-simul, Athanassoula-2008-pseudo-bulges}), we can write
\begin{align}
\begin{split}
    L &\propto \rho\,{R_e}^3 \Rightarrow \rho\propto{R_e}^{\beta-3}\\
    L &\propto \Sigma\,{R_e}^2 \Rightarrow \Sigma\propto{R_e}^{\beta-2}
    \label{eq-scaling-density}
\end{split}
\end{align}
We emphasize again that, because luminosity is tightly correlated with stellar mass \citep{Quilley-2022-bimodality}, the following remarks remain valid for the stellar mass and stellar mass density within galaxies. Within the low redshift approximation used to derive \eq\ref{eq-mu-scaling}, the surface brightness (defined in \eq\ref{eq-mu-def}) also provides directly the physical surface density of the considered galaxy perpendicular to the line-of-sight, which is, in the case of a weakly inclined disk, close to its surface density $\Sigma$.
We note that the values of $\beta=2$ and $\beta=3$ are critical as they correspond to fixed surface density and fixed volume density respectively. 

The second degree polynomial size-luminosity relation of \eq\ref{eq-size-lum-bulge} (see \fg\ref{binggeli-cmap-BT}) lies mostly between both linear relations of \citet{Binggeli-1984}, with $\alpha$ varying from $-0.1$ to $-0.3$ from late to early Hubble types (and from small to large $B/T$ values). We therefore compute the tangents to this polynomial curve at key values of $\beta$ and $\alpha$, and derive the corresponding values of $M_{\mathrm{bulge},g}$:
\begin{align}
\begin{split}
\beta&= 4.0 \;\mathrm{for}\; \alpha=-0.10 \;\mathrm{at}\; M_{\mathrm{bulge},g}=-17.1\\
\beta&= 3.0 \;\mathrm{for}\; \alpha=-0.13 \;\mathrm{at}\; M_{\mathrm{bulge},g}=-17.8\\
\beta&= 2.0 \;\mathrm{for}\; \alpha=-0.20 \;\mathrm{at}\; M_{\mathrm{bulge},g}=-19.1\\
\beta&= 1.3 \;\mathrm{for}\; \alpha=-0.30 \;\mathrm{at}\; M_{\mathrm{bulge},g}=-21.2\\
\label{eq-scaling-diffuse-bulge-tangents}
\end{split}
\end{align} 
In the following, we make the correspondence between each of  the above magnitudes and a $B/T$ value (justified by the fact that $B/T$ varies continuously along the size-luminosity relation shown in \fg\ref{binggeli-cmap-BT}): we compute the geometric mean of $B/T$ for galaxies whose bulge magnitude $M_{\mathrm{bulge},g}$ is within 0.05 mag of the values listed in \eq\ref{eq-scaling-diffuse-bulge-tangents}.

Along our proposed evolutionary path of the joint $B/T$, $R_e$ and $h_e$ growth, as galaxies move backwards the Hubble sequence, the value of $\beta$ in \eq\ref{eq-scaling-diffuse-bulge-tangents} allows one to track how the densities of the bulges of spirals and lenticulars scale with effective radius $R_e$. Bulges with $M_{\mathrm{bulge},g} > -17.8$ (hence $B/T\lesssim0.08$), have the highest values $\beta > 3$, so spheroids would have their volume density still growing with the radius, and disky systems would see their surface density increase faster than $R_e$. Brighter bulges in the range $-19.1 < M_{\mathrm{bulge},g} < -17.8$ (and $0.08\lesssim B/T \lesssim 0.15$) become more diffuse in volume density as they grow in $R_e$, with a dilution factor $\le R_e$, as $\beta$ decreases from 3 to 2. In contrast, a disky bulge in that magnitude interval would still have an increasing surface density as it grows in radius, but with an exponent lower than 1 in $R_e$. As galaxies continue to evolve by growing their bulges to $B/T\gtrsim 0.15$ and $M_{\mathrm{bulge},g} \lesssim -19.1$, they reach $\beta<2$, meaning that both spheroidal and disky systems would see their volume density and surface density, respectively, decrease as they grow in radius. The brightest of these bulges, with $M_{\mathrm{bulge},g} \lesssim -19.1$, hence $\beta\le1.3$, become diffuse faster for larger $R_e$, nearly reaching the strong dilution factor of the ellipticals (\eq\ref{eq-scaling-diffuse}).

We can relate these trends in the variations of the density of bulges to the distinction between classical and pseudo bulges, which, as discussed in \sct\ref{sct-discussion-bulge-types}, has not yet been defined unequivocally from galaxy photometric properties. Given that the EFIGI disk galaxies analyzed here are weakly inclined, \eq\ref{eq-scaling-density} indicates that a disk-like system such as a pseudo-bulge (discarding the more complex boxy-peanut bulges), would have a surface density $\Sigma$ that would increase with radius, as it would scale as ${h_e}^{\beta-2}$ with $\beta\ge3$ for $M_{\mathrm{bulge},g} > -17.8$. This would imply that the central accumulation of stars that they harbor within the disk would not only extend in radius, but strongly increase its surface brightness, and equivalently its surface density.

In contrast, classical bulges, with the brightest $M_{\mathrm{bulge},g}$ and highest $B/T$, would correspond to the $M_{\mathrm{bulge},g} \lesssim -19.1$ interval, for which both the volume and surface density decrease with radius. As classical bulges are thought to be spheroidal systems, only the volume density decrease is relevant here. Classical bulges are therefore expected to show a similar behavior as ellipticals, becoming more diffuse as they grow in size. Violent relaxation in these massive systems may be responsible for their puffing up with increasing size.

In the intermediate interval $-19.1 < M_{\mathrm{bulge},g} < -17.8$ (and also $0.08\lesssim B/T \lesssim 0.15$), in which the surface density still increases but the volume density decreases as bulges grow, both behaviors could be accounted for, whether the bulge is disky or spheroidal: the surface density would be nearly constant up to an increase as $R_e$, or the volume density would be nearly constant up to a decrease as $1/R_e$. We therefore propose that the $-19.1 < M_{\mathrm{bulge},g} < -17.8$ magnitude range corresponds to the transition region between classical and pseudo-bulges. The associated $0.08<B/T<0.15$ interval is in agreement with the transition near $B/T\sim 0.1$ discussed in \sct\ref{sct-discussion-bulge-types} (see also \fg\ref{kormendy-cmap}).

\subsubsection{Surface density of disks of lenticulars and spirals
\label{sct-discussion-diffuse-D}}

We also interpret the size-luminosity relation for disks in terms of density using the formalism introduced in \sct\ref{sct-discussion-diffuse-B}. \fg\ref{binggeli-disk} shows that the small disks of the late-type spirals grow on average as $\log h_e \simeq -0.140 M_{\mathrm{disk},g}$ (see \eq\ref{eq-size-lum-disk2}), which corresponds to $\beta = 2.9$ (see \eq\ref{eq-scaling-diffuse-bulge}). Therefore, for a weakly inclined disk, \eq\ref{eq-scaling-density} implies that the surface density $\Sigma$ (and the surface brightness $\mu_0$) increases as ${h_e}^{0.9}$. This behavior is analogous to that for the pseudo-bulges (in \sct\ref{sct-discussion-diffuse-B}), that are also disky structures. For early disks, we fit in \fg\ref{binggeli-disk} $\log h_e \simeq -0.208 M_{\mathrm{disk},g}$ (see \eq\ref{eq-size-lum-disk1}), yielding $\beta =1.92$. This means that the total disk flux almost grows with disk effective radius as ${h_e}^2$, that is at a nearly constant disk luminosity surface density (\eq\ref{eq-scaling-density}), as well as a nearly constant central (or effective) surface brightness $\mu_0$ (or $\mu_e$), as already mentioned in \sct\ref{sct-results-disk-scaling-relations}.

Consequently, as disk galaxies merge, their disks first grow both in size and surface density: light and matter are redistributed across the disk during the mergers of either irregular or very late spirals, as expected for major mergers. Then, for the larger disks of early spirals and lenticulars resulting from mergers of the late spirals, the size growth occurs at nearly constant surface density, maybe under the effects of flybys or minor mergers. As mentioned in \citet{Quilley-2022-bimodality}, the spiral arms and bars may play a role in this evolution.

\begin{table*}[ht]
\centering
\resizebox{\linewidth}{!}{%
\begin{tabular}{ c c c c c c c c c c }
\hline\hline
Type & Component & Band & \multicolumn{3}{c}{Size-magnitude relation} & \multicolumn{3}{c}{Dispersion in residual}   \\
&  &  & \multicolumn{3}{c}{Polynomial fit$^{(a)}$} & \multicolumn{3}{c}{Polynomial fit$^{(b)}$}  \\
&  &  & 2$^\mathrm{nd}$ order & 1$^\mathrm{st}$ order  & 0$^\mathrm{th}$ order & 2$^\mathrm{nd}$ order & 1$^\mathrm{st}$ order  & 0$^\mathrm{th}$ order \\
\hline
\multirow{2}{*}{cD} & \multirow{2}{*}{Single-S\'ersic} & $g$ & - & $-0.28^{\pm0.05}$ & $-1.79^{\pm1.03}$ & - & - & $0.18^{\pm0.03}$ \\
& & $i$ & - & $-0.28^{\pm0.04}$ & $-2.06^{\pm0.93}$ & - & - & $0.18^{\pm0.03}$ \\

\multirow{2}{*}{dE} & \multirow{2}{*}{Single-S\'ersic} & $g$ & - & $-0.20^{\pm0.03}$ & $-0.32^{\pm0.49}$ & - & - & $0.24^{\pm0.03}$ \\
& & $i$ & - & $-0.19^{\pm0.03}$ & $-0.15^{\pm0.48}$ & - & - & $0.23^{\pm0.03}$ \\

\multirow{4}{*}{E} & \multirow{4}{*}{Single-S\'ersic} & \multirow{2}{*}{$g$} & - & $-0.37^{\pm0.04}$ & $-3.95^{\pm0.37}$ & - & - & $^{(c)}$$0.24^{\pm0.02}$\\
& & & $0.062^{\pm0.010}$ & $2.27^{\pm0.42}$ & $24.09^{\pm4.46 }$ & - & - & $0.21^{\pm0.02}$ \\
& & \multirow{2}{*}{$i$} & - & $-0.32^{\pm0.02}$ & $-3.16^{\pm0.46}$ & - & - & $0.23^{\pm0.02}$ \\
& & & $0.053^{\pm0.012}$ & $2.04^{\pm0.53}$ & $23.02^{\pm5.85}$ & - & - & $0.22^{\pm0.02}$ \\

\multirow{2}{*}{E} & \multirow{2}{*}{S\'ersic} & $g$ & - & $-0.28^{\pm0.01}$ & $-2.46^{\pm0.24}$ & - & - & $0.15^{\pm0.01}$\\
& & $i$ & - & $-0.24^{\pm0.02}$ & $-1.91^{\pm0.34}$ & - & - & $0.14^{\pm0.01}$ \\

\multirow{2}{*}{E to Sm} & \multirow{2}{*}{S\'ersic or bulge} & $g$ & $0.025^{\pm0.001}$ & $0.76^{\pm0.05}$ & $8.25^{\pm0.50}$ & - & $^{(d)}$$0.027^{\pm0.003}$ & $^{(d)}$$0.76^{\pm0.06}$\\
& & $i$ & $0.026^{\pm0.001}$ & $0.83^{\pm0.05}$ & $9.35^{\pm0.52}$ & - & $^{(d)}$$0.039^{\pm0.002}$ & $^{(d)}$$1.03^{\pm0.05}$ \\

\multirow{2}{*}{E and cD} & \multirow{2}{*}{Exponential}& $g$ & - & $-0.249^{\pm0.012}$ & $-1.12^{\pm0.27}$ & - & - & $0.18^{\pm0.01}$ \\
& & $i$ & - & $-0.243^{\pm0.011}$ & $-1.26^{\pm0.26}$ & - & - & $0.18^{\pm0.01}$ \\

\multirow{2}{*}{S0$^-$ to Scd} & \multirow{2}{*}{Disk} & $g$ & - & $-0.208^{\pm0.004}$ & $-0.43^{\pm0.08}$ & - & - & $0.175^{\pm0.004}$ \\
& & $i$ & - & $-0.214^{\pm0.004}$ & $-0.73^{\pm0.09}$ & - & - & $0.170^{\pm0.004}$ \\

\multirow{2}{*}{Sd to Im} & \multirow{2}{*}{Disk} & $g$ & - & $-0.140^{\pm0.007}$ & $0.98^{\pm0.14}$ & - & $0.019^{\pm0.009}$ & $0.59^{\pm0.16}$ \\
& & $i$ & - & $-0.151^{\pm0.007}$ & $0.71^{\pm0.15}$ & - & $0.018^{\pm0.010}$ & $0.58^{\pm0.18}$ \\

\multirow{2}{*}{E to Im} & \multirow{2}{*}{Exponential, disk or single-S\'ersic} & $g$ & $0.020^{\pm0.002}$ & $0.63^{\pm0.08}$ & $8.29^{\pm0.81}$ & $^{(d)}$$0.0043^{\pm0.0013}$ & $^{(d)}$$0.18^{\pm0.05}$ &  $^{(d)}$$2.17^{\pm0.45}$\\
& & $i$ & $0.032^{\pm0.002}$ & $1.17^{\pm0.09}$ & $14.34^{\pm0.91}$ & $^{(d)}$$0.013^{\pm0.001}$ & $^{(d)}$$0.55^{\pm0.06}$ &  $^{(d)}$$6.04^{\pm0.57}$\\

\hline
\end{tabular}}
\caption{Coefficients and associated uncertainties of the polynomial fits to the size-magnitude relations for the various components of the single profile fits or bulge and disk  decompositions for different EFIGI morphological subsamples, and dispersion in the residuals from these fits.}
\begin{list}{}{}
\item {\it Notes:}
\item {(a)} Coefficients of the polynomials obtained by ODR fitting of the profile effective radius (in log) versus magnitude, for a sample of EFIGI galaxies defined by their morphological types, in the listed band. Bulge (or S\'ersic) and disk (or exponential) refer to the corresponding component in the bulge and disk decomposition. When indicated, E, cD, dE and Im galaxies are modeled as single S\'ersic profiles (see \sct\ref{sct-methodo-srx}). 
\item {(b)} Coefficients of the polynomials providing the \rms dispersion (in dex) around the size-magnitude relation fit. 
\item {(c)} The center and $\chi^2$ of the Gaussian fit to the residuals of the elliptical size-luminosity relation appear in the descriptive text in \sct\ref{sct-results-binggeli}. 
\item {(d)} The Gaussian fits to the residuals of the bulge and disk 2nd degree polynomial size-luminosity relations are shown in the right panels of \fgs\ref{dispersion-Re} and \ref{dispersion-he}, respectively, and their centers and $\chi^2$ values appear in the descriptive texts in \sct\ref{sct-results-binggeli-bulge} and \ref{sct-results-disk-scaling-relations}, respectively.  
\end{list}
\label{tab-fits-size-lum}
\end{table*}

%When the dispersion around the fit varies significantly with magnitude, we compute it in several magnitude intervals and model the variations with a polynomial (as explained in \sct\ref{sct-results-binggeli-bulge} and shown in \fgs\ref{dispersion-Re} and \ref{dispersion-he}), whose coefficients are listed here.

\subsection{Multiband scaling relations for mock images     \label{sct-discussion-mock}}

In \fgs\ref{binggeli-cmap-BT} and \ref{binggeli-disk}, we have shown linear and polynomial fits for the size-luminosity relations of galaxy bulges and disks. These fits may play a key role in building realistic mock images as they allow one to deduct a size from a luminosity. Current softwares creating galaxy images such as Stuff \citep{Bertin-2009-skymaker} use size-luminosity relations from \cite{Binggeli-1984} for spheroids and \cite{de-Jong-Lacey-2000-spiral-galaxies-functions} for disks. These previous works have been very useful but remain limited by their sample size or the quality of the available data at that time. For instance, \cite{Binggeli-1984} uses data collected with photometric plates that suffer from nonlinear effects. Updating the size-luminosity relation thus appears as a key lever of action to improve the generation of mock images. 

We provide in Table~\ref{tab-fits-size-lum} the parameters of the size-luminosity relations obtained for the bulges and disks of various morphological types or type groupings of the EFIGI sample, in the $g$ and $i$ bands, for cD, E, dE and Im galaxies modeled as a single S\'ersic profile (see \sct\ref{sct-methodo-srx}), and for the bulges (or S\'ersic components) and the disks (or exponential components) of all types from E to Sm modeled as the sum of a S\'ersic profile and an exponential one (see \sct\ref{sct-methodo}). The relations in the $g$ band are those obtained and commented throughout the article, those in the $i$ band are presented here for the first time. All these fits are polynomial, either linear or of second degree, and were obtained using the ODR package (\sct\ref{sct-methodo-odr}), after correcting for the systematic trend in the relative error in the disk effective radii (\sct\ref{sct-methodo-uncertainties}). All values are given along with their associated uncertainty. 

We also calculated and provide in Table~\ref{tab-fits-size-lum} for each fit its residuals defined as the difference between the actual $\log R_e$ or $\log h_e$ values and the ones predicted from the polynomial fit using the absolute magnitude values $M_{\mathrm{bulge},g}$ or $M_{\mathrm{disk},g}$ for the considered subsample. If, when binning the residuals in magnitude intervals (see for example \fg\ref{dispersion-Re} in \sct\ref{sct-results-binggeli}, and \fg\ref{dispersion-he} in \sct\ref{sct-results-disk-scaling-relations}), we noticed a systematic variation in the \rms deviation of these residuals, we fitted it with a polynomial of degree 1 or 2, and list the corresponding coefficients in the columns labeled ``1st order'' and when appropriate ``2nd order'' of Table \ref{tab-fits-size-lum} under ``Dispersion in residual''. Otherwise, we list in column labeled ``0th order'' the \rms deviation over the entire absolute magnitude interval of the size-magnitude relation. The residual from the scaling relations that were fitted by a Gaussian are labeled in Table \ref{tab-fits-size-lum}, and the parameters can be retrieved in the text in the mentioned sections.

Table \ref{tab-fits-size-lum} shows that the size-luminosity relations in the $g$ and $i$ band have similar slopes within the error bars (for linear relations), as well as similar second degree coefficients, except for the second degree fit for the disks (or exponential components) of all Hubble types. In contrast, the zero-points of linear relations show expected shifts between the two bands, on the order of a few $\sigma$, similarly to the shifts of the $\mu_0$ versus $\log h$ relation among the $g$, $r$, and $i$ filters (see \sct\ref{sct-results-disk-scaling-relations}). The various coefficients of the dispersion in the residuals listed in Table \ref{tab-fits-size-lum} are also mostly within $1\sigma$ between the two bands.

As far a generating mock distributions of nearby galaxies is concerned, the ideal way to proceed would be to know the quadri-variate luminosity functions (in the observing band) of galaxies as a function of absolute luminosity and physical effective radii for both the bulges (or S\'ersic component) and the disks (or exponential component). These would then implicitly include the size-magnitude relations of both components, as well as the luminosity functions of the bulges and disks (or both components) of the various morphological types. In the following, we only mention bulge and disk, but every statement also applies for the S\'ersic or exponential components of E and Im types. 

\begin{table}[ht]
\centering
\resizebox{\linewidth}{!}{%
\begin{tabular}{ c c c c c c c c c c }
\hline\hline
Type & Component & Band & \multicolumn{2}{c}{Effective radius versus $B/T$} & Dispersion   \\
&  &  & \multicolumn{2}{c}{Linear fit$^{(a)}$} &   in residual \\
&  &  & Slope & Intercept &  \\
\hline
\multirow{4}{*}{E to S0a} & \multirow{2}{*}{S\'ersic or bulge} & $g$ & $3.17^{\pm0.17}$ & $4.34^{\pm0.06}$ & $0.55^{\pm0.02}$ \\
& & $i$ & $3.58^{\pm0.20}$ & $4.48^{\pm0.07}$ & $0.58^{\pm0.02}$ \\
& \multirow{2}{*}{Exponential or disk} & $g$ & $2.69^{\pm0.13}$ & $4.79^{\pm0.04}$ &  $0.59^{\pm0.02}$ \\
& & $i$ & $2.41^{\pm0.11}$ & $4.70^{\pm0.03}$ & $0.48^{\pm0.02}$ \\

\multirow{4}{*}{Sa to Sm$^{(b)}$} & \multirow{2}{*}{Bulge} & $g$ & $0.61^{\pm0.02}$ & $3.57^{\pm0.02}$ & $0.31^{\pm0.01}$ \\
& & $i$ & $0.67^{\pm0.02}$ & $3.57^{\pm0.02}$ & $0.30^{\pm0.01}$ \\
& \multirow{2}{*}{Disk} & $g$ & $0.22^{\pm0.01}$ & $4.27^{\pm0.02}$ & $0.220^{\pm0.007}$ \\
& & $i$ & $0.26^{\pm0.01}$ & $4.25^{\pm0.02}$ & $0.218^{\pm0.007}$ \\

\multirow{2}{*}{E to Sm} & Exponential-to-S\'ersic & $g$ & $-0.60^{\pm0.01}$ & $0.50^{\pm0.01}$ & $0.291^{\pm0.008}$ \\
& or disk-to-bulge ratio$^{(c)}$  & $i$ & $-0.62^{\pm0.02}$ & $0.47^{\pm0.01}$ & $0.338^{\pm0.008}$ \\

\hline
\end{tabular}}
\caption{Coefficients and associated uncertainties of the polynomial fits to the effective radii of both components of the profile decompositions in the $g$ and $i$ bands for different EFIGI morphological subsamples, versus the bulge-to-total light ratio $B/T$ in the corresponding band, and dispersion in the residuals from these fits.}
\begin{list}{}{}
\item {\it Notes:}
\item {(a)} Coefficients of the linear polynomials obtained by ODR fitting of the profile effective radius (in log) versus $B/T$ (in log), for a sample of EFIGI galaxies defined by their morphological type, in the listed band. Bulge (or S\'ersic) and disk (or exponential) refer to the corresponding component in the bulge and disk decomposition.
\item {(b)} The center and $\chi^2$ of the Gaussian fits to the residuals of  the bulge and disk effective radius versus $B/T$ relations appear in the descriptive text in \sct\ref{sct-results-size-evol-rad-vs-bt}.
\item {(c)} Coefficients of the linear fits to the effective radii ratio of the exponential to S\'ersic components for E types, and the disk to bulge components for all lenticular and spiral types.
\end{list}
\label{tab-fits-BT}
\end{table}

In the absence of this quadri-variate luminosity function, one should use the  luminosity function of galaxies of the various morphological types to draw the total luminosity distributions of these galaxies. Then using a probability distribution function (PDF) of $B/T$ (or an average value and a dispersion around it) for each morphological type would provide pairs of bulge and disk luminosities for each generated galaxy, and then yield effective radii using the respective size-magnitude relations for bulge and disks (and the dispersion in their residuals). Because the luminosity functions per morphological types are not yet known even for nearby galaxies, one could instead use the luminosity function of galaxies over all galaxy types. In that case it would be crucial to use a realistic PDF of $B/T$ over the whole magnitude range, as this will carry the information about the fractions of galaxies in the various morphological types. The $B/T$ PDF cannot be derived from the EFIGI sample as this sample with visual morphology was designed to have several hundreds of galaxies of each Hubble type, being therefore non representative of a complete magnitude limited sample of the nearby Universe. We intend to measure this $B/T$ PDF using the MorCat magnitude-limited survey of nearby galaxies, of which EFIGI is a subsample (in Quilley \& de Lapparent, \textit{in prep{.}}). On the contrary, the fact that all morphological types are nearly equally densely populated in EFIGI implies that the size-magnitude scaling relations can be considered as reliable for generating mock catalogs (if one excludes systematic deviations caused by some subpopulations of objects within some Hubble type). The larger MorCat sample that contains about $\sim4$ times more galaxies than EFIGI may however improve the statistics of the size-magnitude relations. 

We also provide in Table~\ref{tab-fits-BT} the relations between the $B/T$ of EFIGI galaxies and the effective radii of their bulges and disks, as well as their ratio, as derived in \sct\ref{sct-results-size-evol}, for the lenticular and spiral types in the $g$ and $i$ bands. The relations in the $g$ band are those obtained and commented throughout the article, those in the $i$ band are presented here for the first time. All these fits are polynomial, either linear or of second degree, and were obtained with the ODR package (\sct\ref{sct-methodo-odr}), after correcting for the systematic trend in the relative error in the disk effective radii with radii (\sct\ref{sct-methodo-uncertainties}). We list in column labeled ``Dispersion in residual'' of Table \ref{tab-fits-BT} the \rms deviation of the residuals from the corresponding size-$B/T$ relation. All values are given along with their associated uncertainty. For Sa to Sm disks, these residuals were fitted by a Gaussian (labeled as \textit{(b)}) and the parameters can be retrieved in the text in the mentioned sections.

These derived relations between the $B/T$ of galaxies and $h_e$, $R_e$, or their ratio $h_e/R_e$, could also be used to generate mock distributions. Using the galaxy total luminosity function of galaxies as well as a realistic distribution function of $B/T$ would here be necessary. As we pointed out in \sct\ref{sct-results-size-evol-ratio}, the behavior of the size versus $B/T$ relation for lenticular galaxies is such that generating separately a bulge and a disk can easily lead to nonphysical radii ratio, as $\sim1$ dex of both $h_e$ and $R_e$ is spanned across the same $B/T$ interval, but small bulge radii actually match small disk radii, and vice versa. For all galaxy types, we therefore advise to generate one of the two radii $R_e$ or $h_e$ using the scaling relations and their dispersion listed in Table \ref{tab-fits-BT}, and then to generate a ratio using the corresponding scaling relation and its dispersion in order to derive the other radius, so as to circumvent the issue of unrealistic combinations of bulges and disks. It should however be checked by deriving bulge and disk absolute magnitudes from each value of $B/T$, that the $R_e$ and $h_e$ yield size-magnitude relations that are consistent with those in \eqs\ref{eq-size-lum-bulge} and \ref{eq-size-lum-disk} for bulges and disks, respectively, of lenticular and spiral galaxies (and for S\'ersic or exponential components of E and Im types).

At last, we make a comparison of $h_e/R_e$ with the values obtained from Stuff \citep{Bertin-2009-skymaker} using the default configuration file, which uses the size-luminosity relations from \citet{Binggeli-1984} not only for ellipticals, but also for bulges of lenticulars and spirals. Stuff therefore applies for bright bulges the size-luminosity relation for E galaxies, and that for dE galaxies for intermediate and faint bulges, which are  shown in the right panel of \fg\ref{binggeli-cmap-BT} (\sct\ref{sct-results-binggeli-bulge}) to overestimate the bulge radii by factors of $\sim2-3$. Stuff also uses for all bulge and disk galaxy types the bi-variate luminosity function derived for disks of Sb to Sdm types by \citet{de-Jong-Lacey-2000-spiral-galaxies-functions}, which we showed in \fg\ref{histograms-he-mag} and \sct\ref{sct-results-disk-scaling-relations} to underestimate the frequency of large disk radii for Sb-Sdm and Sm-Im types at disk magnitudes $M_{i,\mathrm{disk}}\ge-20.5$, whereas this function overestimates the frequency of large disks for S0 to Sab types. Among the generated lists of synthetic Stuff galaxies, those with $0.05\le B/T_g\le0.6$ have $h_e/R_e\simeq0.5-6$, whereas we measure $h_e/R_e\simeq5-50$ in the right panel of \fg\ref{D-B-radius-ratio} for EFIGI galaxies within this ${B/T}_g$ interval. Also, synthetic Stuff galaxies with $0.6\le B/T_g\le1$  have $h_e/R_e\simeq0.2-2.5$, compared to $h_e/R_e\simeq2-8$ (\fg\ref{D-B-radius-ratio}). EFIGI galaxies therefore  appear $\sim1$ order of magnitude larger in their disk-to-bulge effective radius ratio than those generated by Stuff. Moreover, the bi-variate luminosity function of disks is modeled in the $i_C$ Cousins band by \citet{de-Jong-Lacey-2000-spiral-galaxies-functions}, whereas the reference pass-band in the default Stuff configuration is Couch $B_J$, which may contribute to the unrealistic values of $h_e/R_e$ in the default Stuff setup (see \fg\ref{histograms-he-mag} and \sct\ref{sct-results-disk-scaling-relations} for a comparison with EFIGI).

We emphasize that using for the bulge and disk scaling relations that originate from different studies based on different data and methodologies is likely to lead to discrepancies. To generate realistic galaxies as the sum of a disk and a bulge, consistent relations between these two components are needed, which require that bulges and disks of the same sample of galaxies are actually measured and consistently matched together. This is what we perform here with the EFIGI sample. 

\section{Conclusions}

In this article, we examine the relations between size, luminosity, and surface brightness for the bulges and disks of the 3106 weakly inclined nearby galaxies from the EFIGI morphological catalog. By controlled profile modeling with the SourceXtractor++ software, we performed bulge and disk decomposition of SDSS images simultaneously in the $g$, $r$, and $i$ bands to obtain the aforementioned parameters, using a S\'ersic and an exponential profile for the bulge and disk, respectively, for all types from E and cD to Sm. We also modeled E, cD, dE, and irregular (Im) galaxies as single S\'ersic profiles, with the Im being best fitted by nearly exponential profiles (see \sct\ref{sct-methodo-srx}).

All linear or higher order polynomial fits to the derived relations take uncertainties on both axes into account, using the total least square method available in the Orthogonal Distance Regression \textit{Python} package (see \sct\ref{sct-methodo-odr}). The quality of the fits also benefits from eliminating the overall systematic decrease in the relative uncertainties of the disk radii, caused by the survey selection effects in angular diameter versus distance for the sample galaxies. 

We first remeasured the relation between the mean effective surface brightness and the effective radius of E galaxies that was found by \cite{Kormendy-1977-II-kormendy-relation}, as well as those between the effective radius and the absolute magnitude established by \cite{Binggeli-1984} for E and dE galaxies, by modeling both  types as single S\'ersic profiles. With the improved statistics and profile modeling provided by the EFIGI sample, we obtained the linear fits of \eqs\ref{eq-kormendy}, \ref{eq-binggeli} and \ref{eq-binggeli-dE}. In the present analysis, we also adopted the nonconventional approach of applying the bulge and disk decompositions to E and cD types, as kinematic studies of ellipticals have shown evidence for stellar disk components, and a two-component profile allows one to model both the steep central light concentrations and large envelopes of cD galaxies. The S\'ersic and exponential components of E types can then be considered together with the bulges and disks, respectively, in both scaling relations. 

Regarding the relation between the surface brightness and the logarithm of the effective radius, we show that the slope of the linear \cite{Kormendy-1977-II-kormendy-relation} relation for E types is also valid for the S\'ersic components of E types as well as for bulges of S0$^-$ to Sb types. In contrast, there is a gradual departure from the relation toward the fainter and smaller bulges of later spiral types, which also have smaller bulge-to-total ratios ($B/T$) and S\'ersic indices (\fg\ref{kormendy-g-per-type-cmap-BT}). 

Regarding the relation between the logarithm of the effective radius and  magnitude, the so-called size-luminosity relation (always considered in a log-log plane), we measured a steeper relation than that obtained by \cite{Binggeli-1984} for E galaxies when using single-S\'ersic profiles, and show that a second degree polynomial provides a better size estimation for both the faint and bright ends of the elliptical population. Flatter size-luminosity relations than for EFIGI E types were also derived for the EFIGI cD and dE types, also using single-S\'ersic profiles (left panel of \fg\ref{binggeli-cmap-BT}), with the latter types nevertheless yielding a steeper relation than that measured by \citet{Binggeli-1984}. 

It is remarkable that the S\'ersic components of E galaxies as well as bulges of lenticular and spiral galaxies can be considered altogether rather than for each morphological type separately, as they form a continuous size-luminosity relation, along which the $B/T$ ratio varies from measured values of $\lesssim 1$ to $0.01$ (see right panel of \fg\ref{binggeli-cmap-BT}). Moreover, the bulge size-luminosity relation is convex, with a curvature leading to $\sim2$ dex larger radii than interpolated from the E galaxies modeled as a single-S\'ersic profile (and $\sim1$ dex larger radii than interpolated from the E S\'ersic components) at the faintest bulge magnitudes ($\sim-15$ in $g$ band). Bulges have a surprising Gaussian distribution of the logarithm of their effective radii around this relation (see \fg\ref{dispersion-Re}), whose rms deviation increases from 0.17 dex to 0.33 dex from bright to faint bulges (compared to 0.21 dex for E types at the bright end).

We highlight that both scaling relations are projections of the planar relation (at null redshifts) between the surface brightnesses, radii, and absolute magnitudes in 3D space (\fg\ref{3D-figure}). We also show that the position of bulges within this plane is driven by the bulge-to-total ratio $B/T$, and by the S\'ersic index $n_{\text{S\'ersic}}$ to a lesser extent.

We interpret these changes in the bulge characteristics across the Hubble sequence as a progressive and continuous transition from ``pseudo-bulges'' to ``classical bulges.'' This gradual shift is observed in terms of $B/T$, $n_{\text{S\'ersic}}$ and the position within the $\langle\mu\rangle_e - M - R_e$ 3D space, and equivalently in the location along both of its projections that are studied. At one end of this transition lie the small and faint bulges with rather smooth profiles that account for a very small fraction of the total luminosity, which we interpret as pseudo-bulges, and they are found in late-type spirals. At the other end, there are larger and brighter bulges with steeper profiles that account for around half the total galaxy light, which we interpret as classical bulges, and they are found in lenticulars as well as early spirals.

A similar study was then performed on the disks of the EFIGI galaxies and we derived analogous scaling relations for the surface brightness versus effective radius, and the size-luminosity relation (\fgs\ref{de-jong-g-per-type} and \ref{binggeli-disk}). The latter relation for EFIGI lenticular and spiral disks (as well as the exponential components of E galaxies) is also convex, with $0.5$ dex larger radii of giant disks at $\sim-16$ absolute magnitude in the $g$ band than interpolated from the bright lenticular disks, and with again a Gaussian dispersion that increases from 0.18 dex at the bright end to 0.32 dex at the faint end. The curvature of the size-luminosity relation for disks is such that disks become larger and denser as they grow in luminosity from Im galaxies and very late spirals to Scd types. Then for types earlier than Scd, the average variation in effective radius with magnitude occurs at a constant mean central surface brightness.

The second degree polynomial fits to the size-luminosity relations (in log-log) that we derived for the S\'ersic components of elliptical galaxies and the bulges of all types of lenticulars and spirals, as well as for the exponential components of the elliptical, the single S\'ersic profiles of irregulars, and the disks of all types of lenticulars and spirals all appear critical in deriving a physically meaningful size for all of these components of galaxies. Indeed, single linear fits would lead to a systematic underestimation of effective radius in some intervals of absolute magnitude, when generating simulated galaxy parameters. 

We further discuss the changes in bulge and disk structure as galaxies merge and evolve along the Hubble sequence in terms of bulge and disk volume and surface density, respectively (\sct\ref{sct-discussion-diffuse}). The size-magnitude relations derived for this study did indeed allow us to estimate the absolute magnitudes and $B/T$ critical values at which a bulge or disk luminosity scale as either $R^2$ or $R^3$. From our derived size-magnitude relations, we detected a change of regime from increasing volume and surface density of the small bulges to more prominent bulges that are more diffuse as they grow in size, which we interpret as related to the change of dynamics from pseudo to classical. We propose that the change from pseudo- to classical bulges occurs in the interval $-19.1 < M_{\mathrm{bulge},g} < -17.8$ (corresponding to $0.08 \lesssim B/T \lesssim 0.15$), in which the surface density of disky bulges increases with increasing radius, but the volume density of spheroidal bulges decreases. We show that at the extreme end of this bulge decreasing volume density trend lie elliptical galaxies, which become more and more diffuse as they grow in size; they are therefore not scale invariant. Regarding the disks, we show that they grow with luminosity both in size and surface density in late-type spirals, whereas they grow at a constant surface brightness on average in earlier spiral types and lenticulars.

We also obtained the unprecedented result that the bulge and disk effective radii vary as power laws of the galaxy $B/T$, which is a key parameter characterizing galaxies along the Hubble sequence \citep{Quilley-2022-bimodality}. The $B/T$ ratio grows along the reverse Hubble sequence (\sct\ref{sct-results-size-evol-bt}), together with the bulge effective radius, whereas if the disk effective radius grows as the irregulars and late spirals merge to form earlier spirals, it stagnates or even decreases for lenticulars (\sct\ref{sct-results-size-evol-rad-vs-type}). We provide the linear scaling relations between the logarithm of the effective radii of both the bulges and the disks, as well as their ratio, as a function of $B/T$, with 0.31 dex and 0.22 dex dispersion for the bulge and disk, respectively. In particular, we show that there is a significant increase in bulge radii for increasing $B/T$ across all spiral types (from Sm to Sa types), as well as an even steeper increase in the bulge radii for lenticular galaxies. For the spiral types, the increase in the disk radii is flatter than for the bulges; however, for lenticulars, both components have a much steeper and similar rate of growth. It is noticeable that the Sa types, which mark the transition between the scaling of spirals and lenticular radii with $B/T$, correspond to the core of the Green Valley \citep{Quilley-2022-bimodality}.

These scaling relations propagate into a single scaling relation for the ratio of disk-to-bulge effective radii, $h_e/R_e$, across two orders of magnitude in $B/T$ for all types. The mean ratio of radii ranges from $\sim4$ to $6$ for lenticulars and ellipticals to $\sim14$ to $16$ for Sbc to Scd types. A larger statistical sample, as well as higher angular resolution images (in order to resolve their small bulges) are needed for measuring the bulge radii of Sd to Sm types. There is a remarkably small dispersion in both $B/T$ (between 0.1 and 0.2 dex) and $h_e/R_e$ (between 0.2 and 0.3 dex) for the lenticular galaxies, indicating that lenticulars with small bulges also have a small disk and vice versa.

We emphasize that in the present article we confirm and complement the galaxy evolution scenario inferred from the color-mass diagram of EFIGI galaxies in \cite{Quilley-2022-bimodality}, in which irregulars and late-type spirals merge to form more massive and earlier-type spirals, which then become passive as they merge into lenticulars and ellipticals. These mergers, along with the consumption of the gas reservoir of galaxies, explain the growth in stellar mass across the full reversed Hubble sequence, from the small irregulars to the giant elliptical galaxies, with a three order of magnitude increase in stellar mass or absolute luminosity across EFIGI spiral galaxies, and the corresponding three order of magnitude increase in their disk luminosity and mass -- at the bright end of the Blue Cloud, where $B/T\lesssim0.2$ \citep{Quilley-2022-bimodality}. Our proposed scenario of galaxy evolution in the previous analysis is complemented by the present results that galaxy growth occurs with a moderate increase in disk and total galaxy size by a factor of $\sim7$, but with a massive increase in $B/T$ by a factor of $\sim100$ and of the bulge radius by a factor of $\sim20$ along the Blue Cloud. 

Therefore, as galaxies merge along the spiral sequence, there are three parallel processes occurring: (1) an $\sim1$ order of magnitude increase in the disk effective radius (right panels of \fg\ref{B-and-D-radius-distrib} or \fg\ref{B-and-D-radius-fits} in \sct\ref{sct-results-size-evol}); (2) a redistribution of matter over the disks that boosts their projected central surface brightness by $\sim4$ magnitudes (\fg\ref{binggeli-disk} in \sct\ref{sct-results-disk-scaling-relations}); (3) a factor of $\sim3$ increase in the bulge effective radius (left panels of \fg\ref{B-and-D-radius-distrib} or \fg\ref{B-and-D-radius-fits} in \sct\ref{sct-results-size-evol}), along with an $\sim2$ order of magnitude growth in $B/T$, and an $\sim5$ order of magnitude growth in bulge absolute luminosity (\fg\ref{binggeli-cmap-BT} in \fg\ref{sct-results-binggeli}). The various phases of bulge and disk growth occurring at different rates across morphological types are likely to be intertwined and to result from the different dynamics between both components.

Finally, all the relations derived here are useful to build more realistic mock catalogs. By providing the relations between size and luminosity, and further relating them to Hubble type and $B/T$, it is possible to derive -- from a given luminosity of a bulge or a disk -- a set of parameters describing its light profile, and thus leading to improved simulated images. We provide in Table \ref{tab-fits-size-lum} the parameters of the power-law fits to the size-luminosity relations performed for the S\'ersic or bulge components, as well as the exponential or disk components of E, cD, dE, lenticular, and spiral galaxies.  In Table \ref{tab-fits-BT}, the parameters of the power-law fits to the bulge and disk effective radii as a function of $B/T$ are also listed. In both Tables, the dispersion in the residuals from the fits are also provided, whether constant or variable along the relations.

As we have shown in \cite{Quilley-2022-bimodality} and in the present analysis via the coherent variations of the sizes and luminosities of bulges and disks of galaxies along the Hubble classification scheme, we intend to explore in a forthcoming analysis a physical and quantitative characterization of the morphological sequence in terms of these various bulge and disk parameters, using a complete magnitude-limited sample of the nearby Universe (Quilley \& de Lapparent, \textit{in prep{.}}), the MorCat completion of the EFIGI sample to $g\le15.5$. We also plan to use MorCat to calculate the quadri-variate luminosity functions of the bulges and the disks of galaxies as a function of absolute luminosities and physical effective radii of both types of components, as these are essential for generating realistic mock images and catalogs of galaxies in the nearby Universe. Generating mock catalogs at larger distances will require one to measure the evolution of the sizes of bulges and disks with redshift for all morphological types that may exist.

\section{Acknowledgments}

We are very grateful to Emmanuel Bertin for initiating and leading the EFIGI project, and for maintaining the computer cluster on which all calculations were performed. We also thank the anonymous referee for their insightful comments and suggestions.

This research made use of {\tt SourceXtractor++}\footnote{\url{https://github.com/astrorama/SourceXtractorPlusPlus}}, an open source software package developed for the Euclid satellite project. This research made use of the VizieR catalog access tool, CDS, Strasbourg, France (DOI: 10.26093/cds/vizier). The original description of the VizieR service was published in A$\&$AS 143, 23. This research also made use of NASA's Astrophysics Data System.

This work is based on observations made with the NASA Galaxy Evolution Explorer. GALEX is operated for NASA by the California Institute of Technology under NASA contract NAS5-98034. Funding for the SDSS and SDSS-II has been provided by the Alfred P. Sloan Foundation, the Participating Institutions, the National Science Foundation, the U.S. Department of Energy, the National Aeronautics and Space Administration, the Japanese Monbukagakusho, the Max Planck Society, and the Higher Education Funding Council for England. The SDSS Web Site is http://www.sdss.org/.
The SDSS is managed by the Astrophysical Research Consortium for the Participating Institutions. The Participating Institutions are the American Museum of Natural History, Astrophysical Institute Potsdam, University of Basel, University of Cambridge, Case Western Reserve University, University of Chicago, Drexel University, Fermilab, the Institute for Advanced Study, the Japan Participation Group, Johns Hopkins University, the Joint Institute for Nuclear Astrophysics, the Kavli Institute for Particle Astrophysics and Cosmology, the Korean Scientist Group, the Chinese Academy of Sciences (LAMOST), Los Alamos National Laboratory, the Max-Planck-Institute for Astronomy (MPIA), the Max-Planck-Institute for Astrophysics (MPA), New Mexico State University, Ohio State University, University of Pittsburgh, University of Portsmouth, Princeton University, the United States Naval Observatory, and the University of Washington.

\bibliographystyle{aa}
\bibliography{biblio}

\end{document}